\documentclass[letterpaper,11pt]{article}
\pdfoutput=1 

\usepackage{jheppub} \usepackage{amsmath,amssymb,color,hyperref,xcolor,slashed,psfrag}

\usepackage[T1]{fontenc} \usepackage{xspace}
\usepackage[separate-uncertainty = true,multi-part-units=single, range-units=single]{siunitx}
\usepackage[parfill]{parskip} \usepackage{microtype}
\usepackage{csquotes}
\usepackage[capitalize]{cleveref}

\usepackage{subcaption}

\preprint{\begin{minipage}[t]{8cm}\begin{flushright}
      FERMILAB-PUB-23-028-T,\\
      IPPP/23/05
\end{flushright}\end{minipage}}

\title{Jet-veto resummation at \NNNLL$_\text{p}$+\NNLO{} in boson production processes}

\author[a]{John M. Campbell,}
\emailAdd{johnmc@fnal.gov}

\author[b]{R. Keith Ellis,}
\emailAdd{keith.ellis@durham.ac.uk}

\author[c]{Tobias Neumann,}
\emailAdd{tneumann@bnl.gov}

\author[d]{Satyajit Seth}
\emailAdd{seth@prl.res.in}

\affiliation[a]{Fermilab, PO Box 500, Batavia IL 60510-5011, USA}
\affiliation[b]{Institute for Particle Physics Phenomenology, Durham University, Durham, DH1 3LE, UK}
\affiliation[c]{Department of Physics, Brookhaven National Laboratory, Upton, New York 11973, USA}
\affiliation[d]{Physical Research Laboratory, Navrangpura, Ahmedabad - 380009, India}
\date{\today}

\usepackage{scalefnt}

\def\etacut{y_\text{cut}}
\def\lnz{\ln(-4z-\img 0)}

\def\cB{{\cal B}}

\newcommand{\LLR}{\text{LL$_R$}\xspace}
\newcommand{\ep}{\epsilon}

\newcommand{\ord}[1]{{\mathcal O}(#1)}

\newcommand{\Lperp}{L_\perp}
\newcommand{\cusp}{\mathrm{cusp}}

\newcommand{\img}{\mathrm{i}}
\newcommand{\nn}{\nonumber}
\newcommand{\MSbar}{$\overline{\text{MS}}$\xspace}
\newcommand{\as}{\frac{\alpha_s}{4 \pi}}
\def\beq{\begin{equation}}
\def\eeq{\end{equation}}
\def\Pone{P^{(1)}}
\def\Rone{R^{(1)}}
\def\Ptwo{P^{(2)}}

\newcommand{\beqn}{\begin{eqnarray}}
\newcommand{\eeqn}{\end{eqnarray}}

\newcommand{\veto}{{\rm veto}}
\newcommand{\Ibar}{\bar{I}}
\newcommand{\pTveto}{p_T^{\hspace{-0.3mm}\rm veto}\hspace{-0.3mm}}
\newcommand{\ptveto}{\pTveto}
\newcommand{\doneveto}{d_1^{\veto}}
\newcommand{\dtwoveto}{d_2^{\veto}}
\newcommand{\dthreeveto}{d_3^{\veto}}
\newcommand{\dfourveto}{d_4^{\veto}}
\newcommand{\abbrev}{\scalefont{.9}}
\newcommand{\MCFM}{\text{\abbrev MCFM}}

\newcommand{\CuTeMCFM}{\text{\abbrev CuTe-MCFM}}
\newcommand{\SCET}{\text{\abbrev SCET}}

\newcommand{\NLL}{\text{\abbrev NLL}}

\newcommand{\NNLL}{\text{\abbrev NNLL}}

\newcommand{\NNNLL}{\text{\abbrev N$^3$LL}}
\newcommand{\NNNLLpart}{\text{\abbrev N$^3$LL$_\text{p}$}}

\newcommand{\NNLO}{\text{\abbrev NNLO}}

\newcommand{\NLO}{\text{\abbrev NLO}}

\newcommand{\RadISH}{\text{\abbrev RadISH}}
\newcommand{\JetVHeto}{\text{\abbrev JetVHeto}}
\newcommand{\LHC}{\text{\abbrev LHC}}
\newcommand{\ATLAS}{\text{\abbrev ATLAS}}
\newcommand{\CMS}{\text{\abbrev CMS}}
\newcommand{\QCD}{\text{\abbrev QCD}}
\newcommand{\JVE}{\text{\abbrev JVE}}
\newcommand{\RG}{\text{\abbrev RG}}
\newcommand{\RGE}{\text{\abbrev RGE}}
\newcommand{\PDF}{\text{\abbrev PDF}}

\renewcommand{\d}{\mathrm{d}}

\newcounter{notecount}

\abstract{Vetoing energetic jet activity is a crucial tool for suppressing
  backgrounds and enabling new physics searches at the \LHC{}, but the
  introduction of a veto scale can introduce large logarithms that
  may need to be resummed.  We present an implementation of jet-veto
  resummation for color-singlet processes at the level of \NNNLLpart{}
  matched to fixed-order \NNLO{} predictions.  Our public code \MCFM{}
  allows for predictions of a single boson, such as $Z/\gamma^*$,
  $W^\pm$ or $H$, or with a pair of vector bosons, such as $W^+W^-$,
  $W^\pm Z$ or $ZZ$.  The implementation relies on recent calculations
  of the soft and beam functions in the presence of a jet veto over
  all rapidities, with jets defined using a sequential recombination
  algorithm with jet radius $R$.  However one of the ingredients that
  is required to reach full \NNNLL{} accuracy is only known
  approximately, hence \NNNLLpart{}.  We describe in detail our
  formalism and compare with previous public codes that operate at the
  level of \NNLL{}.  Our higher-order predictions improve
  significantly upon \NNLL{} calculations by reducing theoretical
  uncertainties. We demonstrate this by comparing our predictions with
  \ATLAS{} and \CMS{} results.}

\begin{document} 
	
\setcounter{tocdepth}{2}
\maketitle

\section{Introduction}

Jet vetoing is a crucial technique in particle physics that is used primarily to suppress 
backgrounds in 
processes involving the production of $W^+W^-$ final states (e.g. directly or in Higgs decays). 
By identifying and removing events that contain energetic hadronic jets (vetoing), the impact of 
the dominant top-quark pair production background is reduced. The concrete jet-veto implementation 
depends 
on factors such 
as the jet algorithm and its parameters, as well as the kinematic selection 
cuts applied to the identified jets. For \LHC{} analyses, the most common jet vetoing scheme is to 
impose a maximum transverse momentum cut $\ptveto{}$ on anti-$k_T$ jets.

The jet veto scale $\ptveto{}$ can induce large logarithms if it is smaller than the hard 
process scale $Q$, which then mandates resummation. 
In this paper we describe a coherent implementation of jet veto resummation
in processes involving the production of a color-singlet boson ($W, Z/\gamma^*$ and $H$ bosons) or a 
pair of bosons ($ZZ$, $W^\pm Z$, and $W^+W^-$). 
Our resummation operates at the level of \NNNLLpart{}\footnote{The 
	last missing ingredient for \NNNLL{} resummation is the exact $\dthreeveto$ (the three-loop 
	rapidity anomalous 
	dimension) which we approximate and take into account with an uncertainty estimate. We discuss 
	this 
	in detail in the subsequent section.} matched to fixed order \NNLO{}.

We build on the pioneering work of previous studies, which have demonstrated the 
effectiveness of resummation methods for a jet veto 
\cite{Berger:2010xi,Stewart:2010tn,Becher:2012qa,Banfi:2012yh,Banfi:2012jm}. 
General purpose implementations include a numerical approach to resummation at 
\NNLO{}+\NNLL{}  \cite{Kallweit:2020gva,Re:2021con} and an automated approach to jet veto studies 
at \NLO{}+\NNLL{} \cite{Becher:2014aya}. Publicly available codes operating at \NNLL{} and 
addressing the same issue are, \JetVHeto{}~\cite{JetHveto}, the code {\abbrev 
MCFM-RE}~\cite{MCFM-RE} which is derivative of both \MCFM{} and \JetVHeto, and {\abbrev 
MATRIX+RadISH}~\cite{Matrix+Radish}. Both \JetVHeto{} and \RadISH{} implement the same analytic 
resummation formula of ref.~\cite{Banfi:2012jm}.

Our research extends and improves upon these earlier results through detailed phenomenological 
studies of specific final states, including Higgs boson production 
\cite{Banfi:2012jm,Becher:2013xia,Stewart:2013faa,Banfi:2015pju}, $W^+W^-$ production 
\cite{Dawson:2016ysj,Arpino:2019fmo}, and $ZZ$ and $W^\pm Z$ production \cite{Wang:2015mvz}. 
Another important aspect of our study is the performance of the resummation at \NNNLLpart{} 
accuracy, which has not always been the case in previous work.
We also describe our approximation of the missing $\dthreeveto$ that 
would be necessary to reach full 
\NNNLL{} accuracy. Finally, we include our results in \MCFM{}, a publicly distributed code, which 
allows users to easily perform studies in practice.

Resummation of jet-veto logarithms has a close relationship with the resummation of transverse 
momentum logarithms. In the latter, one is interested in transverse momenta all the way down to 
zero $p_T$, so the logarithms can be larger than in jet-veto processes where $\pTveto$ in the 
range \SIrange{25}{30}{\GeV} is used experimentally. In this paper we explore which jet-veto 
processes actually 
require resummation at these values of $\pTveto$, supply the best predictions for those processes 
where 
it is warranted, and confront our theoretical predictions with experimental data where it is 
available.

In \cref{sec:factorization} we discuss the jet-veto factorization theorem including its ingredients 
that result in the resummation. 
We describe our setup for phenomenology including our uncertainty procedure in \cref{sec:setup},
compare with the public code \JetVHeto{} in \cref{sec:jetvheto}, and study the phenomenological 
implications for a
wide range of processes in \cref{sec:pheno}. We conclude in \cref{sec:conclusions}.

\section{Jet-veto factorization and resummation}
\label{sec:factorization}

We consider processes where jets have been defined using
sequential recombination jet algorithms \cite{Salam:2010nqg} with distance measure
\begin{equation}\label{jetdef}
	d_{ij} = \mbox{min}(k_{Ti}^{2n},k_{Tj}^{2n})\,
	\frac{\Delta y_{ij}^2+\Delta\phi_{ij}^2}{R^2} \,, \qquad
	d_{iB} = k_{Ti}^{2n} \,,
\end{equation}
where the choice $n=-1$ is the anti-$k_T$ algorithm \cite{Cacciari:2008gp},
$n=0$ is the Cambridge-Aachen algorithm \cite{Dokshitzer:1997in,Wobisch:1998wt},
and $n=1$ is the $k_T$ algorithm \cite{Catani:1993hr,Ellis:1993tq}.
$k_{Ti}$ denotes the transverse momentum of (pseudo-)particle $i$ with respect to the beam 
direction,
and $\Delta y_{ij}$ and $\Delta\phi_{ij}$ are the rapidity and azimuthal angle differences of
(pseudo-)particles $i$ and $j$.

To describe the resummation method we focus on the simplest case of quark-antiquark induced 
Drell-Yan production of a lepton pair of invariant mass $Q$ and rapidity $y$. The case of gluon initiated processes is 
structurally the same, but with different ingredients that we give below and in the appendices.
In the presence of a jet veto over all rapidities we have a
factorization formula~\cite{Becher:2012qa,Becher:2013xia,Stewart:2013faa},
\begin{align}
	\frac{d^2\sigma(\pTveto)}{dQ^2 dy } &= \frac{d \sigma_0}{dQ^2} \,\left| C^V(-Q^2,\mu) 
	\right|^2 \nonumber \\
	&\times \Big[ \cB_q(\xi_1,Q,\pTveto,R,\mu, \nu)\,\cB_{\bar{q}}(\xi_2,Q,\pTveto,R, \mu, \nu)\,  
	{\cal S}(\pTveto,R,\mu,\nu) \Big] +\mathcal{O}\left(\frac{\pTveto}{Q}\right)
\label{eqn:BBS}
\end{align}
where $\xi_{1,2}=(Q/\sqrt{s})\,e^{\pm y}$ and,
\begin{equation}\label{lowestorder}
	\frac{d\sigma_0}{dQ^2} = \frac{4 \pi \alpha^2}{3 N_c Q^2 s} \, .
\end{equation}
In this equation $C^V$ is a matching coefficient whose square is the hard coefficient function
that corrects the lowest order cross-section, see
\cref{lowestorder}. $\cB_q$ and $\cB_{\bar{q}}$ are the quark beam functions which describe
the emission of radiation collinear to the two beam directions in the presence of a jet veto, and 
${\cal S}$
describes the emission of soft radiation in the presence of a jet veto.
The quantity $\nu$ is a supplementary scale necessitated by the rapidity 
divergences present
in beam and soft functions. The main process-independent ingredients are the beam and soft 
functions for both incoming quarks and gluons which have 
been published recently at the two-loop level \cite{Abreu:2022zgo,Abreu:2022sdc}.
The hard function is process specific. We have used the existing two-loop fixed order implementations
in \MCFM{}. 

Overall the factorization theorem achieves a separation of scales. The hard function contains 
logarithms of the ratio $Q^2/\mu^2$, which can be minimized by setting $\mu^2 = \mu_h^2 \sim Q^2$. 
However, inside the beam and soft functions, it is natural to choose $\mu = \ptveto$ to avoid large 
logarithms. The resummation of large logarithms is achieved by choosing $\mu \sim Q$ in the hard 
function and evolving it down to the resummation scale $\mu \sim \ptveto$ using the renormalization group (\RG{}). For the 
hard function the evolution is solved analytically, see \cref{sec:rengroup}.

In \RG{}-improved power counting the logarithms $L_\perp=2\log(\mu_h/\ptveto)$,
where $\mu_h$ is of order $Q$, are assumed to be of 
order $1/\alpha_s$.
With this definition the counting of powers of $\alpha_s$ and of the large logarithm $L_\perp$ is 
shown in \cref{tab:counting}. 
\begin{table}
	\begin{center}
		\caption{Counting of orders in the resummation, adapted from
			ref.~\cite{Becher:2006mr}. The second column indicates the nominal order when counting 
			$L_\perp \sim 1/\alpha_s$. The third column states which logarithms are included. The 
			last three columns show the necessary additional anomalous dimensions and hard function corrections in each successive order.
                        The requisite anomalous dimensions are provided in \cref{sec:betaanom}.\label{tab:counting}}
		\begin{tabular}{cccccc}
			Approximation &  Nominal order & Accuracy $\sim \alpha_s^n L_\perp^k$ & 
			$\Gamma_{\text{cusp}}$ &
			$\gamma_{\text{coll.}}$ & $H$ \\ \hline
			LL              & $\alpha_s^{-1}$ & $2n \geq k \geq n+1$ \hfill  & $\Gamma_0$ & tree  & 
			tree \\
			NLL+LO          & $\alpha_s^0$    & $2n \geq k \geq n$ \hfill    & $\Gamma_1$, & $\gamma_0$ & tree \\
			N${}^2$LL+NLO   & $\alpha_s^1$    & $2n \geq k \geq \text{max}(n-1,0)$ & $\Gamma_2$ & 
			$\gamma_1$ & 1-loop \\
			N${}^3$LL +NNLO & $\alpha_s^2$    & $2n \geq k \geq \text{max}(n-2,0)$ & $\Gamma_3$ & 
			$\gamma_2$ & 2-loop \\
		\end{tabular}
		
	\end{center}
\end{table}
The non-logarithmic terms that the resummation does not provide are easily accounted for by adding 
the matching corrections. The matching corrections are a finite contribution and add the effect of 
fixed-order corrections while removing the logarithmic overlap through a fixed-order expansion of 
the resummation.

\subsubsection{Soft function}
The jet veto soft function has been
calculated using an exponential regulator~\cite{Li:2016axz} in Ref.~\cite{Abreu:2022sdc}.
The calculation is divided into the sum of the soft function for a reference
observable and a correction factor,
\beq
S(\pTveto,R,\mu,\nu)=S_\perp(\pTveto,\mu,\nu) + \Delta S(\pTveto,R,\mu,\nu)\,.
\eeq
In Ref.~\cite{Abreu:2022sdc} the reference observable is the transverse momentum of the color 
singlet system denoted by $S_\perp$.
$S_\perp$ can be derived from the expression given
in Refs.~\cite{Vladimirov:2016dll,Li:2016ctv} after performing the Fourier transform to momentum 
space (see, for
instance, the rules given in Table 1 of Ref.~\cite{Billis:2019vxg}).
$\Delta S$ depends on the jet algorithm and contributes for two or more emissions.
It thus depends only on double real emission diagrams.

\subsubsection{Refactorization and reduced beam functions}
For consistency with the transverse momentum resummation framework in \CuTeMCFM{} 
\cite{Becher:2020ugp} we cast the 
factorization theorem in terms of the collinear anomaly framework. In this framework the rapidity 
logarithms are exponentiated directly instead of resummed by solving rapidity \RG{} equations 
\cite{Chiu:2012ir,Chiu:2011qc}. For this we 
rewrite the square bracket in \cref{eqn:BBS} as follows,
\begin{align}
	& \cB_q(\xi_1,Q,\pTveto,R,\mu, \nu)\,\cB_{\bar{q}}(\xi_2,Q,\pTveto,R,\mu, \nu)
	{\cal S}(\pTveto,R,\mu,\nu) \nonumber \\
	&= \left( \frac{Q}{\pTveto} \right)^{-2F_{qq}(\pTveto,R,\mu)} e^{2h^{F}(\pTveto,\mu)}\,
	\bar B_q(\xi_1,\pTveto,R,\mu)\,\bar B_{\bar{q}}(\xi_2,\pTveto,R,\mu) \,.
	\label{eq:BNR}
\end{align}
The $\nu$ dependence vanishes in this combination of beam and soft functions.

We have factored out $e^{h^{F/A}(\pTveto,\mu)}$ from each beam function, resulting in the 
reduced beam functions $\bar{B}$. By construction 
$h^{F/A}$ are solutions of the \RGE{} equation,
\begin{equation}\label{eq:RGforH}
	\frac{d}{d\ln\mu}\,h^{F/A}(\pTveto,\mu)= 2\Gamma_{\rm cusp}^{F/A}(\mu)\,\ln\frac{\mu}{\pTveto} 
	- 2\gamma^{q/g}(\mu)\, ,
\end{equation}
with boundary condition $h^{F/A}(\mu,\mu)=0$.
The superscripts $F$ or $A$ signify whether the color 
is treated in the fundamental ($F$) or adjoint ($A$) representation, corresponding to a quark 
initiated process or a gluon initiated process, respectively.
The exact form of $h^{F/A}(\pTveto,\mu)$, determined by solving Eq.~(\ref{eq:RGforH}), is given in 
Appendix~\ref{exponenth}.
In terms of the reduced beam functions the jet-vetoed cross-section is now given by,
\begin{equation}\label{eq:Refactorized}
	\frac{d^2\sigma(\pTveto)}{dQ^2 dy } = \frac{d \sigma_0}{dQ^2} \,
	\bar H(Q,\mu,\pTveto)
	\bar B_q(\xi_1,\pTveto,R,\mu)\,\bar B_{\bar{q}}(\xi_2,\pTveto,R,\mu) +\mathcal{O}(\pTveto/Q) \,,
\end{equation}
The choice of $h^{F/A}$ in \cref{eq:RGforH} divides \cref{eqn:BBS} 
into two separately RG invariant pieces, the product of the two reduced beam functions ($\bar B_q \, \bar B_{\bar{q}}$), and the hard function, ($\bar H$)
\begin{eqnarray}\label{Hbardef}
	\bar H(Q,\mu,\pTveto) = \left| C^V(-Q^2,\mu) \right|^2 \left( \frac{Q}{\pTveto} 
	\right)^{-2F_{qq}(\pTveto,R,\mu)}\,
	e^{2h^F(\pTveto,\mu)} \, .
\end{eqnarray}
For quark-initiated processes the functions $C^V$ and $F_{qq}$ obey the following \RG{} equations.
\begin{align}
\label{eq:RGforC}
\frac{d}{d\ln\mu}\,C^V(-Q^2,\mu) &= \Big[\Gamma_{\rm cusp}^F(\mu)\,\ln\frac{-Q^2}{\mu^2} + 
2\gamma^q(\mu) \Big] C^V(-Q^2,\mu) \,,\\
\label{eq:RGforF}
\frac{d }{d \ln \mu} F_{qq}(\pTveto,R,\mu) &=  2 \Gamma_\cusp^F(\mu) \,.
\end{align}
\cref{eq:RGforC,eq:RGforF} are structurally the same for the gluon case with different anomalous dimensions.

The function $\bar{H}$ is \RG{} invariant due to the \RGE{}'s satisfied by $C^V$ and 
$F_{qq}$ and $h_F$:
\beq
\frac{d}{d\mu} \bar{H}(Q,\mu,\pTveto)=\mathcal{O}(\alpha_s^3)\,.\nonumber
\eeq
Consequently, the remaining product of reduced beam functions is also \RG{} invariant, up to the 
order calculated. In our 
case, 
\beq \label{eq:RGinvariantBprod}
\frac{d}{d\mu}  \bar B_q(\xi_1,\pTveto,R,\mu)\,\bar B_{\bar{q}}(\xi_2,\pTveto,R,\mu) = 
\mathcal{O}(\alpha_s^3) \,.
\eeq
The confirmation of \cref{eq:RGinvariantBprod} and the confirmation of the $R$-dependence of the collinear anomaly
given in the next section are two simple checks of the results of Refs.~\cite{Abreu:2022zgo,Abreu:2022sdc}. Full details
of the formulas needed to perform this check are given in Appendix~\ref{beam_function_ingredients}.

If the scale $\pTveto$ is in the perturbative domain, 
the reduced beam function can be written in terms of the matching kernels $\bar{I}$ as
\begin{equation*}\label{pdfmatch}
	\bar B_i(\xi,\pTveto,R,\mu)
	= \sum_{j=g,q,\bar q} \int_\xi^1\!\frac{dz}{z}\,
	\Ibar_{ij}(z,\pTveto,R,\mu)\,\phi_{j/P}(\xi/z,\mu) \,,
\end{equation*}
where $\phi$ denotes the usual collinear parton distribution of a parton of flavor $j$ in a proton 
$P$. The 
matching coefficients $\Ibar$
are extracted from $I$, the two-loop beam and soft functions of Refs.~\cite{Abreu:2022zgo,Abreu:2022sdc} as,
\beq
\bar{I}_{ij}(z,\pTveto,R,\mu)=e^{-h^{F/A}(\ptveto,\mu)} \, I_{ij}(z,\pTveto,R,\mu) \, .
\eeq
The coefficients in Ref.~\cite{Abreu:2022zgo} are presented as a Laurent expansion in the jet radius parameter
$R$. Analytic expressions are presented for all flavor channels except for a set of 
$R$-independent non-logarithmic
terms which are presented as numerical grids. For our purposes we have interpolated the numerical
grids using a spline fit.
We give further details on the reduced beam functions in \cref{sec:reducedbeamfunctions}.

\subsection{The collinear anomaly coefficient and its approximations}
\label{sec:collinearanomaly}

The missing ingredient for a complete \NNNLL{} resummation is the three-loop
collinear anomaly coefficient and therefore warrants a longer discussion. This limitation has been discussed in the 
literature and approximated in various ways. Here we discuss the uncertainty associated with the 
approximations and how we take it into account in our phenomenological predictions.

As presented in \cref{eq:RGforF} the collinear anomaly coefficients obey the \RG{} equations,
\begin{align} 
  \frac{d}{d\ln \mu} F_{qq}(\pTveto,R,\mu)&=2 \Gamma^F_{\cusp}(\mu)\, , \label{eq:Fggevol} \\
  \frac{d}{d\ln \mu} F_{gg}(\pTveto,R,\mu)&=2 \Gamma^A_{\cusp}(\mu)\, ,
  \label{eq:Fqqevol}
\end{align}
where, for example, $F_{qq}$ has the expansion,
\begin{align}
  F_{qq}(\pTveto,R,\mu)&=\as F_{qq}^{(0)}(\pTveto,R,\mu)
  +\left(\as\right)^2 F_{qq}^{(1)}(\pTveto,R,\mu) \nonumber \\
  &+\left(\as\right)^3 F_{qq}^{(2)}(\pTveto,R,\mu)
  +\left(\as\right)^4 F_{qq}^{(3)}(\pTveto,R,\mu) + \ldots
  \label{eq:Fevolexp}
\end{align}

While the logarithmic structure is given by the \RG{} equations, the constant boundary parts 
$d_k^\veto(R,B)$ where $B=F$ or $A$ need to be determined by separate calculations and are also 
referred to as the rapidity anomalous dimensions in the framework of 
Refs.~\cite{Chiu:2012ir,Chiu:2011qc}:
\begin{eqnarray}
	\label{Fqqexpansion}
	F_{qq}^{(0)}(\pTveto,R,\mu_h)&=& \Gamma_0^F \Lperp+d_1^\veto(R,F) \, , \nn \\
	F_{qq}^{(1)}(\pTveto,R,\mu_h)&=& \frac{1}{2} 
	\Gamma_0^{F}\*\beta_0\*\Lperp^2+\Gamma_1^{F}\*\Lperp+\dtwoveto(R,F) \, , \nn \\
	F_{qq}^{(2)}(\pTveto,R,\mu_h)&=& \frac{1}{3}\Gamma_0^{F}\*\beta_0^2\*\Lperp^3+\frac{1}{2} 
	(\Gamma_0^{F}\*\beta_1+2\*\Gamma_1^{F}\*\beta_0)\*\Lperp^2 \nn\\
	&+&(\Gamma_2^{F}+2\*\beta_0\*\dtwoveto(R,F))\*\Lperp+\dthreeveto(R,F) \nn \, ,\\
	F_{qq}^{(3)}(\pTveto,R,\mu_h)&=&
	\frac{1}{4}\*\beta_0^3\*\Gamma_0^{F}\*\Lperp^4+(\Gamma_1^{F}\*\beta_0^2+\frac{5}{6}\*\Gamma_0^{F}\*\beta_0\*\beta_1)\*\Lperp^3\nn
	 \\
	&+&(\frac{1}{2}\*\Gamma_0^{F}\*\beta_2+\Gamma_1^{F}\*\beta_1+\frac{3}{2}\*\Gamma_2^{F}\*\beta_0+3\*\dtwoveto(R,F)\*\beta_0^2)\*\Lperp^2
	 \nn \\
	&+&(\Gamma_{3}^{F}+3\*\dthreeveto(R,F)\*\beta_0+2\*\dtwoveto(R,F)\*\beta_1)\Lperp+\dfourveto(R,F)
	 \, . 
\end{eqnarray}
The analogous expression for gluons ($F \to A$) is given in Eq.~(\ref{Fggexpansion}).
The coefficients in the expansion of the cusp anomalous dimension, $\Gamma_k^F$,  are given in Appendix~\ref{Cusp_Anom_Dim}.

For single gluon emission $d_1^\veto(R,B)=0$. The function $\dtwoveto$ is defined below in \cref{d2vetofinal}. 
There is only partial information on $\dthreeveto$ from 
Refs.~\cite{Alioli:2013hba,Dasgupta:2014yra,Banfi:2015pju}, and we have to rely on an 
approximation. To estimate the validity of this approximation we first study similar approximations 
of $d_2^\veto$.

The function $\dtwoveto$ is given by~\cite{Becher:2013xia},
\begin{align} \label{d2vetofinal}
\dtwoveto(R,B) & = d_2^B -32  C_B\,f(R,B)\,,\text{ where }\nonumber \\
   d_2^B & = C_B \left[ \left(\frac{808}{27}  - 28\zeta_3 \right) C_A
    - \frac{224}{27}\,T_F n_f\right].
\end{align}
The function $f(R,B)$, which gives the dependence on the jet radius $R$, is known as an expansion 
about $R=0$ up to terms including $R^4$,
\begin{eqnarray}
\label{fRBexpansion}
  f(R,B) &=& C_B \Big(-\frac{\pi^2 R^2}{12}+\frac{R^4}{16}\Big) \nonumber \\
   &+& C_A\,\Big( c_L^A \ln R + c_0^A + c_2^A R^2 + c_4^A R^4 + \dots \Big) \nonumber \\
  &+&T_F n_f\,\Big( c_L^f \ln R + c_0^f + c_2^f R^2 + c_4^f
  R^4 + \dots \Big) \,.
\end{eqnarray}
The terms on the first line are due to independent emission,
whereas the terms on the second and third lines are due to correlated emission~\cite{Banfi:2012yh}.
The expansion coefficients are given in \cref{app:rapidityanomdim} in analytic and numerical form.

\subsubsection{Approximations for $\dtwoveto$}
Using \cref{d2vetofinal,fRgluon} we have for the
gluon case in the limit $n_f \to 0$ and retaining only
logarithmic and constant terms in $R$,
\begin{align} \label{leadingcolor}
  \dtwoveto(R,A) &= -32 C_A^2 \Big[-\frac{1}{32 C_A^2}d_2^A+c_L^A \ln R+c_0^A  \Big] \nonumber  \\
  &\simeq -32 C_A^2 \Big[-1.096259 \ln R + 0.7272641] \nn \\
  &\sim 32 C_A^2 \times \ln \Big(\frac{R}{2}\Big) \,.
\end{align}
This result was used as a basis for an approximation to $\dthreeveto$ in ref.~\cite{Becher:2013xia}.
However, the leading color ($n_f=0$) approximation is rather poor.
With full $n_f$ dependence, but retaining only logarithmic and constant terms in $R$ and setting 
$n_f=5$ we have
\begin{eqnarray}
\dtwoveto(R,B) &=& 32 C_B C_A \Big[(1.096+0.0295 n_f)\*\ln R -(0.72726+0.12445 n_f)\Big] \nonumber \\
                  &\sim&   32 C_B C_A \Big[ 1.2435 \ln\Big(\frac{R}{2.96}\Big)\Big].
\end{eqnarray}

\begin{figure}
	\centering

	\begin{minipage}{.47\textwidth}
\begin{center}
    \includegraphics[angle=270,width=1.1\textwidth]{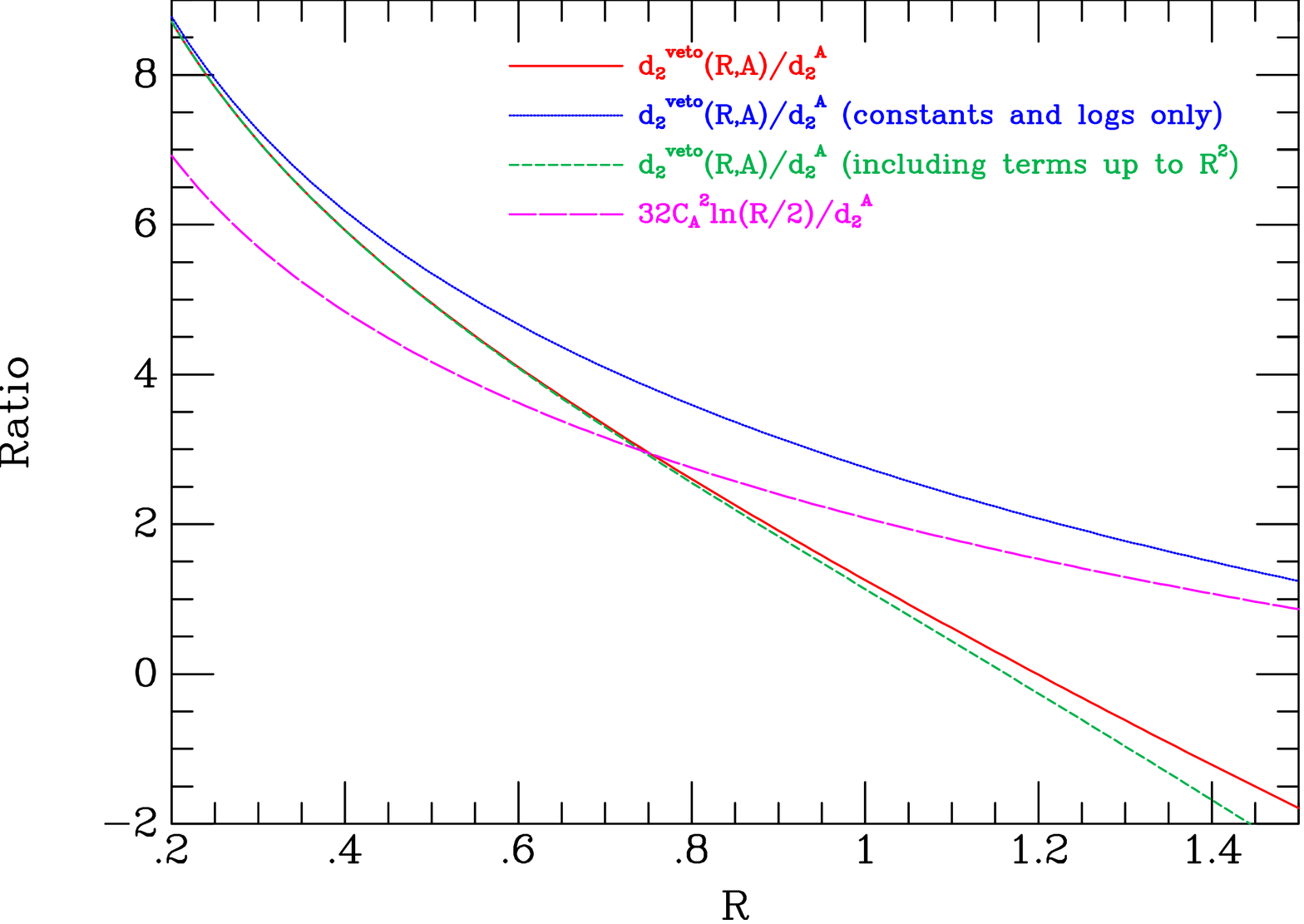}
    \vspace{-3em}
    \captionof{figure}{Approximations of $\dtwoveto(R,A)$ scaled by the constant $d_2^A$.
The full result, Eq.~(\ref{d2vetofinal}) is plotted in red.
The approximation retaining only constant terms and logarithms of $R$ is shown in blue.
The approximation retaining constant terms and logarithms of $R$ and $R^2$ terms is shown in green.
The leading color ansatz, Eq.~(\ref{leadingcolor}), derived setting $n_f=0$, is $32 C_A^2 \ln(R/2)$
and is shown in magenta.
The red, blue and green curves are all plotted for $n_f=5$.
}
\label{fig:Rplot}
\end{center}
\end{minipage}
\hfill
	\begin{minipage}{.47\textwidth}
	\begin{center}
		\includegraphics[width=0.9\textwidth,angle=-90]{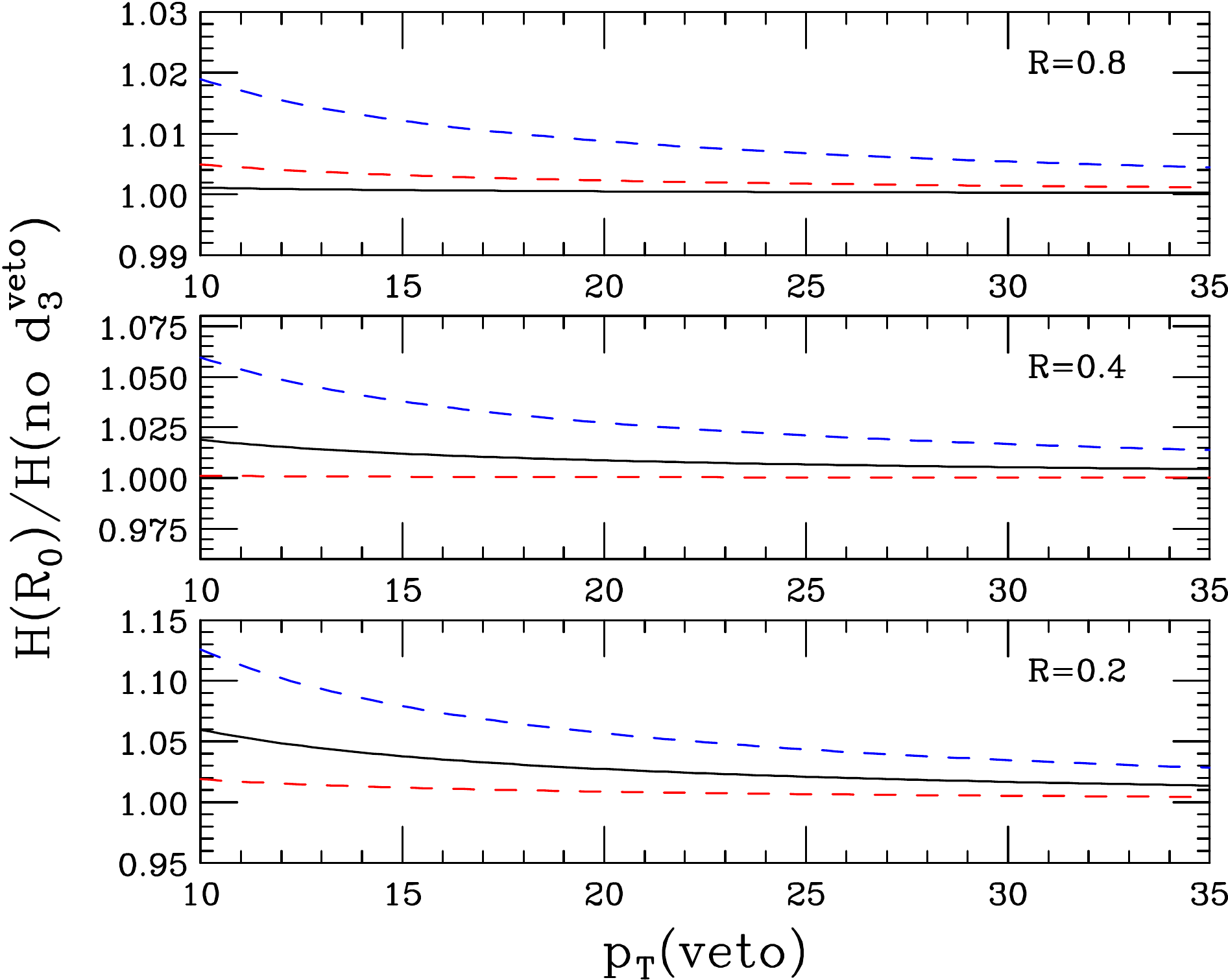}
		\captionof{figure}{Effect of $R_0$ variation in $d_3^{\rm veto}$
			as given by Eq.~(\ref{eq:banfi}) with $Q=125$~GeV
			and $n_f=5$,
			compared to the case $\dthreeveto=0$:
			$R_0 = 1$ (black), $R_0 = 0.5$ (red, dashed), $R_0 = 2$ (blue, dashed).
			\label{fig:R0dep}}
	\end{center}
\end{minipage}
\end{figure}

In \cref{fig:Rplot} we show $\dtwoveto(R,A)$ and its approximations in units of $d_2^A$ as 
a function of the jet radius $R$. As a reminder, $d_2^A$ is the 
non-$R$ dependent part of $d_2$, see \cref{d2vetofinal}. We first compare the full result (red) 
with the 
inclusion of terms up order $R^2$ (green). This shows that the $R$ expansion converges quickly and 
it is sufficient to consider only terms up to $R^4$ for practical applications. Including only 
the logarithm and the constant (blue) gives a reasonable approximation for sufficiently small 
$R$, with percent-level deviations around $R=0.4$. The leading color approximation (magenta) works 
only crudely as a first guess and could be used in the absence of any better estimate.

\subsubsection{The function $\dthreeveto$}
\label{sec:dthreeveto}

While the complete $\dthreeveto$ is unknown so far, we can extract the leading 
logarithmic term from results in the literature. Given that this approximation works 
reasonably well for $d_2^\text{veto}$ for $R\sim0.4$, it is reasonable to expect a similar behavior 
for $\dthreeveto$. We further estimate the uncertainty associated with such an approximation.

From \cref{eq:Fevolexp} the collinear anomaly coefficient at $\mu=\pTveto$ is given by,
\begin{equation}
  F_{gg}(\pTveto,R,\pTveto)=\left(\as\right)^2 \dtwoveto(R,A)+\left(\as\right)^3 \dthreeveto(R,A) + \ldots
\end{equation}

Therefore, expanding the collinear anomaly we have that
\begin{align}
  \Big(\frac{Q}{\pTveto}\Big)^{-2 F_{gg}(\pTveto,\pTveto)} =&
   1-2\left(\frac{\alpha_s(\pTveto)}{4 \pi}\right)^2 \ln\left(\frac{Q}{\pTveto}\right) 
   \dtwoveto(R,A)  \nn \\
  &-2 \ln \left(\frac{\alpha_s(\pTveto)}{4\pi}\right)^3 \ln\left(\frac{Q}{\pTveto}\right) 
  \dthreeveto(R,A) \, + {\cal O}(\alpha_s^4) .
\end{align}
At order $\alpha_s^3$ the leading term in the limit $R \to 0$
can be extracted from Eq. (C.2) of Ref.~\cite{Banfi:2015pju} which reads,
\begin{align}
\label{eq:Fcorrel}
  &\mathcal{F}^\text{correl}_{\LLR,31}(R) =
  \left(\frac{\alpha_s}{4\pi}\right)^3 \ln\Big(\frac{Q}{\pTveto}\Big) \cdot
  128 C_A \ln^2\frac{R}{R_0}\nn \\
  &\times \Big[
        1.803136 C_A^2 - 0.589237 n_f 2 T_R C_A
        + 0.36982 C_F n_f 2 T_R - 0.05893 n_f^2 4 T_R^2 \Big]\,.
\end{align}
Comparing the third-order coefficient in the two equations we thus have for a general color 
representation
\begin{align}
\dthreeveto(R,B)&= -64 C_B \ln^2 \Big(\frac{R}{R_0}\Big) (1.803136 C_A^2+0.36982 C_F n_f -0.589237 
C_A n_f -0.05893 n_f^2) \nn \\
&= -8.38188 \times 64 C_B \ln^2 \Big(\frac{R}{R_0}\Big)~\mbox{for}~n_f=5 \,.
\label{eq:banfi}
\end{align}
Hence, the sign of the leading term in the small $R$ limit is known.
In this limit $\dthreeveto$ leads to an increase in the cross-section.
This approximation only gives the leading $R$ behavior, and it has been suggested
that one may plausibly take $\frac{1}{2} <R_0 < 2$ as an uncertainty envelope 
\cite{Banfi:2015pju}.

Since $\dthreeveto$ enters through the collinear anomaly as an overall factor, we consider the 
impact of varying $R_0$ in \cref{fig:R0dep}. For typical values of $\ptveto=\SI{30}{\GeV}$ (as 
considered in this paper for the comparison with experimental studies) there is an effect of less 
than two percent for $R=0.4$. This is in agreement with the deviations we found for 
$d_2^\text{veto}$ for this approximation.

We take into account this variation in our uncertainty 
estimates, see \cref{sec:uncertainties}. A definitive statement on this issue will have to await
an exact calculation of $\dthreeveto$.

\section{Setup for phenomenology }
\label{sec:setup}

Before discussing phenomenological results, we list our input parameters, the method for matching to 
fixed order, and the approach for estimating uncertainties at fixed order and at the resummed level.

\subsection{Input parameters}
The input values used in our numerical studies are shown in \cref{parameters}.
As indicated in the table we use the complex mass scheme for the $W$ and
$Z$ boson masses.
The number of light quarks, $n_f$, is set equal to five, except for the case of
$W^+W^-$-production where $n_f=4$. 
We use the \PDF{} distribution \texttt{NNPDF31\_nnlo\_as\_0118} except for $W^+W^-$
where we use \texttt{NNPDF31\_nnlo\_as\_0118\_nf\_4} \cite{NNPDF:2017mvq}. Note that we use these
\NNLO{} parton distributions even in our lower order predictions.

\begin{table}
  \begin{center}
	\caption{Input and derived parameters used for our numerical estimates.\label{parameters}}
\label{inputparameters}
\begin{tabular}{|l|l|l|l|}
\hline
$M_W$ & 80.385~GeV                & $\Gamma_W$ & 2.0854~GeV\\
$M_Z$ & 91.1876~GeV               & $\Gamma_Z$ & 2.4952~GeV \\
$G_\mu$ & $1.166390\times10^{-5}$~GeV$^{-2}$   &   &\\
 $m_t$          & 173.2~GeV   &  $m_h$          & 125~GeV\\
\hline
\hline
\renewcommand{\baselinestretch}{1.8}
$m_W^2 = M_W^2-i M_W \Gamma_W $ & \multicolumn{3}{l|}{$(6461.748225 - 167.634879\, i) $~GeV$^2$} \\
$m_Z^2 = M_Z^2-i M_Z \Gamma_Z $ & \multicolumn{3}{l|}{$(8315.17839376 - 227.53129952\, i)$~GeV$^2$} 
\\
$\cos^2\theta_W={m_W^2}/{m_Z^2}$          & \multicolumn{3}{l|}{$(0.7770725897054007 + 
0.001103218322282256\, i)$}\\
$\alpha = \frac{\sqrt{2}G_\mu}{\pi} M_W^2 (1-\frac{M_W^2}{M_Z^2})$ & 
\multicolumn{3}{l|}{$7.56246890198475\times10^{-3}$ giving $1/\alpha\approx 132.23\ldots$}\\
\hline
\end{tabular}
\end{center}
\end{table}
\renewcommand{\baselinestretch}{1}

In the cases of $WW$ and $ZZ$ production, at ${\cal O}(\alpha_s^2)$ the
cross-section receives contributions from processes with two gluons in the initial state.
When performing the resummed calculations we include such contributions at \NLL{}
relative to the leading order, which is of order $\alpha_s^2$.
For the complete process the terms included are of order $\alpha_s^n L^k$
with $2n-4 > k > \max(n-2,0)$ and hence they contribute at $\NNNLL$.
Because of the large flux of gluons, one might worry that this formal
counting is not appropriate.
However, these contributions only represent about 3\% of the cross-section for $\ptveto=10$~GeV,
rising to about 6--8\% for $\ptveto=60$~GeV.  Therefore, neglecting higher order corrections to
these contributions, which are not implemented in our code, is justified.

We match the resummation and fixed-order {\abbrev N${}^k$LO} corrections using a naive additive 
scheme as follows,
\begin{align}
	\sigma^{\text{N$^{(k+1)}$LL+N$^{(k)}$LO}}(\pTveto)
	& = \sigma^{\text{N$^{(k+1)}$LL}}(\pTveto) + \sigma^{\Delta,k}(\pTveto)\,,\text{ where}
	\label{eq:naive} \\
	\sigma^{\Delta}(\pTveto) & = \sigma^{\text{N$^{k}$LO}}(\pTveto) -
	\d\sigma^\text{N$^{(k+1)}$LL}(\pTveto)\bigg|_\text{exp. to {\text{N$^{k}$LO}}}.
\end{align}  
The matching correction $\sigma^{\Delta}(\pTveto)$ is defined as a function of $\pTveto$, using 
the 
difference
between the fixed-order contribution and the resummed result expanded to the same fixed order.
The limit $\ptveto\to0$ of $\sigma^{\Delta}(\pTveto)$ is finite, which also allows its use as a 
higher-order subtraction scheme.

The use of a naive matching without a transition mechanism that switches off the resummation at 
large $\ptveto$  is justified since the matching 
corrections for all considered cases in this paper are small; even in the most extreme case they are less 
than $20\%$. In other words, the 
resummation alone provides a good description of the cross-sections and does not need to be 
switched off. Any transition 
function to turn off the resummation at large $\ptveto$ would have a very small effect. This is 
in contrast to transverse-momentum resummation where a transition function is 
necessary \cite{Becher:2020ugp}.

\subsection{Uncertainty estimates at fixed order}
\label{sec:founcertainties}

Ultimately the resummed predictions should offer a practical advantage compared to the fixed-order 
predictions. In many cases, the quantity $\log(Q/\ptveto)$ is not very large, and it may not seem 
worthwhile to use resummed results. However, as we will show, the resummation works remarkably well 
on its own and has matching corrections of only up to around 20\%, often much less. 
The clear separation of scales and the resummation then allow for smaller and more reliable 
uncertainty estimates. To set the stage, we first examine perturbative convergence and 
uncertainties at fixed order for quark and gluon induced boson processes, as well as for $WW$ and 
$ZZ$ production.

Constructing jet-vetoed cross-sections at fixed order requires the combination of different 
cross-sections. However, if we naively subtract the jet cross-section from the inclusive result, it 
can result in underestimated uncertainties and narrowing uncertainty bands. To avoid 
this, different methods have been proposed in the literature, of which we compare the following two.

One strategy, which we term the "two-scale" approach, is to consider the different relevant scales $Q$ and $\pTveto$ of the vetoed 
cross-section $\sigma_0$, and include both of them in the uncertainty estimate through a 
multi-point variation around both scales \cite{Becher:2014aya}. To compute this uncertainty, we  
separately vary the renormalization scale $\mu_r$ and the 
factorization scale $\mu_f$ over the values $\{ \mu_h, 2\mu_h, \mu_h/2, \ptveto, 2\ptveto, 
\ptveto/2 \}$, where $\mu_h$ depends on the process under consideration. An estimate of the 
uncertainty 
is then obtained by adding in quadrature the maximum deviations from $\mu_r = \mu_f = \mu_h$, from 
$\mu_r$ and $\mu_f$ variation separately.

Another approach, advocated by 
Refs.~\cite{Banfi:2013eda,Banfi:2015pju}, takes the jet-veto efficiency (\JVE{}) as the central 
quantity, which is the ratio of jet-vetoed cross-section to total cross-section. By combining the 
uncertainties of these two quantities in quadrature, one obtains a more robust estimate of the 
uncertainty in the jet-vetoed cross-section.  This is because the uncertainties are considered 
uncorrelated: the uncertainties in the jet-veto efficiency are typically due to non-cancellation of 
real and virtual contributions, while those in the total cross-section are connected with large 
corrections from higher orders \cite{Banfi:2015pju}.

For our \JVE{} approach, we follow the simplest formulation (``scheme (a)'' of Ref.~\cite{Banfi:2015pju})
to compute a \JVE{}-based uncertainty.  For this we consider variation
over the scales $\{ \mu_h, 2\mu_h, \mu_h/2 \}$ of $\sigma_\text{incl}$ and combine in quadrature 
the uncertainty
from the calculation of the $0$-jet efficiency ($\sigma_0/\sigma_\text{incl}$)
and the uncertainty from the inclusive calculation.  Our final fixed-order uncertainty 
band is the envelope of the two-scale and \JVE{} approaches.

\begin{figure}
	\centering
	\begin{subfigure}[t]{0.45\textwidth}
		\centering
    	\includegraphics[width=1.1\textwidth,angle=-90]{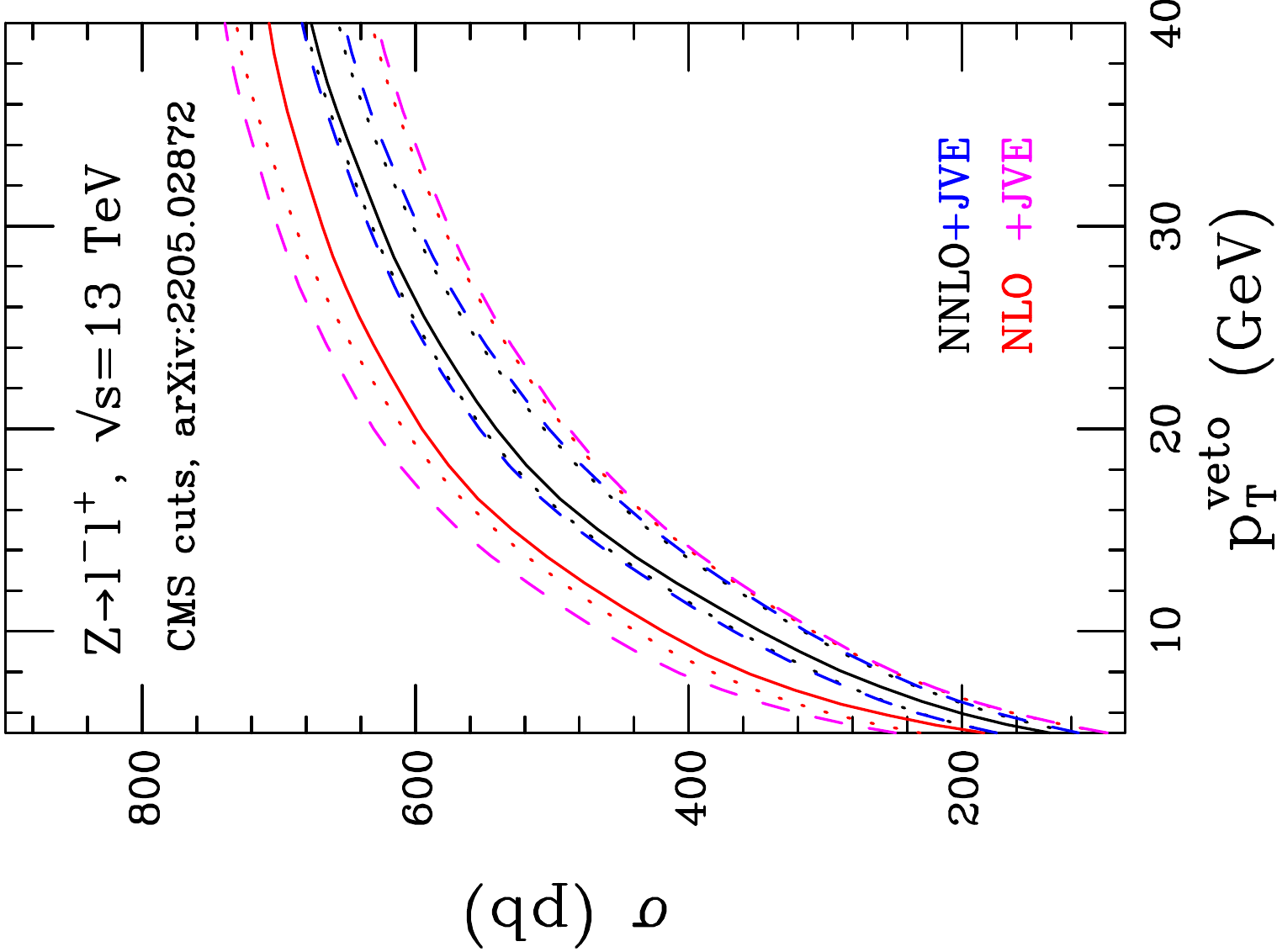}
    	\caption{$Z$ production using the setup	of ref.~\cite{CMS:2022ilp}. }
	\end{subfigure}
    \hfill
    \begin{subfigure}[t]{0.45\textwidth}
		\centering
    	\includegraphics[width=1.1\textwidth,angle=-90]{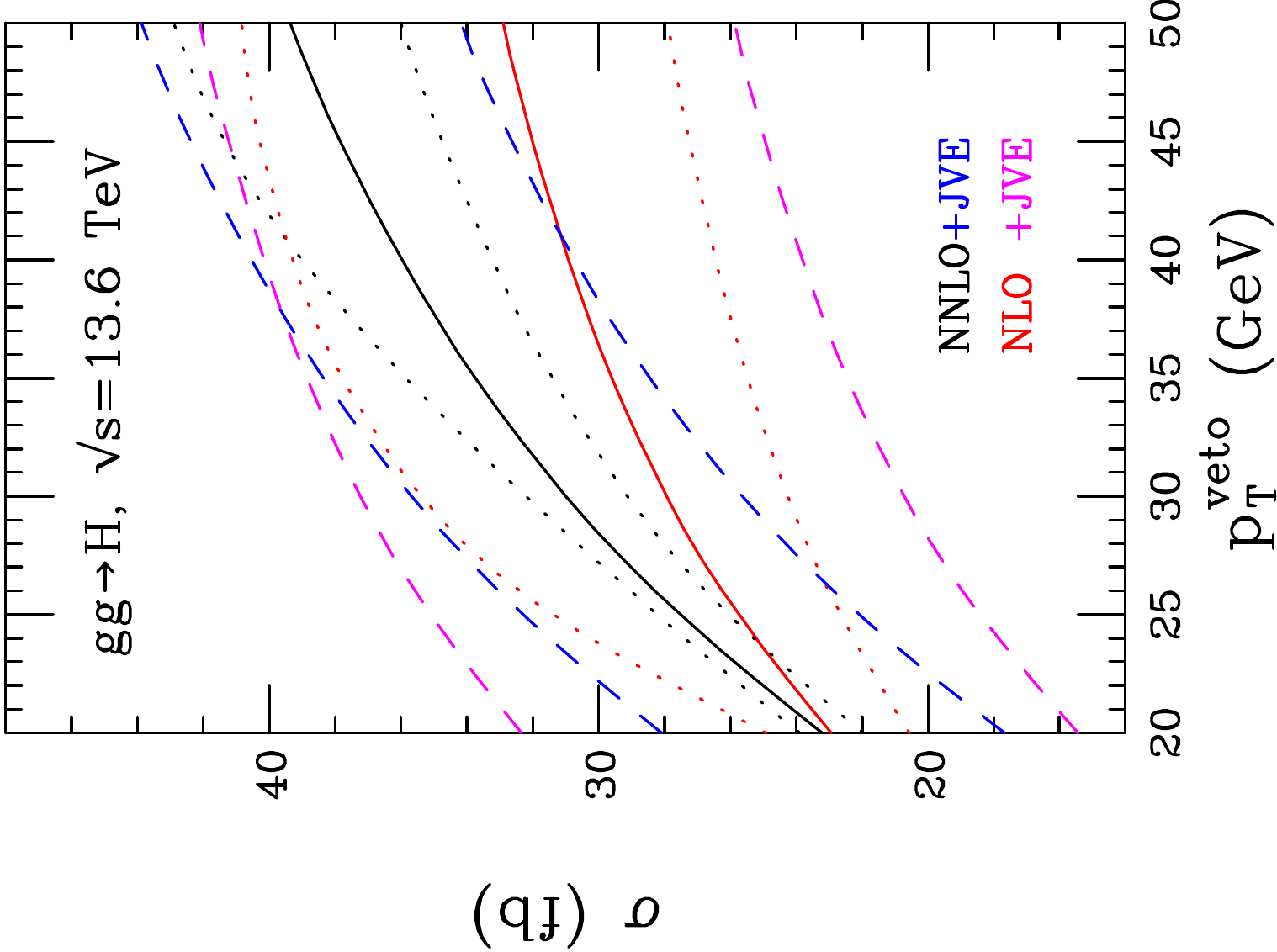}
    	\caption{$H$ production.}
    \end{subfigure}
    
    \caption{Comparison of \NLO{} and \NNLO{} fixed order predictions as a function of the jet veto.
    Central predictions solid, uncertainty estimates using either the two-scale approach
    (dotted) or the envelope of that and the JVE approach (dashed).
   \label{fig:ZHfixedorder}}
\end{figure}

With these procedures, our fixed-order results for $Z$ and $H$ production are shown in
Fig.~\ref{fig:ZHfixedorder}.  For $Z$ production we use the canonical choice
$\mu_h = Q$, where $Q$ is the invariant mass in the final state.
For Higgs production we use  $\mu_h = Q/2$, guided by the
calculation of the inclusive cross-section where such a choice results in
markedly-improved perturbative convergence.  We observe that for $Z$ production
the \NNLO{} uncertainty band is wholly contained within the \NLO{} one, while
for the Higgs case the bands at least overlap somewhat throughout the range.
For Higgs production following the combined two-scale and \JVE{} approach results in a significantly larger uncertainty
at both \NLO{} and \NNLO{},
especially at smaller values of $\ptveto$. On the other hand, for $Z$ production
the additional uncertainty from the \JVE{} approach is very small and negligible at \NNLO{}.

\begin{figure}
	\centering
	\begin{subfigure}[t]{0.48\textwidth}
		\centering
    	\includegraphics[width=1.1\textwidth,angle=-90]{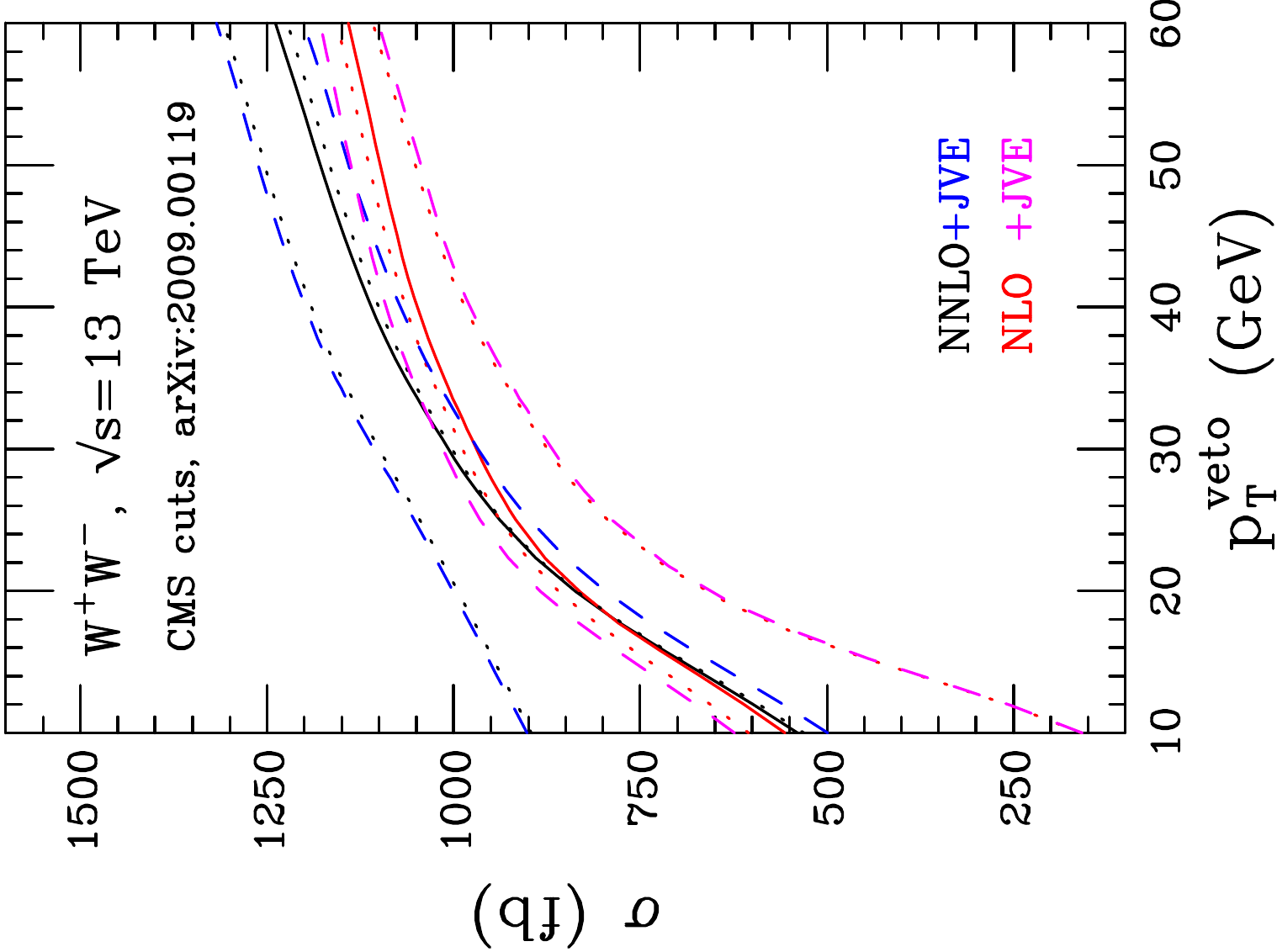}
    	\caption{$WW$ production using the setup
    		of ref.~\cite{CMS:2020mxy}.}
   	\end{subfigure}
    \hfill
    \begin{subfigure}[t]{0.48\textwidth}
    	\centering
    	\includegraphics[width=1.1\textwidth,angle=-90]{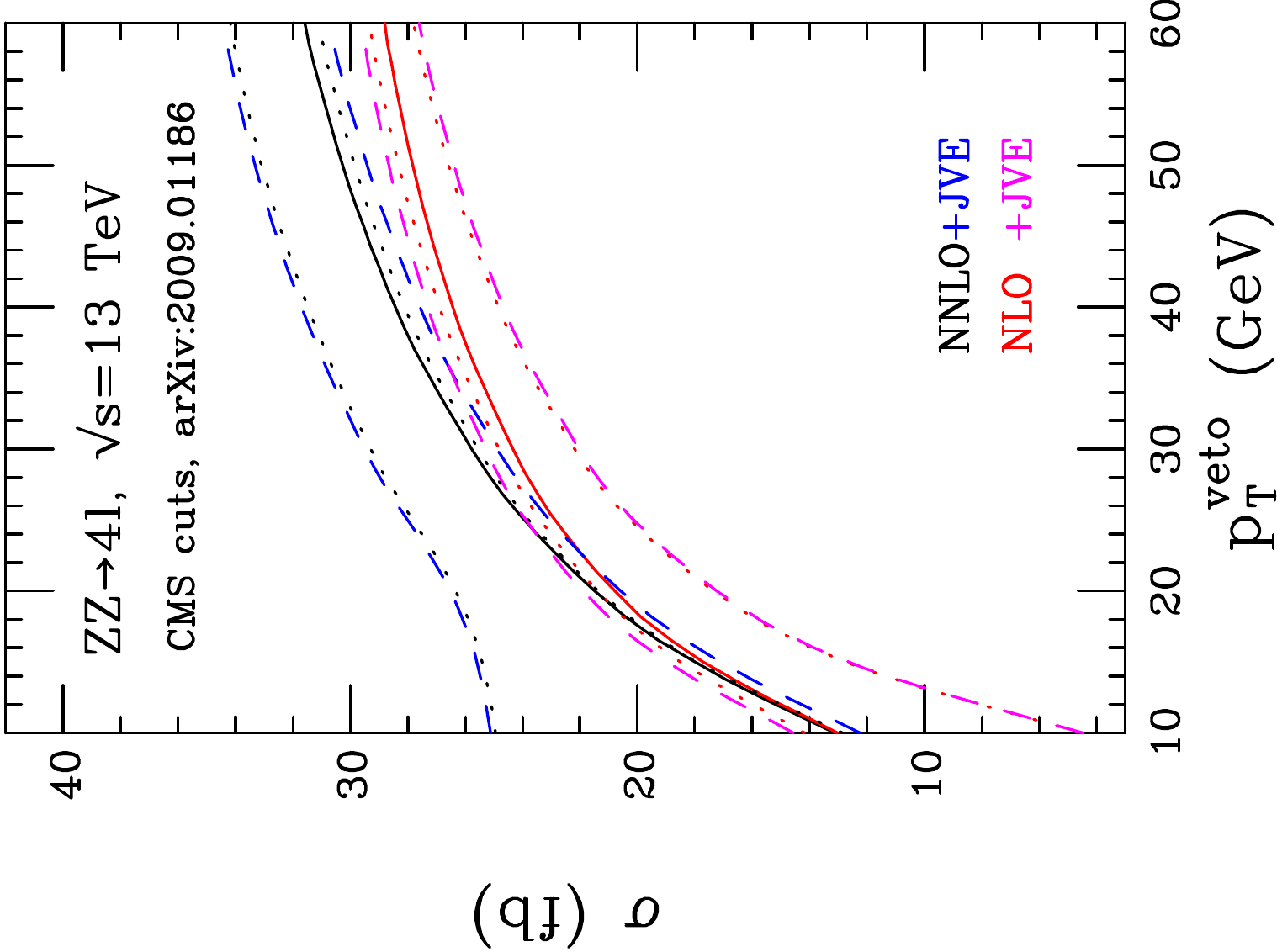}
    	\caption{$ZZ$ production using the setup
    		of ref.~\cite{CMS:2020gtj}. }
    \end{subfigure}
    \caption{Comparison of \NLO{} and \NNLO{} fixed order predictions as a function of the jet veto.
    Central predictions solid, uncertainty estimates using either the two-scale approach
    (dotted) or the envelope of that and the JVE approach (dashed).
    \label{fig:WWZZfixedorder}}
\end{figure}

Predictions for $WW$ and $ZZ$ production (with $\mu_h = Q$) are shown in \cref{fig:WWZZfixedorder}.
The limited overlap between the \NLO{} and \NNLO{} bands indicates that uncertainties are 
underestimated, even with the generous scale uncertainty procedure that we follow.
The additional uncertainty resulting from the \JVE{} procedure is small, especially at \NNLO{}, 
because the scale uncertainty of the inclusive cross-sections is very small.

\subsection{Uncertainty estimates at the resummed and matched level}
\label{sec:uncertainties}

For our central predictions, we set the resummation and factorization scales to $\mu = \ptveto$ and 
the hard scale (corresponding to the renormalization scale) to $\mu_h = Q$, where $Q$ is the 
invariant mass of the color-singlet final state. The exception is Higgs production, where we choose 
$\mu_h=Q/2$ as previously discussed. For the collinear anomaly coefficient $\dthreeveto$, we use 
the form given in \cref{eq:banfi} \cite{Banfi:2015pju} with $R_0 = 1$.

Complications arising at fixed order, described in \cref{sec:founcertainties},
are not present in the resummed
case and therefore we can follow a simpler approach where we 
vary all scales in our formalism and take the envelope, as detailed below. While the matching of 
resummed predictions to fixed-order could still introduce a 
complication, the matching corrections are not dominant. The bulk of the cross-section comes from the 
resummation and it allows us to follow the simple procedure of varying all scales in the naively 
obtained (without \JVE{}) jet-veto cross-section too.

The small and narrowing uncertainty bands 
at fixed order would typically appear in regions where the resummation is found to be dominant, 
i.e. where fixed-order contributes very little through the matching corrections. In practice we 
observe that the size of uncertainties are overall uniform in both the resummation and large 
$\ptveto$ fixed-order regions, as can be seen in all of our following predictions. This supports 
the conclusion that our procedure is sufficient.

Overall, our procedure for estimating uncertainties is as follows.
\begin{enumerate}
	\item For the resummation (fixed-order) parts we vary both the resummation 
	(factorization) and hard (renormalization) scales
	by a factor of two about their central values, adding the excursions in quadrature to
	obtain the total scale uncertainty.
	\item For the resummation we re-introduce the rapidity scale in Eq.~(\ref{eq:BNR}) by 
	re-writing the
	collinear anomaly factor as follows~\cite{Becher:2013xia,Jaiswal:2015nka}:
	\begin{equation}
		\bigg( \frac{Q}{\pTveto} \bigg)^{-2F_{ii}(\pTveto,R,\mu)}
		= \bigg( \frac{Q}{\nu} \bigg)^{-2F_{ii}(\pTveto,R,\mu)}
		\bigg( \frac{\nu}{\pTveto} \bigg)^{-2F_{ii}(\pTveto,R,\mu)} \,.
	\end{equation}
	For $\nu \sim \pTveto$ the second factor can be expanded since it does not contain a large 
	logarithm.
	We vary the rapidity scale $\nu$
        in the range $[\ptveto/2, 2\ptveto]$ for gluon-initiated processes
        and
        in the range $[\ptveto/6, 6\ptveto]$ for quark-initiated processes.
        The large variation for quark-initiated processes ensures overlapping uncertainty 
	bands at \NNLL{} and \NNNLLpart{}; this is
	achieved by the range given above, as demonstrated explicitly in \cref{sec:jetvheto,sec:pheno}.
	\item The parameter $R_0$ in $\dthreeveto$ is varied between $0.5$ and 2.
\end{enumerate}
We first combine the scale uncertainties
(1 and 2) in quadrature and then, to obtain our total uncertainty,
add the variation of $R_0$ (3) linearly.

\subsection{Effects of cuts on rapidity at fixed order}
The usual jet veto resummation described so far imposes no cut on the jet rapidity. This is in 
contrast to 
experimental analyses, see \cref{tab:jetvetocross}, which impose such a cut because of limited 
detector acceptance and to diminish
the effect of pileup.
Ref.~\cite{Michel:2018hui} identifies three different regimes, depending on $p_t, Q$ and $\etacut$.
\begin{itemize}
	\item For $\pTveto/Q \gg \exp(-\etacut)$ standard jet veto resummation should apply, effects 
	due to 
	the
	rapidity cut are corrections power suppressed by $Q\exp(-\etacut)/\pTveto$.
	\item For $\pTveto/Q \sim \exp(-\etacut)$ the effects of a rapidity cut must be treated as a 
	leading power
	correction.
	\item For $\pTveto/Q \ll \exp(-\etacut)$ the logarithmic structure is changed already at 
	leading 
	log level,
	and non-global logarithms appear.
\end{itemize}

\begin{table}
	\begin{center}
		\caption{Jet rapidity cuts applied in the experimental studies examined later in this 
		paper.}
		\label{tab:jetvetocross}
		\begin{tabular}{l|l|l}
			Process      & Ref. & $\etacut$ \\ \hline
			Higgs	     & -- & no study \\
			$Z$ (CMS)    & \cite{CMS:2022ilp}   & $2.4$ \\
			$W$ (ATLAS)  & \cite{ATLAS:2017irc} & $4.4$ \\
			$WW$ (CMS)   & \cite{CMS:2020mxy}   & $4.5$ \\
			$WZ$ (ATLAS) & \cite{ATLAS:2019bsc} & $4.5$ \\
			$WZ$ (CMS)   & \cite{CMS:2021icx}   & $2.5$ \\
			$ZZ$ (CMS)   & -- & no study \\
		\end{tabular}
	\end{center}
\end{table}

We estimate the practical impact of experimentally used jet rapidity cuts at fixed order.
Including the rapidity cut in the resummation requires large changes and ingredients, which are 
also only available a low order so far \cite{Michel:2018hui}.

\begin{figure}
	\centering
	\begin{subfigure}[t]{0.48\textwidth}
		\centering
		\includegraphics[width=1.1\textwidth,angle=-90]{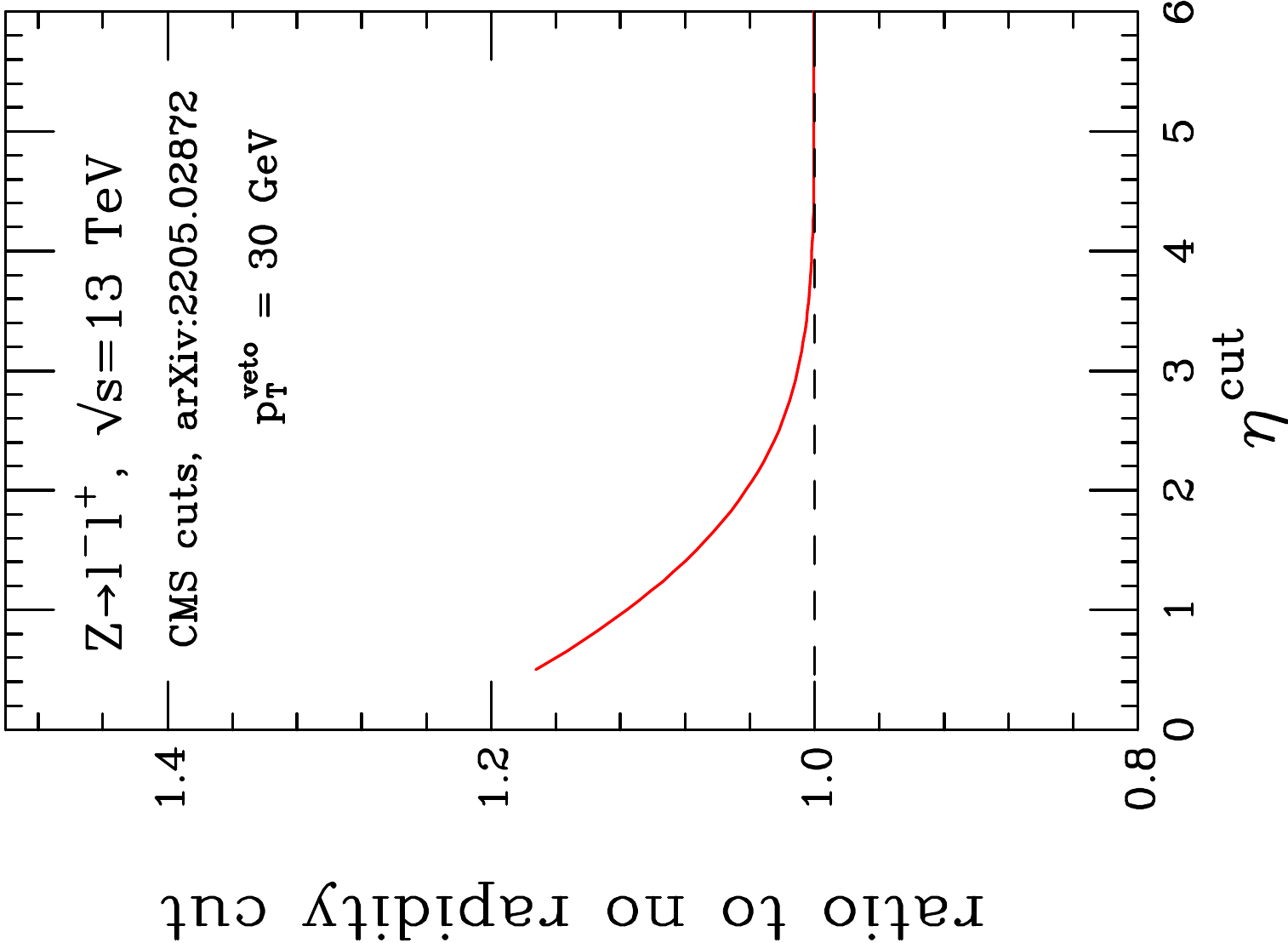}
		\caption{$Z$ production following the setup of ref.~\cite{CMS:2022ilp}.}
	\end{subfigure}
	\hfill
	\begin{subfigure}[t]{0.48\textwidth}
		\includegraphics[width=1.1\textwidth,angle=-90]{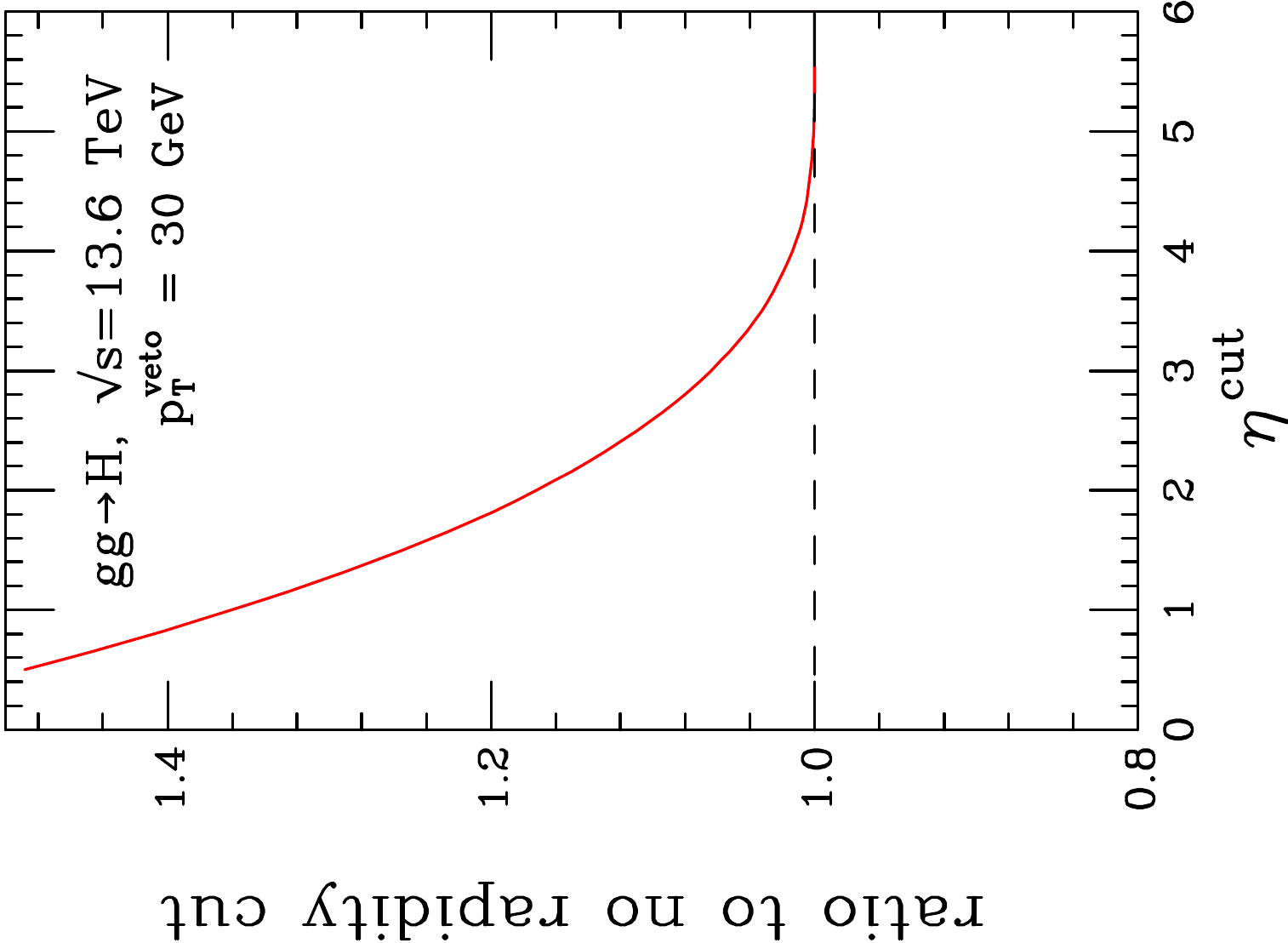}
		\caption{$H$ production.}
	\end{subfigure}
	
	\caption{Effect of the jet rapidity cut
		at \NNLO{} with $\ptveto=30$~GeV.
		\label{fig:ZH136rapstudy}}
	
\end{figure}

The effect of the jet rapidity cut for the $Z$ and Higgs production cases
is illustrated in Fig.~\ref{fig:ZH136rapstudy}.  These calculations
are performed at \NNLO{} for $\ptveto=30$~GeV.  The rapidity cut plays a bigger
role for Higgs production: for example for $\etacut=2.5$ the
cross-section is 11\% larger than the result with no rapidity cut, compared to only 2\% for $Z$ 
production. This is due to the larger logarithm ($\log(m_H/\ptveto)/\log(m_Z/\ptveto) \approx 
1.28$) and the larger color prefactor ($C_A/C_F$ = 2.25) in Higgs production.
However, for $\etacut=4.5$ the effect of the rapidity cut
is negligible in both cases.

\begin{figure}
	\centering
	\begin{subfigure}[t]{0.48\textwidth}
		\centering
		\includegraphics[width=\textwidth,angle=-90]{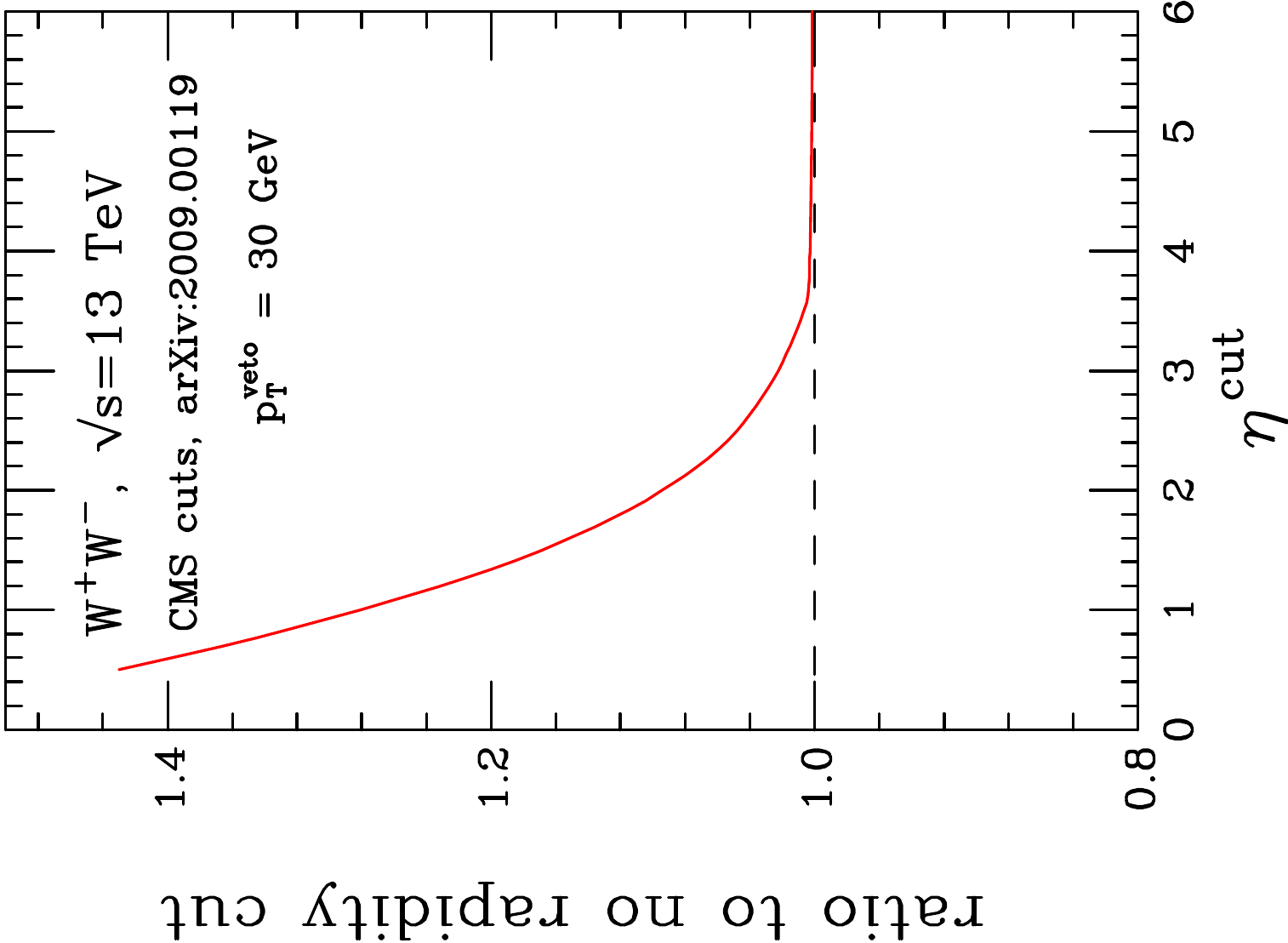}
		\caption{$WW$ production.}
	\end{subfigure}
	\hfill
	\begin{subfigure}[t]{0.48\textwidth}
		\centering
		\includegraphics[width=\textwidth,angle=-90]{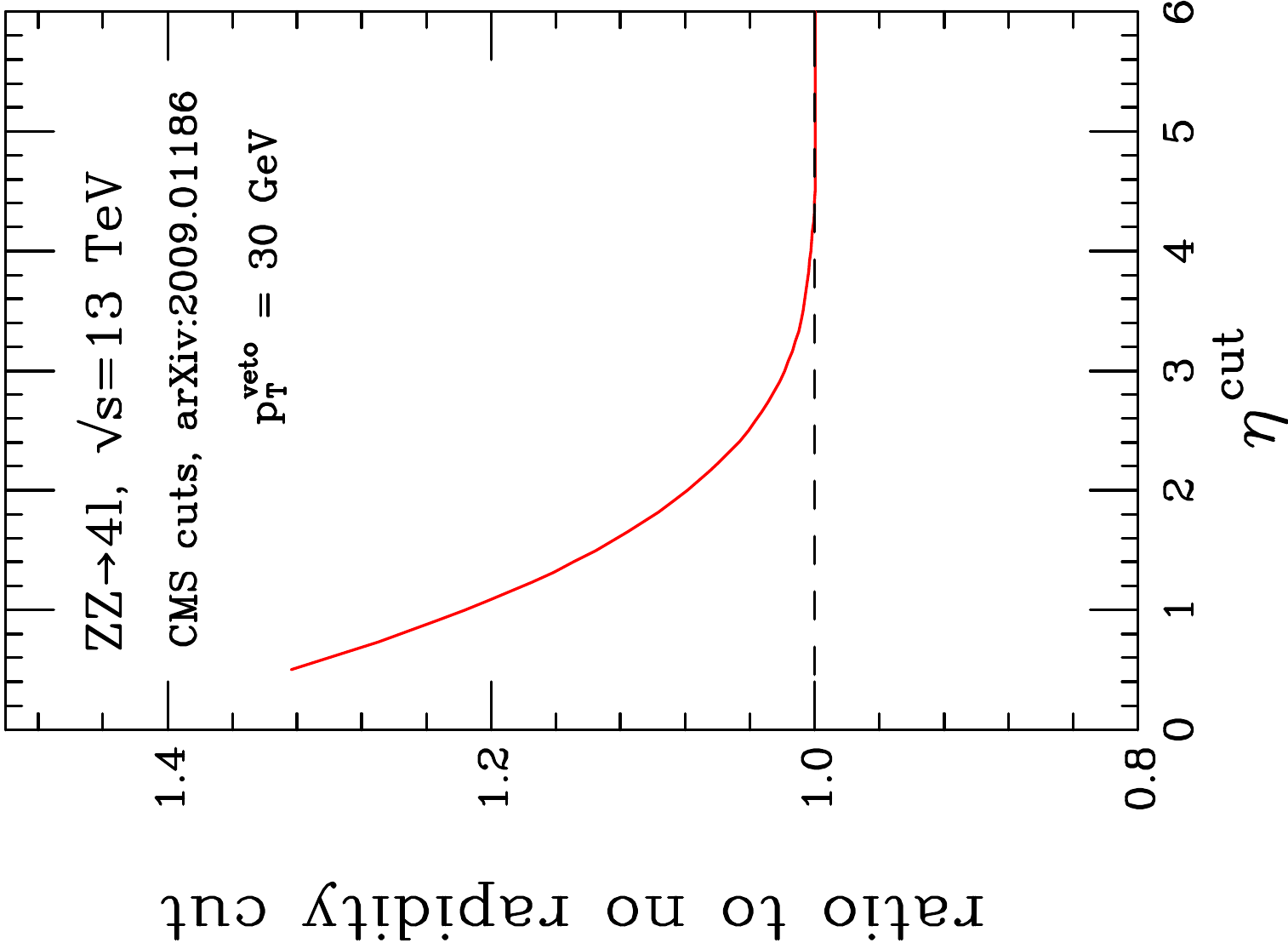}
		\caption{$ZZ$ production.}
	\end{subfigure}
	\caption{Effect of the jet rapidity cut
		at \NNLO{} with $\ptveto=30$~GeV.
		\label{fig:WWZZ13rapstudy}}
\end{figure}

The corresponding results for diboson processes are shown in Fig.~\ref{fig:WWZZ13rapstudy}.
In this case, the disparity between $Q$ and $\ptveto$ is much larger, so the
rapidity cut can play a crucial role, although the effect is still not as important
as for Higgs production.
For $\etacut=2.5$ the $WW$ and $ZZ$
cross-sections 4\% larger than the results with no rapidity cut, and
the effect of $\etacut=4.5$ is negligible.

\section{Comparison with \JetVHeto{}}
\label{sec:jetvheto}

While jet-veto resummed phenomenology has been extensively studied in the literature, the only 
public codes that permit detailed predictions use \JetVHeto{} or \RadISH{}. 
For jet-veto resummation \RadISH{} implements the analytic \JetVHeto{} resummation formula 
\cite{Banfi:2012jm}. The codes rely on the formalism of the 
{\abbrev CAESAR} approach \cite{Banfi:2004yd,Banfi:2012yh} extended to \NNLL{} \cite{Banfi:2012jm}.
An extension of the \RadISH{} code has been used to perform joint jet-veto and boson 
transverse momentum resummation \cite{Monni:2019yyr}.

For our comparisons we use \RadISH{} version 3.0.0 \cite{Bizon:2017rah,Monni:2016ktx} and 
\JetVHeto{} version 3.0.0 \cite{Banfi:2015pju,Banfi:2012jm,Banfi:2013eda} including small-$R$ 
resummation \cite{Dasgupta:2014yra,Banfi:2012yh} as part of {\abbrev MCFM-RE} \cite{Arpino:2019fmo}.
Both codes operate at the level of \NNLL{} and we have checked that they give indeed the same 
results.

In our comparison, we would like focus on the differences in the resummation part, since the 
fixed-order part is identical in each calculation. We explore how central values and uncertainties 
compare at \NNLL{} to our results and in how far \NNNLLpart{} results improve the perturbative 
convergence.
However, the matching to fixed-order is handled differently in each formalism.
Different matching schemes (e.g. additive or 
multiplicative schemes of various types) probe higher-order effects. It has also been advocated to match at the level 
of jet-veto efficiencies \cite{Banfi:2015pju}. 
Fortunately, matching corrections are generally small for jet-veto scales of \SIrange{30}{40}{\GeV} 
for all considered boson and di-boson processes. We therefore focus on the resummation in our 
comparison.

The \JetVHeto{} formalism considers three scales $\mu_R$, $\mu_F$ and $Q$ that are all similar in 
magnitude to the hard scale.
To ensure that the resummation switches off for $\ptveto{} \gtrsim Q$, the resummed logarithms are  
modified through the prescription $\log(Q/\ptveto{}) \to 1/p \,
\log((Q/\ptveto{})^p + 1)$.  For \JetVHeto{} $p$ has a default value of $5$ \cite{Banfi:2015pju}, 
while for \RadISH{} the 
default choice is $4$. For comparison purposes we use $p=5$ in both cases. It is evident that for 
sufficiently small $\ptveto{}$ the precise value of 
$p$ does not matter. Changing this parameter 
has a similar effect to turning off the resummation with a transition function. In principle 
this demands a fully matched 
calculation, but the matching corrections of our considered cases are small and we have checked 
that the effect of 
changing $p$ to $3$ or $4$ is subleading compared to the scale uncertainties. Here we focus on 
those scale uncertainties.

In ref.~\cite{Banfi:2015pju} it has been argued that the $Q$ should be varied by a factor of
$\frac{3}{2}$ around its central value, based on new insights from convergence at {\abbrev N$^3$LO} 
for Higgs production. For simplicity, we use a more conservative variation by a factor of two. We 
independently vary $\mu_R, \mu_F$ and $Q$ by a factor of two around a central scale of 
$m_{\ell\ell}$ for $Z$-boson production 
and around $m_H/2$ for Higgs production. Our uncertainty bands for this comparison are obtained by 
taking the envelope of these results.

\subsection*{$Z$-boson production}

\begin{figure}
	\centering
	\includegraphics[width=0.8\textwidth]{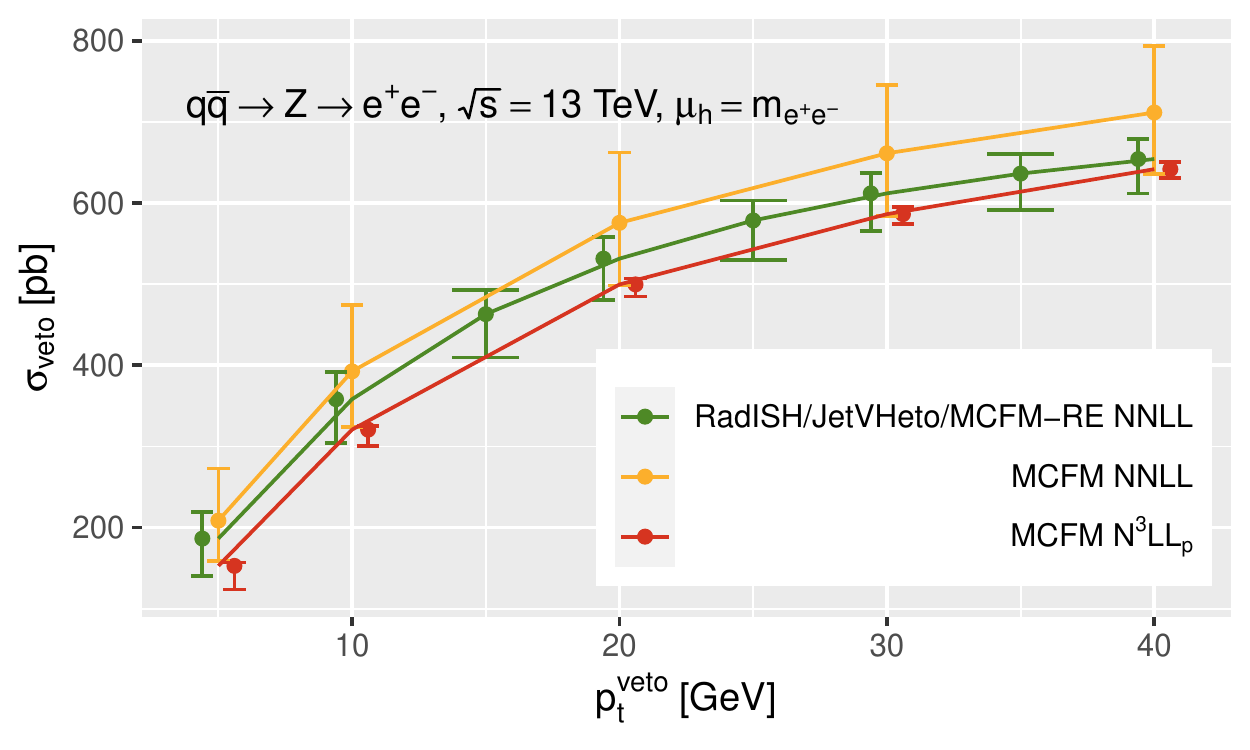}
	\caption{Comparison of \JetVHeto{} \NNLL{} resummation with our \NNLL{} and \NNNLLpart{} 
	results for $Z$ production with cuts as in \cref{tab:zcms}.}
	\label{fig:jetvhetoZ}
\end{figure}

For the comparison of $Z$ production we choose a central hard scale of $m_{\ell\ell}$ with results 
shown in \cref{fig:jetvhetoZ}.
We find that the uncertainty bands of our \MCFM{} \NNLL{} predictions
 mostly contain those obtained by \JetVHeto{} (as estimated according to our
 procedure just described). Furthermore, the uncertainty bands of both \NNLL{} predictions overlap
 with our \NNNLLpart{} results, indicating robust uncertainties.

At \NNNLLpart{} uncertainties decrease dramatically compared to \NNLL{}, but they are quite 
asymmetric, 
which suggests that a symmetrization of uncertainties may be necessary in this case. 
We also observe that without the large uncertainties at \NNLL{}, there would be no overlap between 
the \NNNLLpart{} results and \NNLL{}. This highlights the importance of carefully estimating and 
comparing uncertainties to accurately assess the compatibility of different methods and 
results.

\subsection*{$H$-boson production}

\begin{figure}
	\centering
	\includegraphics[width=0.8\textwidth]{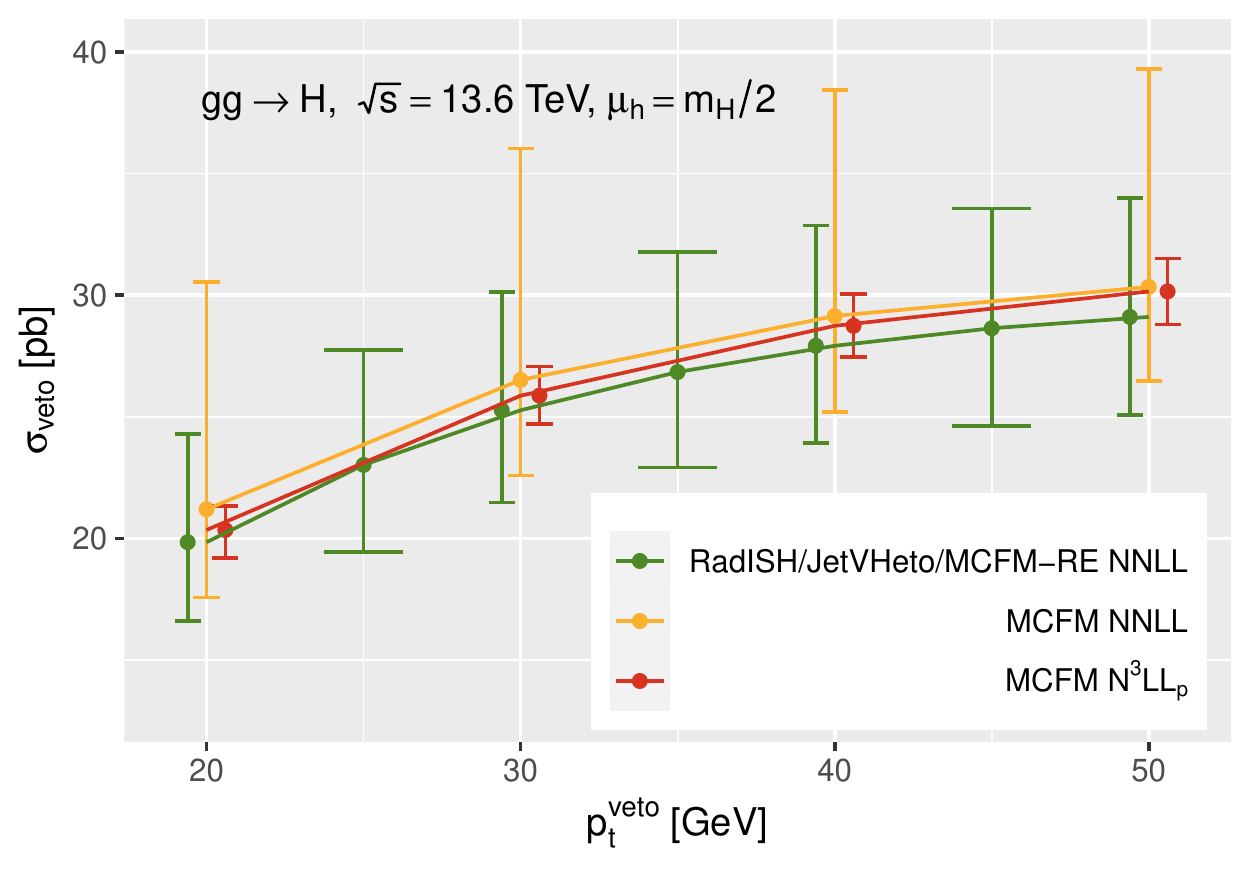}
	\caption{Comparison of \JetVHeto{} \NNLL{} resummation with our \NNLL{} and \NNNLLpart{} 
		results for non-decaying $H$ production.}
	\label{fig:jetvhetoH}
\end{figure}

In our study of Higgs production, we choose a central hard scale of $m_H/2$ and show results in 
\cref{fig:jetvhetoH}. All results are computed in the $m_t\to\infty$ theory and rescaled by a 
factor of $1.0653$ to account for finite 
top-quark mass effects, see \cref{massrescaling}.

The Higgs case is distinct from $Z$ production since it is gluon-gluon 
initiated instead of quark-initiated. In this case, our predictions agree well with the \JetVHeto{} 
results, but our uncertainties at \NNLL{} are again much larger.

Note that we vary the \JetVHeto{} scale $Q$ by a factor of two, while the
\JetVHeto{} authors vary by a factor of $3/2$ in the Higgs case.
This difference in the amount of variation may require some tuning in our formalism, at least at 
the \NNLL{} level. However, the perturbative convergence is again excellent with small 
uncertainties at \NNNLLpart{} and central predictions that agree well with \NNLL{}.

\section{Phenomenological results}
\label{sec:pheno}

In this section, we present the results of our phenomenological studies, which are based on the 
uncertainty procedure, matching to fixed-order, and input parameters described in \cref{sec:setup}. 
We compare our findings with experimental results from the literature and discuss their 
implications.

\subsection{$Z$ and $W$ production}

\label{ZandWproduction}
The process of $Z$ production has already been extensively studied in the literature, thus
enabling a variety of cross-checks of our calculation.  The implementation of the hard
function and its evolution has been verified by comparison with the explicit results
given in Table 1 of ref.~\cite{Becher:2011xn}.
The full machinery of the resummation and matching procedure can also be compared with the results 
of ref.~\cite{Banfi:2012jm}, with which we find excellent agreement within uncertainties, see also 
\cref{sec:jetvheto}.

We first investigate the impact of choosing a time-like hard scale in the resummed result for $Z$ 
production. Previous work has shown that choosing a space-like hard scale ($\mu_h^2 = Q^2$) can 
lead to significant corrections in the perturbative expansion of some processes, while a time-like 
hard scale ($\mu_h^2 = -Q^2$) can resum certain $\pi^2$ contributions~\cite{Ahrens:2008qu} using a 
complex 
strong coupling.

\begin{table}
        \begin{center}
        	        \caption{\label{tab:zcms} Cuts used in the analysis of $Z$ production, adapted 
        	        from
        		ref.~\cite{CMS:2022ilp}.}
                \begin{tabular}{r | l }
                        { lepton cuts} & $q_T^{l_1} > \SI{30}{\GeV}$, $q_T^{l_2} > \SI{20}{\GeV}$,
                        $|\eta^{l}|<2.4$ \\
                        {lepton pair mass}& $\SI{71}{\GeV}<m_{l^-l^+}<\SI{111}{\GeV}$ \\
                        {jet veto}& anti-$k_T$, $R=0.4$, 0-jet events only
                \end{tabular}
        \end{center}

\end{table}

For this comparison we consider purely
resummed results at \NNLL{} and \NNNLLpart{}, only considering uncertainties 
originating from scale variation (items 1 and 2 of our uncertainty procedure in 
\cref{sec:uncertainties}).
We consider the process $pp \to Z/\gamma^* \to \ell^- \ell^+$, i.e. a final state of definite lepton
flavor.  We use the same set of cuts and vetoes as in the $\sqrt s =\SI{13}{\TeV}$ \CMS{} analysis 
\cite{CMS:2022ilp}, but extend the veto to jets of all rapidities, rather than only those with 
$|y|<2.4$. This difference, and the effect of matching to \NNLO{}, is discussed in 
detail in \cref{sec:cmsz}.

\begin{figure}
	\centering
	\begin{subfigure}[t]{0.48\textwidth}
		\centering
		    \includegraphics[width=\textwidth,angle=-90]{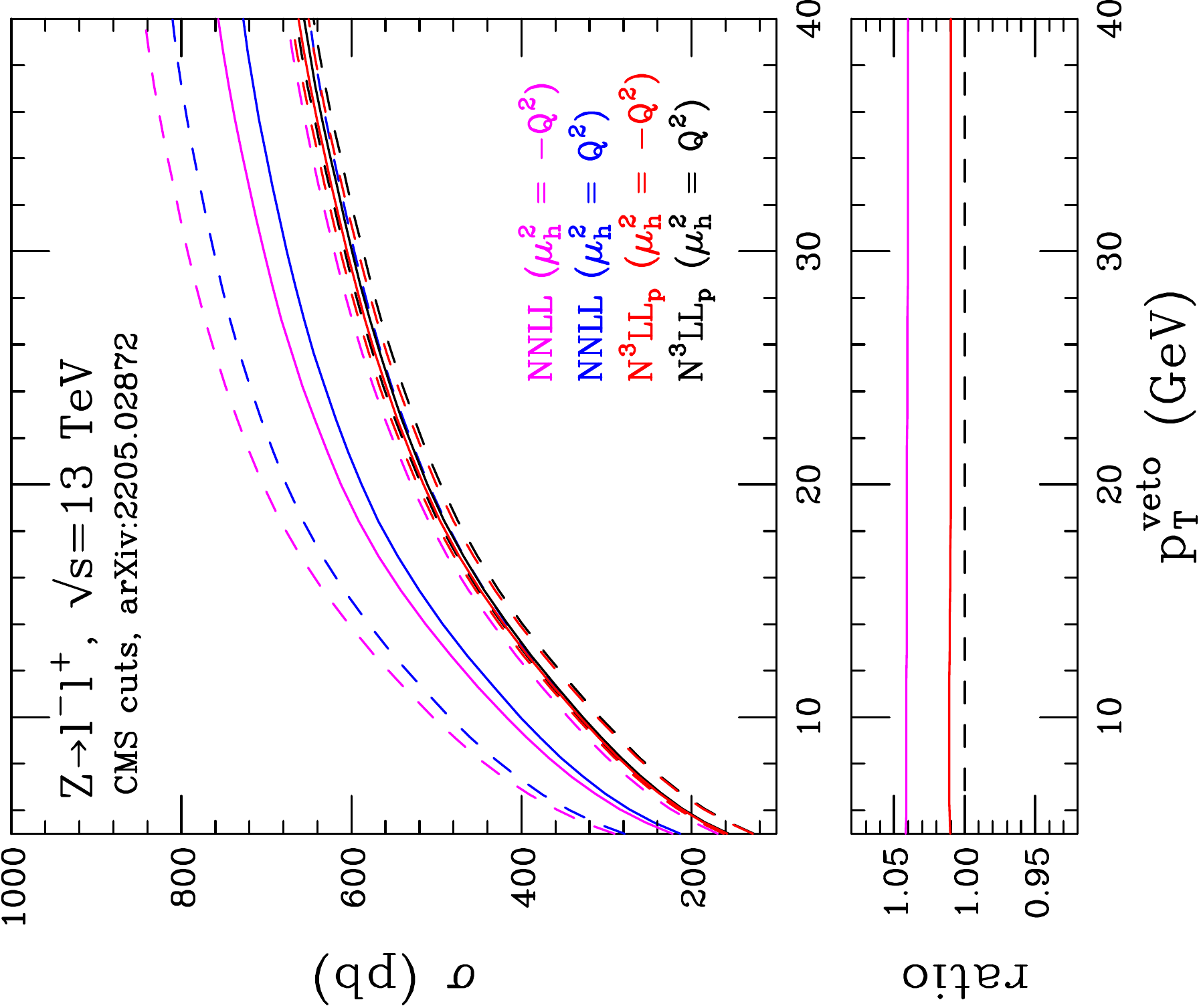}
		\caption{ Predictions are computed using a central
			choice for the hard scale given by either $\mu_h^2 = Q^2$
			or $\mu_h^2 = -Q^2$. The lower panel shows the ratio of the result
			for $\mu_h^2 = -Q^2$ to the one for $\mu_h^2 = Q^2$.
			\label{fig:Z13timestudy}}
	\end{subfigure}
	\hfill
	\begin{subfigure}[t]{0.48\textwidth}
		\centering
		\includegraphics[width=\textwidth,angle=-90]{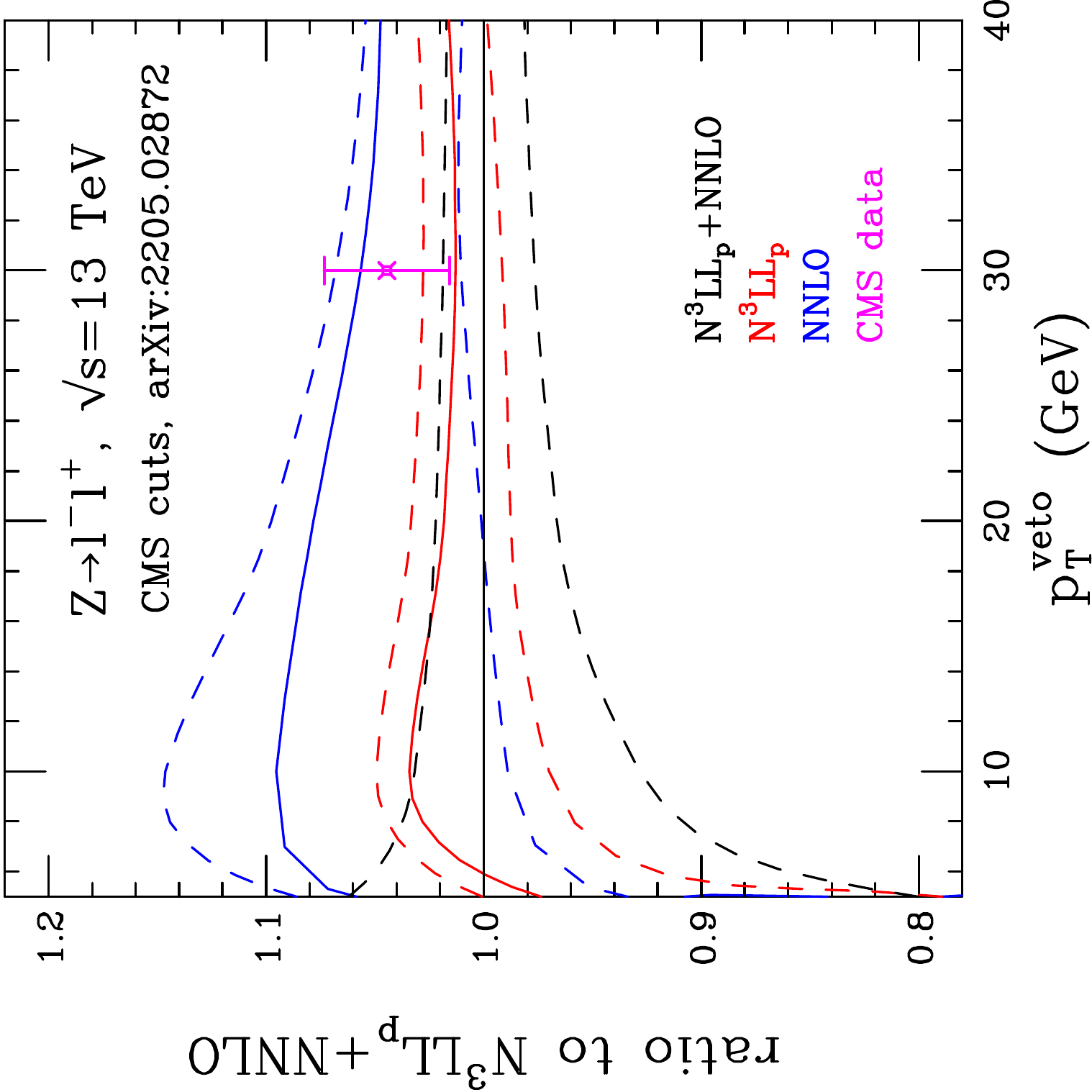}
		\caption{Predictions and \CMS{} measurement as ratio to matched result. 
				\label{fig:z13vetomatched}}
	\end{subfigure}

\caption{Comparison of \NNLL{} and \NNNLLpart{} predictions
	for $Z$ production as a function of the jet veto, using the setup
	of ref.~\cite{CMS:2022ilp} (central predictions solid, uncertainty estimate
	according to the text, dashed). }
\end{figure}

Our results are shown in Fig.~\ref{fig:Z13timestudy} as a function of the value of
the jet veto. 
We observe that the results do not depend strongly on the choice of hard
scale, with a difference of about $4\%$ at \NNLL{} and only $1\%$ at
\NNNLLpart{}. This indicates that resumming the $\pi^2$ terms results in only a small enhancement 
of the cross-section for $W$ and $Z$ production.
Based on these findings, we use the space-like hard scale ($\mu_h^2 = Q^2$) in our subsequent 
studies of $Z$ and $W$ boson production, as it is the more commonly used choice in the literature.

\subsubsection{CMS $Z$ production}
\label{sec:cmsz}

As previously mentioned, the \CMS{} measurement we are comparing to includes a jet rapidity cut of 
$|y|<2.4$. To assess the importance of this restriction, we first compare the \NNLO{} 
predictions with and without the rapidity cut, as a function of the jet veto value. This 
comparison, shown in \cref{tab:Zrapidity}, helps us better understand the limitations of our 
analysis.

 We use the quantity $\epsilon(p_T^\text{veto})$ to 
quantify the increase in the cross-section when the rapidity cut is applied, defined as
\begin{equation} 
  \epsilon(p_T^\text{veto}) =
  \frac{\sigma_{0-\text{jet}}(\etacut = 2.4)}{\sigma_{0-\text{jet}}(\text{no}~\etacut)} - 1\,.
\end{equation}

The experimental measurement we are comparing to uses a jet veto of $p_T^\text{veto} = 
\SI{30}{\GeV}$, 
for which the rapidity cut has only a 3\% effect on the cross-section. This suggests that our 
calculation with an all-rapidity jet veto is appropriate for comparing to the experimental 
measurement. However, as $p_T^\text{veto}$ decreases, the impact of the rapidity cut becomes more 
significant, until at $p_T^\text{veto} = \SI{5}{\GeV}$ it is no longer appropriate to neglect the 
rapidity cut. This is consistent with the arguments of Ref.~\cite{Michel:2018hui}, which suggest 
that the 
standard jet veto resummation formalism should suffice as long as $\ln(Q/p_T^\text{veto}) \ll 
\etacut$. In our case, $\ln(Q/p_T^\text{veto})$ ranges from 0.8 to 2.9 for 
$p_T^\text{veto}$ from 40 down to $5$~GeV, so the standard jet veto resummation should be 
appropriate, albeit with sizeable power corrections,
for $\etacut = 2.4$ except for the smallest values of $\ptveto$.

\begin{table}
\begin{center}
	\caption{The $Z+0$-jet cross-section prediction at NNLO ($\mu=Q$), with and without a jet 
	rapidity cut.}
\label{tab:Zrapidity}
\begin{tabular}{l|ccccc}
$p_T^\text{veto}$ [GeV] & 5 & 10 & 20 & 30 & 40 \\
\hline
$\sigma_{0-\text{jet}}(\text{no}~\etacut)$~[pb]    & 140 & 347 & 539 & 627 & 675 \\
$\sigma_{0-\text{jet}}(\etacut = 2.4)$~[pb]        & 242 & 411 & 569 & 643 & 685 \\
\hline
$\epsilon$ & 0.73 & 0.18 & 0.06 & 0.03 & 0.01
\end{tabular}
\end{center}
\end{table}

\begin{figure}
	\includegraphics[]{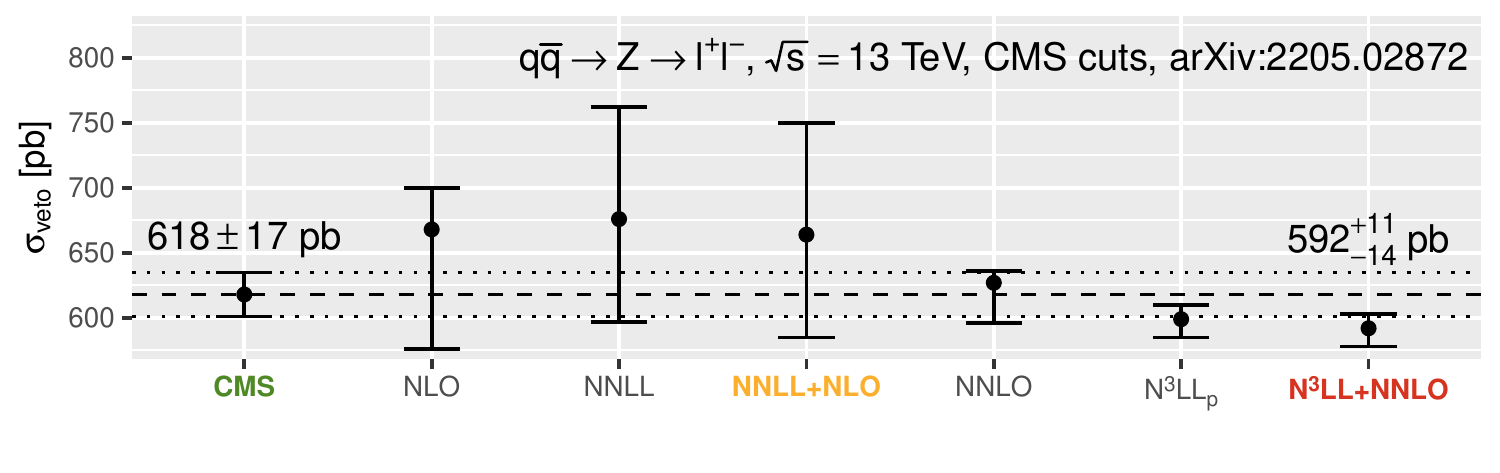}
	\vspace{-1em}
	\caption{Comparison of $Z$-boson jet-vetoed predictions with the \CMS{} \cite{CMS:2022ilp} 
	\SI{13}{\TeV} 
	measurement. Shown are results at fixed-order, purely resummed and matched.}
	\label{fig:Zjetvetocross}
\end{figure}

We now turn to a comparison with the \CMS{} result \cite{CMS:2022ilp}, which uses
a jet threshold of $30$~GeV. Our comparison with fixed-order, purely resummed and matched 
predictions is shown in \cref{fig:Zjetvetocross}.
We find that the fixed-order and resummed results differ by only a few percent, indicating that 
resummation is not necessary for this value of the jet veto. This is because the quantity 
$\ln(M_Z/\ptveto) = 1.1$ is not large enough to require resummation. The \CMS{} measurement yields 
a cross-section of \SI{618 \pm 17}{pb}, while our best prediction is $592^{+11}_{-14}\,${pb}.

We study the production of $Z$ bosons as a function of the jet veto in 
Fig.~\ref{fig:z13vetomatched}. We observe that the difference between the resummed and central 
fixed-order results is small, even for the smallest values of $\ptveto$ considered. However, the 
uncertainties in the fixed-order prediction are larger across the whole range, particularly for 
small $\ptveto$. For values of $\ptveto$ in the range of \SIrange{20}{40}{\GeV}, which are of 
practical interest, the \NNNLLpart{} uncertainty is smaller than the \NNLO{} uncertainty by 
about a factor of 1.5.
 
\subsubsection{ATLAS $W$ production}

We now perform a comparison with $\sqrt{s}=\SI{8}{\TeV}$ \ATLAS{} data on $W$ production 
\cite{ATLAS:2017irc}.  For this study, jets were identified using
the anti-$k_T$ algorithm with $R=0.4$ and must satisfy $p_T>30$~GeV and $|y|<4.4$.
We have checked at fixed order that this large rapidity cut has a negligible impact of a few per 
mille, i.e. results are unchanged within the numerical precision to which we work.

\begin{figure}
	\centering
	\includegraphics[width=0.9\textwidth]{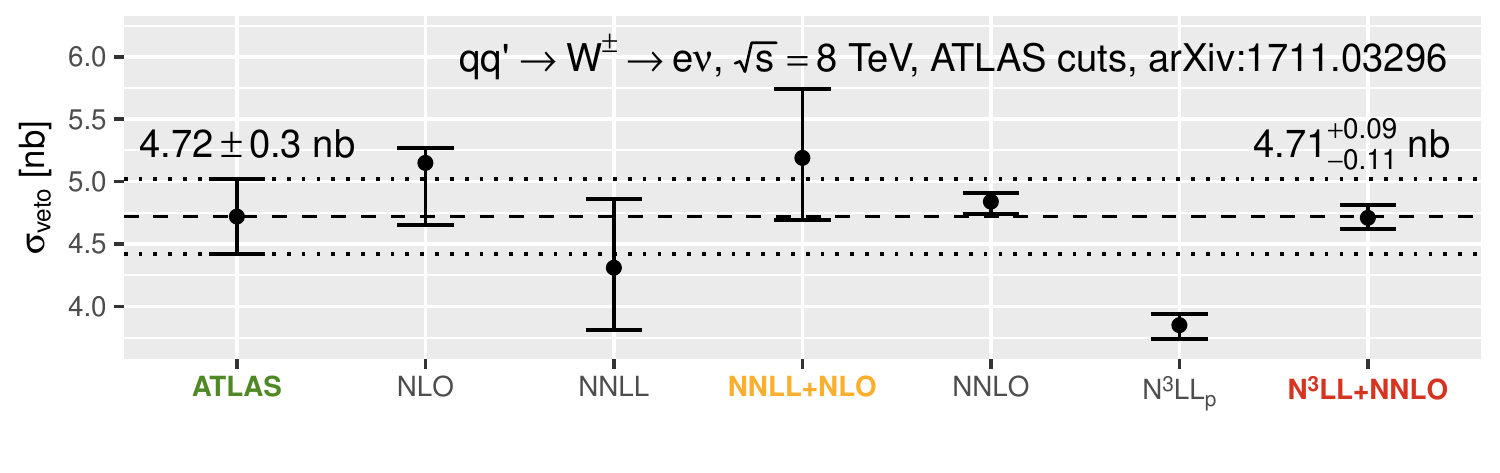}
	\vspace{-1em}
	\caption{Comparison of $W$-boson jet-vetoed predictions with the \ATLAS{} \cite{ATLAS:2017irc} 
		\SI{8}{\TeV} 
		measurement. Shown are results at fixed-order, purely resummed and matched.}
	\label{fig:Wjetvetocross}
\end{figure}

Summing over both $W$ charges and including only the decay into electrons we compare our 
predictions in \cref{fig:Wjetvetocross}. We show results at fixed order, at the resummed level, and 
at the matched level.
The effect of matching is large and we thus conclude that this value for the 
jet veto is outside the sensible range for a purely resummed 
result, unlike for the $Z$ study in the previous subsection.

We observe excellent agreement with the theoretical prediction, albeit with a larger
experimental uncertainty. The experimentally measured cross-section is \SI{4.72 \pm 0.3}{nb} while 
our best prediction is $4.71^{+0.09}_{-0.11}\,${nb}.
Since this measurement corresponds to an integrated luminosity
of only \SI{20}{fb^{-1}} it is clear that the high-luminosity \LHC{} will eventually be able to 
provide a much keener test of perturbative \QCD{} in this process.

\subsection{$W^+ W^-$ production}

Experimental studies of $WW$ production were performed by both 
\ATLAS{}~\cite{ATLAS:2017bbg,ATLAS:2019rob}
and \CMS{}~\cite{CMS:2019ppl,CMS:2020mxy}.  Here we focus on the \CMS{} analysis of 
ref.~\cite{CMS:2020mxy}
since it provides a measurement of the $0$-jet cross-section as a function of the jet $p_T$ veto.
This cross-section measurment corresponds to a sum over both electron and muon decays of the $W$ bosons,
which we denote by the label $pp \to W^- W^+ \to 2\ell 2\nu$.  In order to account for this in our
calculation, we compute the result for $pp \to e^- \mu^+ \bar\nu_e \nu_\mu$ at \NNLO{} and multiply
it by the factor that accounts exactly for all lepton combinations through \NLO{}.
The impact of $ZZ$ contributions in the same-flavor case results in a slight enhancement over 
the na\"ive factor of four.  We find that, independent of the value of the jet veto in the range
that we consider, this factor is equal to $4.15$.

The \CMS{} analysis only imposes a jet rapidity cut of $\etacut = 4.5$, so our expectation is that the 
standard jet veto resummation formalism should be appropriate for $p_T^\text{veto}$ values between 
60 and $10$~GeV, since in this case the logarithm of the ratio of $Q$ to $p_T^\text{veto}$ are in 
the range of 1.3 to 3.1. This expectation is supported by the \NNLO{} analysis in 
Table~\ref{tab:WWrapidity}, which shows only a small 2\%
effect from the rapidity cut for $p_T^\text{veto}=10$~GeV (and none for values above that). 
Unlike the processes considered so far, $Q$ is no longer set by a resonance mass but is instead
a distribution with a peak slightly above the $2 M_W$ threshold. For illustration, we have
used an average value of $Q \sim \SI{220}{\GeV}$.

\begin{table}[b]
	\begin{center}
			\caption{The $pp \to W^- W^+ \to 2\ell 2\nu+$0-jet cross-section at \NNLO{}, with and without a jet 
			rapidity cut.}
			\label{tab:WWrapidity}
		\begin{tabular}{l|cccccc}
			$p_T^\text{veto}$ [GeV] & 10 & 25 & 30 & 35 & 45 & 60 \\
			\hline
			$\sigma_{0-\text{jet}}(\text{no}~\etacut)$~[fb]    & 535 & 963 & 1004 & 1054 & 1145 & 1237 \\ 
			$\sigma_{0-\text{jet}}(\etacut = 4.5)$~[fb]        & 548 & 963 & 1004 & 1054 & 1145 & 1237 \\ 
			\hline
			$\epsilon$ & 0.02 & 0.00 & 0.00 & 0.00 & 0.00 & 0.00
		\end{tabular}
	\end{center}
\end{table}

\begin{figure}
	\centering
	\begin{subfigure}[t]{0.48\textwidth}
		\centering
		    \includegraphics[width=\textwidth,angle=-90]{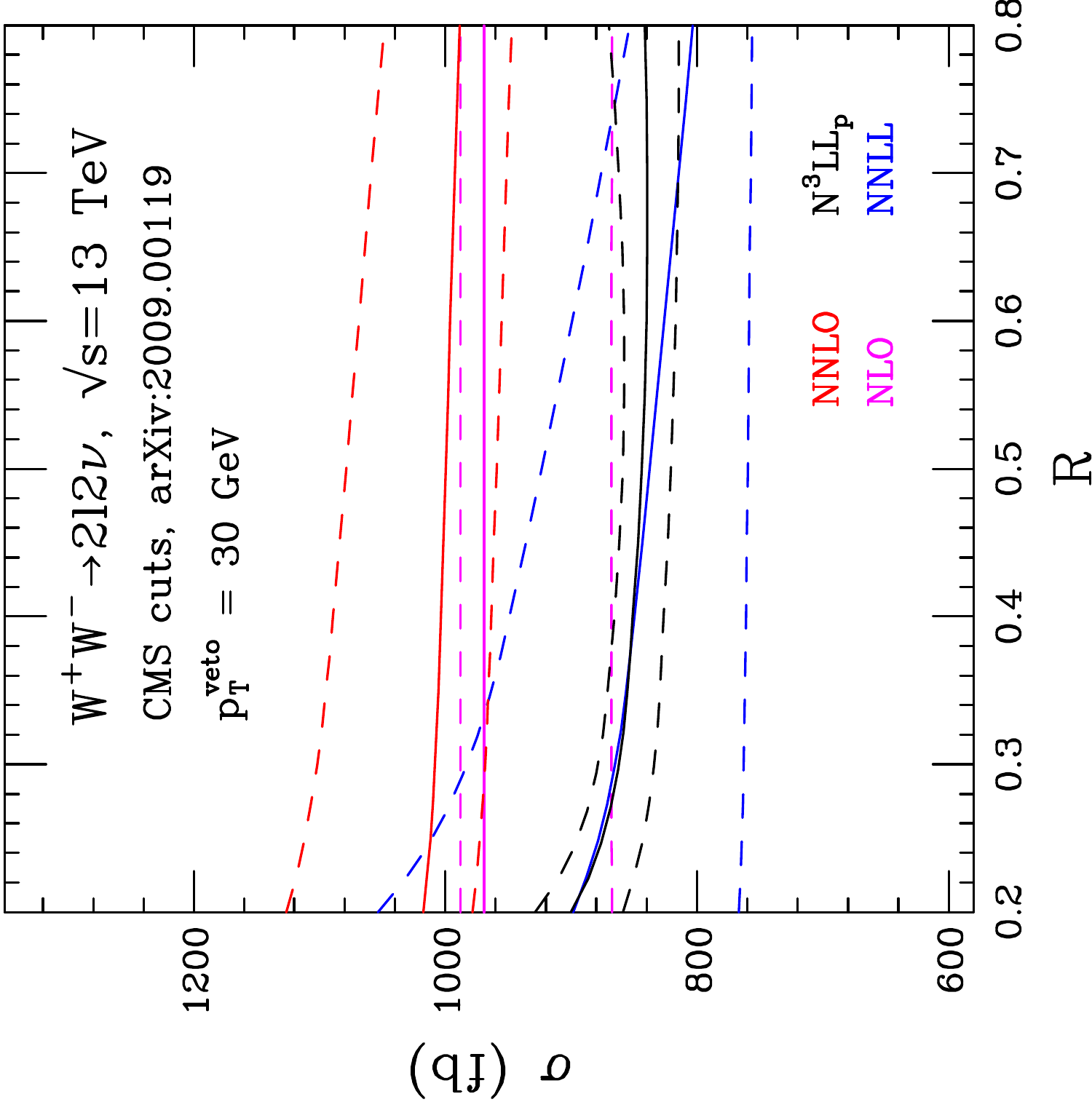}
		\caption{Jet radius $R$ dependence of fixed-order and purely resummed results.
			\label{fig:ww13vetoRstudy}}
	\end{subfigure}
	\hfill
	\begin{subfigure}[t]{0.48\textwidth}
		\centering
	    \includegraphics[width=\textwidth,angle=-90]{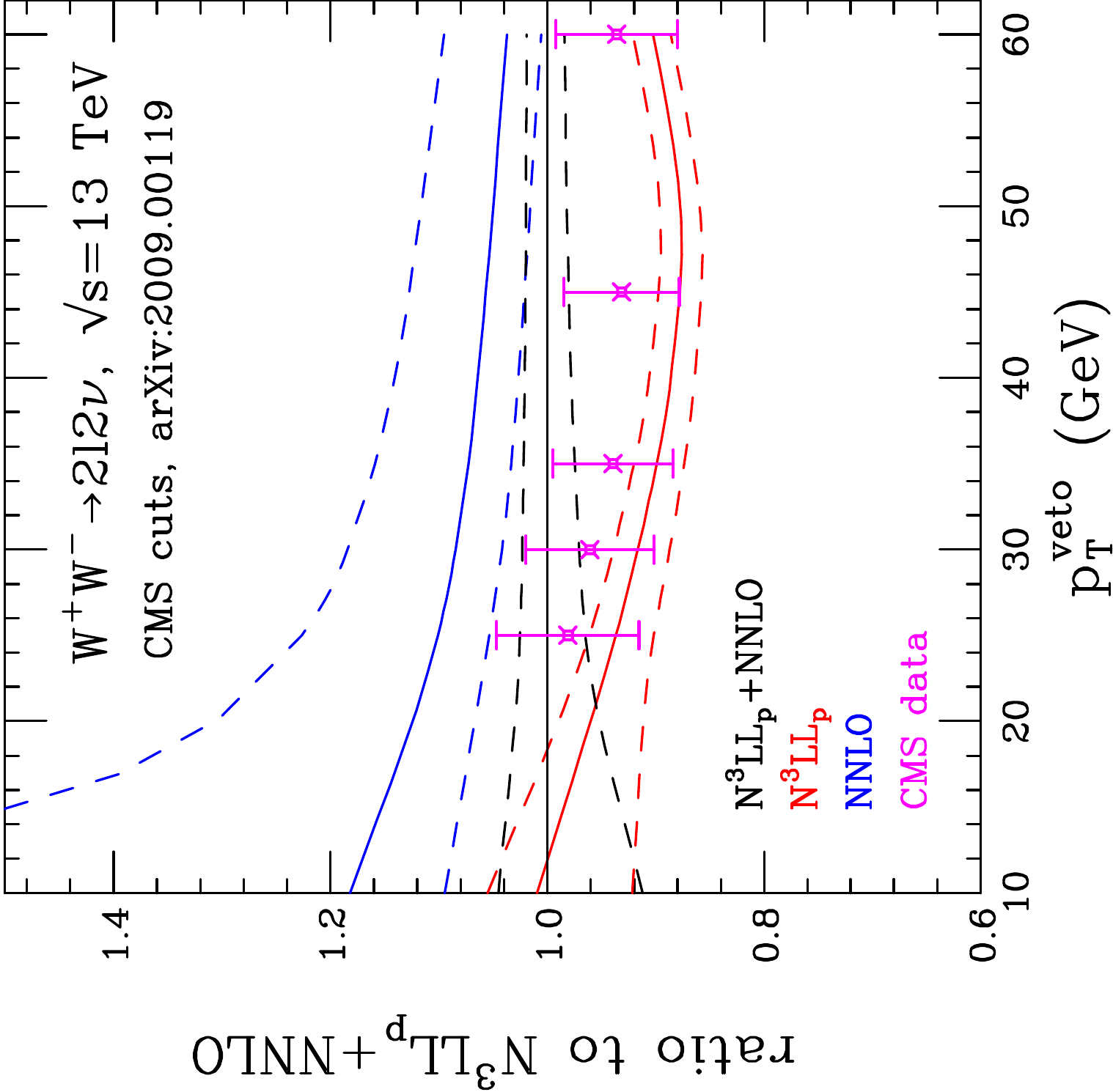}
	\caption{Predictions and \CMS{} measurement as a ratio to the matched result.
		\label{fig:ww13vetomatched}}
	\end{subfigure}

\caption{Comparison of \NNLO{}, \NNNLLpart{} and matched \NNNLLpart{}+\NNLO{}
	results for $W^+W^-$ production. 
}
\end{figure}

We first fix the value of $\ptveto = \SI{30}{\GeV}$ and study the sensitivity of the pure 
fixed-order 
and resummed calculations to the jet-clustering parameter $R$. The results are shown in 
Fig.~\ref{fig:ww13vetoRstudy}.
At \NLO{}, there is at most one additional parton, so the \NLO{} result does not depend on the 
value of $R$. 
However, the \NNLL{} result exhibits a mild dependence on $R$, which is most noticeable in the 
size 
of the uncertainties. These uncertainties are much larger for smaller values of $R$, as was 
previously observed and discussed in the context of Higgs production in Ref.~\cite{Becher:2013xia}.
At \NNLO{}, the fixed-order calculation becomes sensitive to the value of $R$, although the 
dependence is very small. At \NNNLLpart{}, the dependence is reduced compared to \NNLL{}, 
especially at small $R$. Overall, these results suggest that the jet-clustering 
parameter has a mild effect on the predictions of the fixed-order and resummed calculations for 
$WW$ production. We have not investigated the effect of small $R$ resummation~\cite{Banfi:2015pju} on these results.

\begin{figure}
	\centering
	\includegraphics[width=0.5\textwidth,angle=-90]{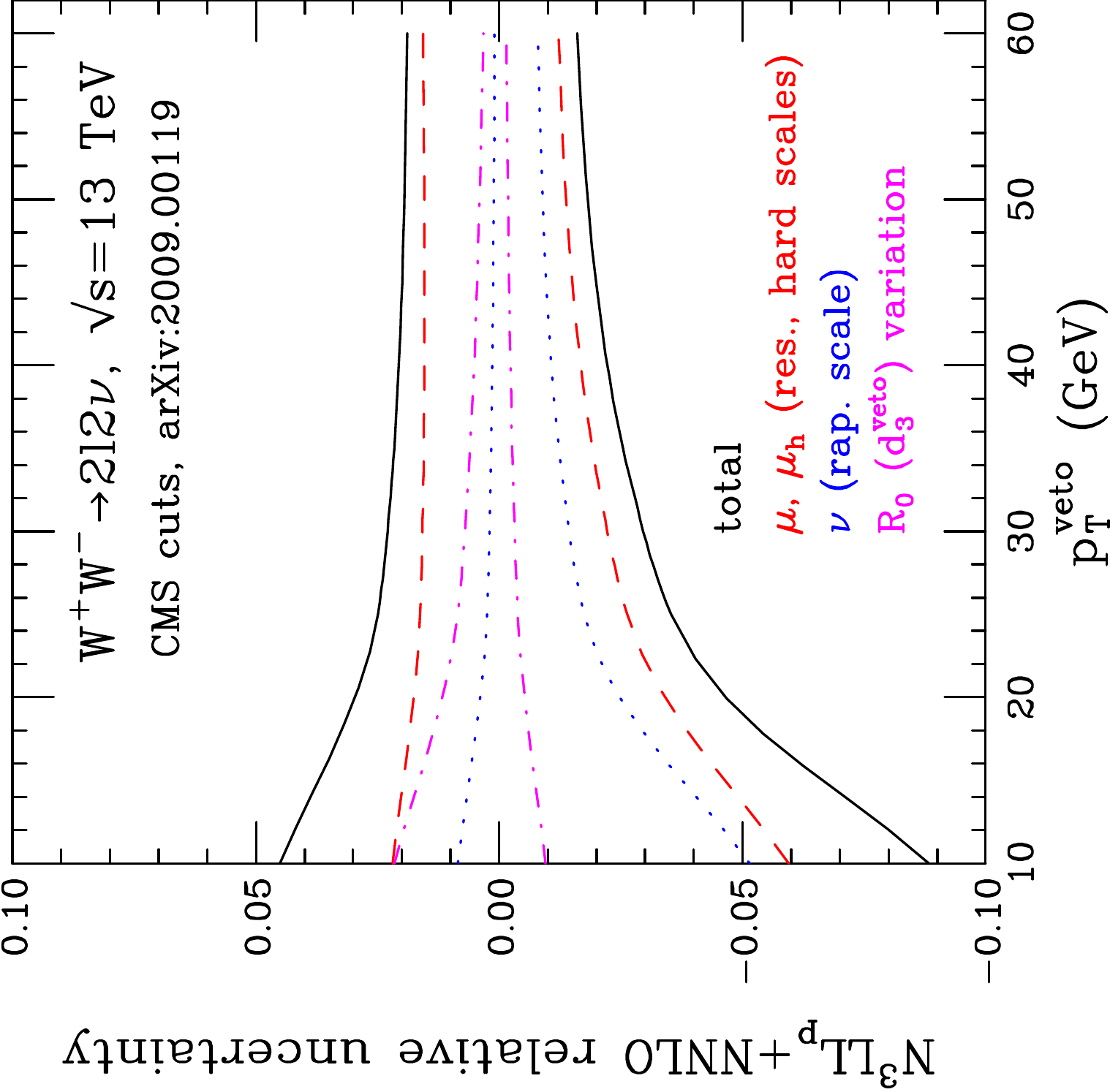}
        \caption{Uncertainty breakdown of the \NNNLLpart{}+\NNLO{}
	results for $W^+W^-$ production.  \label{fig:ww13vetomatched_split}}
\end{figure}

In Fig.~\ref{fig:ww13vetomatched}, we extend our previous analysis of the jet-veto dependence of 
$WW$ production, which was presented in Ref.~\cite{Campbell:2022uzw}. The effect of matching is 
substantial for values of $\ptveto$ greater than \SI{20}{\GeV}, so for typical jet vetoes in the 
range of \SIrange{20}{40}{\GeV}, matched predictions are important.
We find that the fixed-order description is only capable of providing an adequate result for the 
highest value of $\ptveto$ studied here. A comparison with the \CMS{} measurement shows better 
agreement with the matched resummed calculation, although the experimental uncertainties are still 
substantial, corresponding to an integrated luminosity of 36~fb${}^{-1}$.
A breakdown of the estimated uncertainty on the matched \NNNLLpart{}+\NNLO{} prediction,
into the categories described in \cref{sec:uncertainties}, is shown in \cref{fig:ww13vetomatched_split}.
The uncertainty from the variation of the hard (renormalization) and resummation (factorization)
scales dominates, except for the very lowest values of $\ptveto$ where the uncertainty on
$\dthreeveto$ becomes significant.

We eagerly anticipate a measurement with more statistics in order to hone this comparison. Future 
measurements with higher precision and larger data samples will 
provide a more stringent test of the theoretical predictions and help to refine our understanding 
of $WW$ production at the \LHC{}.

\subsection{$W^\pm Z$ production}

\subsubsection{ATLAS}

For $W^\pm Z$ production, we first compare our results with an analysis from the \ATLAS{} 
collaboration at $\sqrt s = \SI{13}{\TeV}$ \cite{ATLAS:2019bsc}. The $0$-jet cross-section is 
measured with jets defined by the anti-$k_T$ algorithm with $p_T>25$~GeV, $|y|<4.5$, and $R=0.4$.

Since $\ln(Q/p_T^\text{veto}) =2.3$ (for $p_T^\text{veto}=\SI{25}{\GeV}$,
using an average $Q$ of about \SI{240}{\GeV}), we expect that standard jet veto resummation 
should be applicable in this case, since $\etacut = 4.5$. We have checked that the effect of the 
rapidity cut is at the per mille level, which is less than our numerical precision.

\begin{figure}
	\centering
	\includegraphics[width=0.9\textwidth]{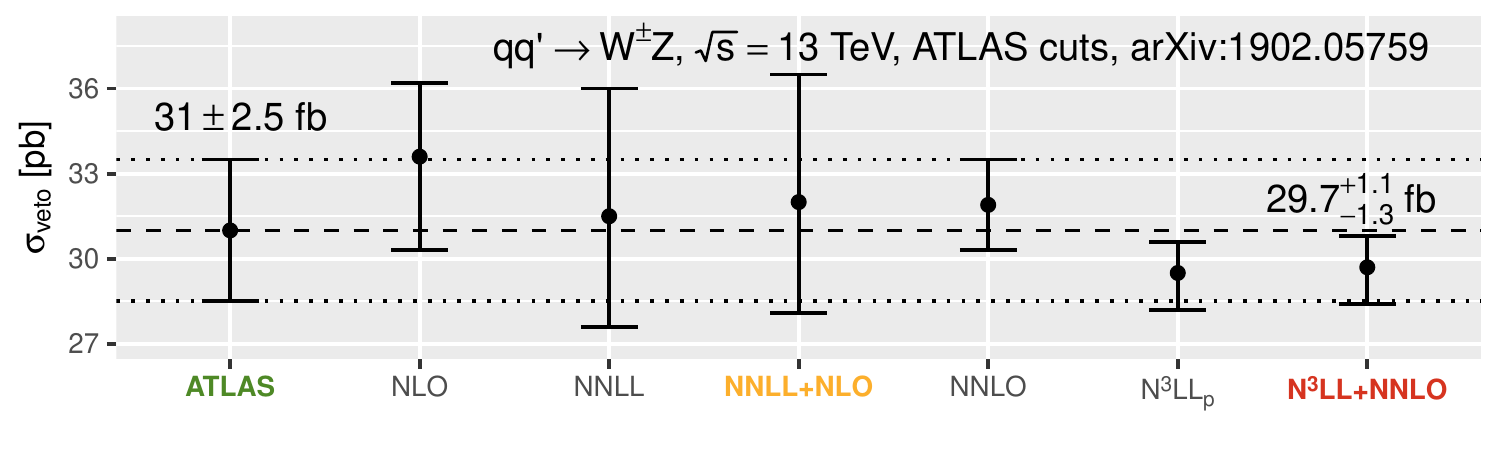}
	\vspace{-1em}
	\caption{Comparison of $W^\pm Z$ jet-vetoed predictions with the \ATLAS{} 
		\SI{13}{\TeV} 
		measurement \cite{ATLAS:2019bsc}. Shown are results at fixed order, purely resummed and 
		matched.}
	\label{fig:WZjetvetocross}
\end{figure}

The \ATLAS{} result is presented for a single leptonic channel and summed over both $W$ charges. 
The corresponding theoretical predictions at fixed order, at the resummed level, and at the matched 
level are shown in \cref{fig:WZjetvetocross}.

Overall, the measurement is in good agreement with both the \NNNLLpart{}+\NNLO{} and \NNLO{} 
predictions, within the mutual uncertainties. Only a more precise measurement would be able to 
definitively support the need for resummation in this case. Since the \ATLAS{} analysis includes 
only \SI{36}{fb^{-1}} of data, it is likely that a more precise measurement will be possible in the 
near future.

\subsubsection{CMS}

We now contrast the \ATLAS{} study of the $W^\pm Z$ process with one from 
\CMS{}~\cite{CMS:2021icx}. In the \CMS{} study, jets are defined by the anti-$k_T$ algorithm with 
$p_T>25$~GeV, $|y|<2.5$, and $R=0.4$.

To assess the applicability of the jet-rapidity inclusive resummation framework, we must compare 
$\ln(Q/p_T^\text{veto}) =2.3$ with $\etacut = 2.5$. This suggests that the standard jet veto 
resummation formalism may not be appropriate in this case, and that the use of $\etacut$-dependent 
beam functions~\cite{Michel:2018hui} may be necessary to provide a reliable theoretical prediction.
Despite this, we still pursue the comparison here, without using $\etacut$-dependent beam 
functions, to examine the limitations of our approach.

\begin{figure}
	\centering
	\includegraphics[width=0.9\textwidth]{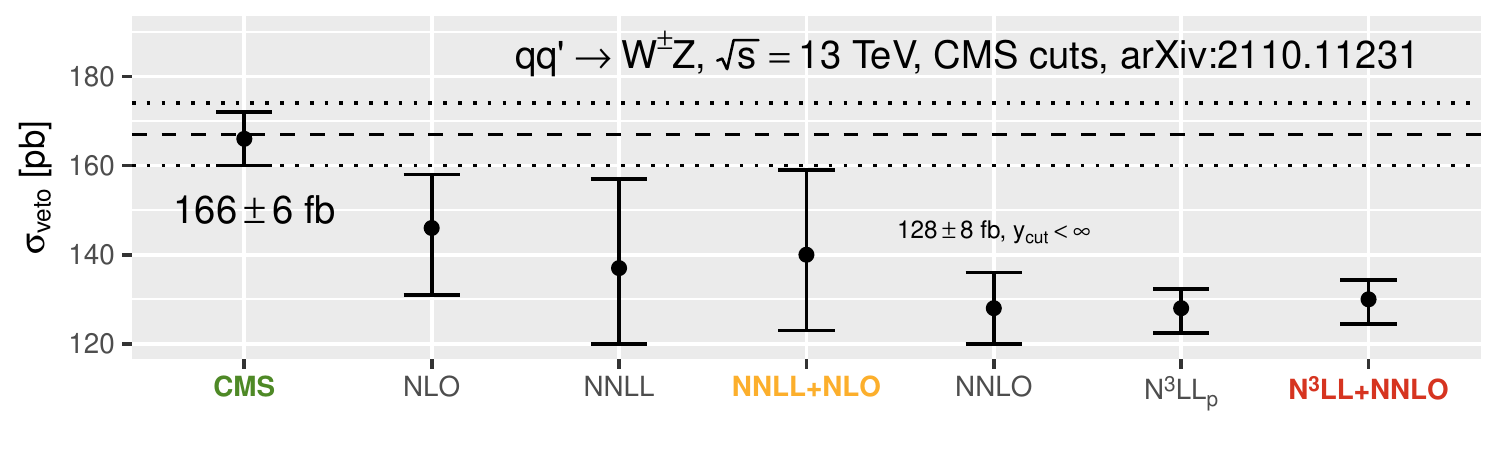}
	\vspace{-1em}
	\caption{Comparison of $W^\pm Z$ jet-vetoed predictions with the \CMS{} \cite{CMS:2021icx} 
		\SI{13}{\TeV} 
		measurement. Shown are results at fixed-order, purely resummed and matched, all 
		without a rapidity cut.}
	\label{fig:WZjetvetocross_CMS}
\end{figure}

The \CMS{} result for $W^\pm Z$ production is presented after summing over all lepton flavors and 
both $W$ charges. On the theoretical side, we perform a similar analysis, but ignore same-flavor 
effects that only enter at the 2\% level.
To construct the jet-vetoed cross-section for the \CMS{} measurement, we combine the differential 
results in Figure 14(c) of Ref.~\cite{CMS:2021icx} with the inclusive cross-sections reported in 
Table~6 of the same reference. Our results are shown in Fig.~\ref{fig:WZjetvetocross_CMS}.

We find that neither the resummed prediction nor the \NNLO{} one are in good agreement with the 
\CMS{} data, even when the \NNLO{} calculation takes the jet rapidity cut into account (increasing 
the \NNLO{} result from \SI{128}{fb} to \SI{137}{fb}). This 
suggests that resummation is required in this case, and that the use of $\etacut$-dependent beam 
functions is necessary to provide a reliable theoretical prediction. Overall, these results 
highlight the importance of using appropriate resummation techniques to 
accurately predict $W^\pm Z$ production at the \LHC{} with a small jet rapidity cut.

\subsection{$ZZ$ production}

In the absence of jet-vetoed cross-sections for comparison, we use the cuts from a recent \CMS{} 
study~\cite{CMS:2020gtj} to investigate our theoretical predictions for $ZZ$ production as a 
function of $\ptveto$.  In the results that follow
we consider a sum over $Z$ decays into both electrons and muons, which we denote by $pp \to ZZ \to 4$~leptons,
and apply the cuts shown in \cref{tab:ZZ-CMS}. 
\begin{table}
	\begin{center}
		\begin{tabular}{r | l }
			{ lepton cuts} & $q_T^{l_1} > \SI{20}{\GeV}$, $q_T^{l_2} > \SI{10}{\GeV}$, \\
			&  $q_T^{l_{3,4}} > \SI{5}{\GeV}$, $|\eta^{l}|<2.5$ \\
			{lepton pair mass}& $\SI{60}{\GeV}<m_{l^-l^+}<\SI{120}{\GeV}$ \\
			{jet veto}& anti-$k_T$, $R=0.5$
		\end{tabular}
	\end{center}
	\caption{\label{tab:ZZ-CMS} Fiducial cuts used for the $ZZ$ analysis, taken
	from the \CMS{} study in Ref.~\cite{CMS:2020gtj}.}
\end{table}

We expect that standard jet veto resummation should provide good predictions for $\etacut = 4.5$, 
since $\ln(Q/p_T^\text{veto})$ is in the range of 1.4 to 3.2 for $p_T^\text{veto}$ values between 
60 and \SI{10}{\GeV}, using an average $Q$ of about \SI{240}{\GeV}.
For $\etacut = 2.5$, we expect larger rapidity effects for the smallest values of $p_T^\text{veto}$.
This is supported by our analysis in Table~\ref{tab:ZZrapidity}, which shows only a very small 
(1\%) effect from a rapidity cut of $\etacut = 4.5$ for $p_T^\text{veto}=\SI{10}{\GeV}$ (and no 
effect for 
higher values). Even for $\etacut = 2.5$, the rapidity cut has a relevant effect only for 
$p_T^\text{veto}$ values below \SI{30}{\GeV}, and is mostly insignificant beyond that.

\begin{table}
\begin{center}
	\caption{The $ZZ+0$-jet cross-section at \NNLO{} ($\mu = Q$), with and without a jet 
	rapidity cut.}
\label{tab:ZZrapidity}
\begin{tabular}{l|cccccc}
$p_T^\text{veto}$ [GeV] & 10 & 20 & 30 & 40 & 50 & 60 \\
\hline
$\sigma_{0-\text{jet}}(\text{no}~\etacut)$~[fb]    & 13.3 & 21.5 & 25.8 & 28.4 & 30.3 & 31.6 \\
$\sigma_{0-\text{jet}}(\etacut = 4.5)$~[fb]        & 13.4 & 21.5 & 25.8 & 28.4 & 30.3 & 31.6  \\
$\sigma_{0-\text{jet}}(\etacut = 2.5)$~[fb]        & 14.9 & 22.4 & 26.3 & 28.8 & 30.6 & 31.8  \\
\hline
$\epsilon(\etacut = 4.5)$ & 0.01 & 0.00 & 0.00 & 0.00 & 0.00 & 0.00  \\
$\epsilon(\etacut = 2.5)$ & 0.12 & 0.04 & 0.02 & 0.01 & 0.01 & 0.01
\end{tabular}
\end{center}

\end{table}

\begin{figure}
\centering
	\begin{subfigure}[t]{0.48\textwidth}
		\centering
    \includegraphics[width=\textwidth,angle=-90]{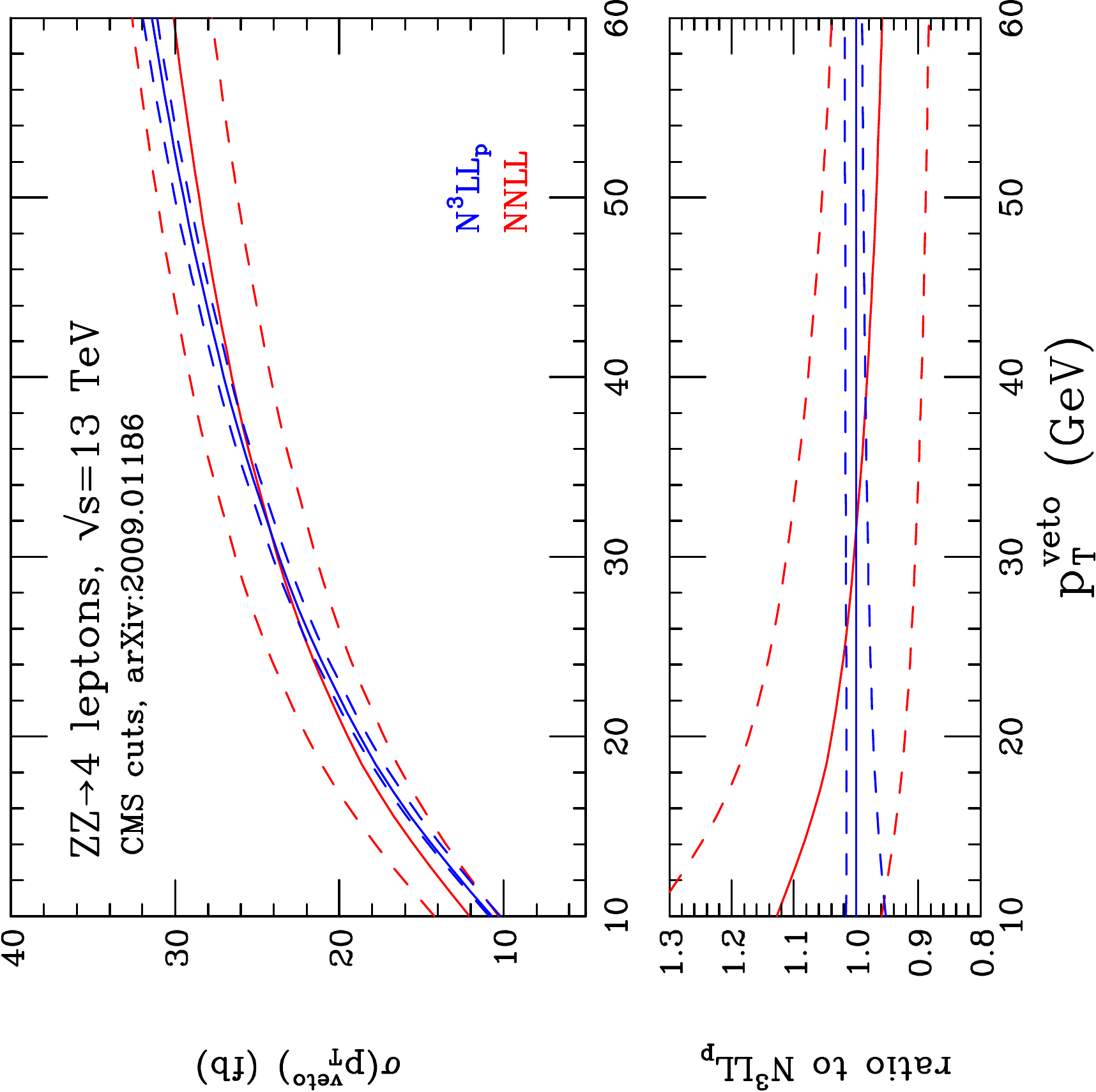}
    \caption{Purely resummed results.
    \label{fig:ZZresorders}}
	\end{subfigure}
	\hfill
	\begin{subfigure}[t]{0.48\textwidth}
		\centering
    \includegraphics[width=\textwidth,angle=-90]{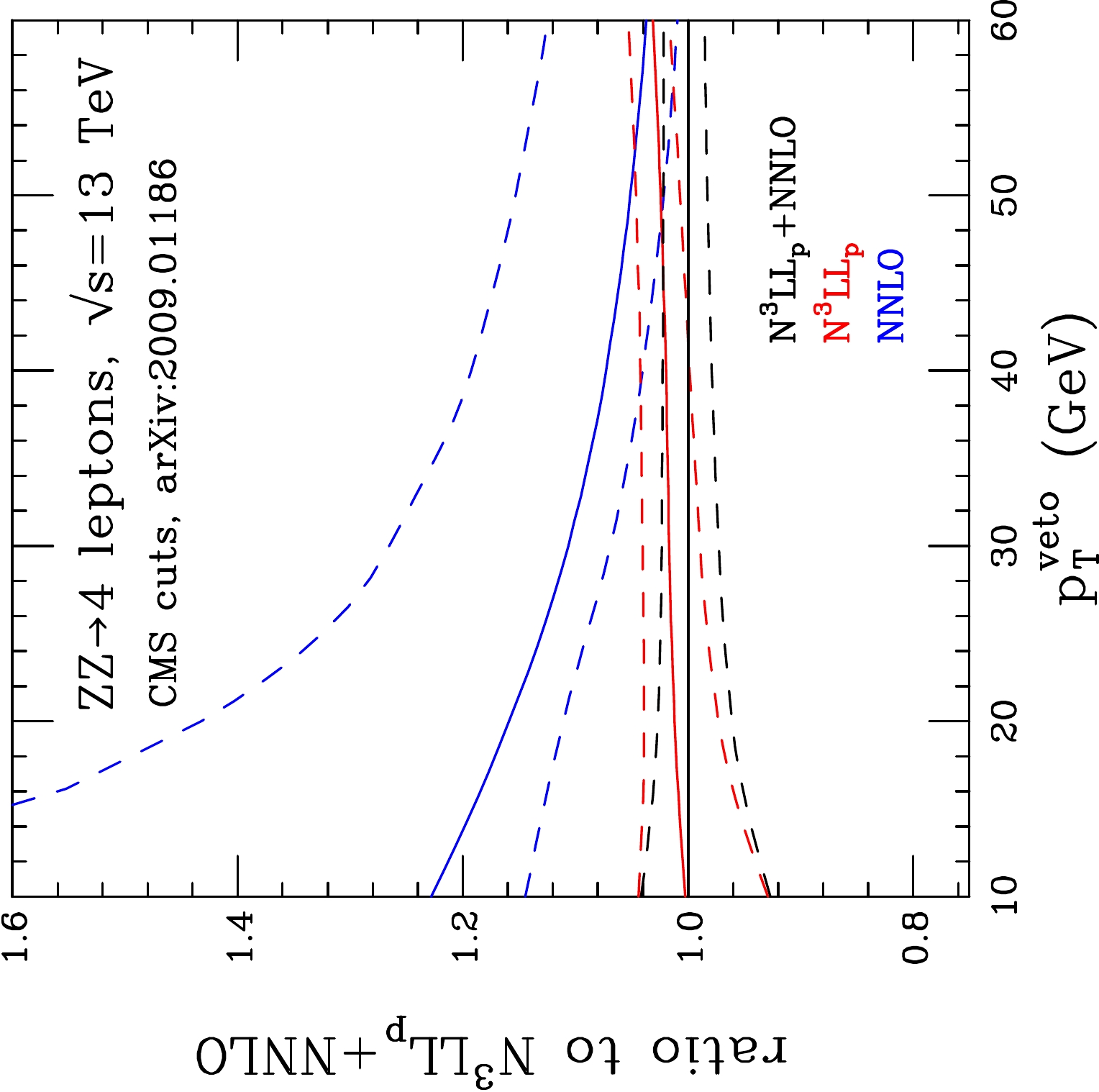}
\caption{Ratio to matched result.
	\label{fig:zz13vetomatched}}
	\end{subfigure}

\caption{
Comparison of \NNLO{}, \NNNLLpart{} and matched \NNNLLpart{}+\NNLO{}
results for $ZZ$ production as a function of the jet veto.
}

\end{figure}
Fig.~\ref{fig:ZZresorders} shows a comparison of the dependence on $\pTveto$  
for purely-resummed results at two different logarithmic orders.
The central predictions are very similar at \NNLL{} and \NNNLLpart{} and are consistent 
within uncertainties for all values of $\ptveto$. Fig.~\ref{fig:zz13vetomatched} compares the 
matched \NNNLLpart{}+\NNLO{} and \NNLO{} results. The \NNLO{} prediction has large uncertainties 
over the whole range of $\ptveto$ and only overlaps with \NNNLLpart{}+\NNLO{} around \SI{40}{\GeV}
and higher.
The difference between the central resummed and 
fixed-order results is significant (around 10\%) for typical values of $p_T^\text{veto}$ around 
\SI{30}{\GeV}. For most relevant values of $\ptveto$ at the \LHC{}, resummation is clearly 
important for providing a precision prediction for this process.

\subsection{Higgs production}
\label{Higgsproduction}
For gluon fusion Higgs production an important topic is the inclusion of finite top-quark mass 
effects. Although at \NNLO{} these could be included exactly~\cite{Bonciani:2022jmb,Czakon:2021yub},
the mass effects are not relevant in the jet-vetoed case \cite{Neumann:2014nha} at the current level of precision.
A simple overall one-loop rescaling factor that takes into account the full mass dependence is 
sufficient to introduce mass effects into $m_t\to\infty$ {\abbrev EFT} predictions.
In the resummation formalism, the coefficient for the matching of Higgs production 
in \QCD{} onto \SCET{} can be calculated in two ways, referred to as one-step and two-step 
procedures.

\subsubsection{One-step and two-step schemes}

The one-step procedure is based on the observation that the ratio $m_H/m_t$ is not large in a
logarithmic sense (c.f. $\rho=m_H^2/m_t^2 \approx 1/2$ and $\alpha_s \log 1/\rho \approx 0.07$). This procedure matches the full \QCD{} 
result, typically obtained at higher orders as an expansion in the parameter $r$, onto \SCET{} at 
the scale $\mu_h \sim m_H$. In this way, terms of order $\rho$ are retained, but logs of $m_t/m_H$ are 
neglected.

In the two-step procedure outlined in 
Refs.~\cite{Idilbi:2005er,Idilbi:2005ni,Ahrens:2009cxz,Mantry:2009qz}, the top quark is first 
integrated out at a scale $\mu_t \approxeq m_t$, and then the \QCD{} effective Lagrangian is 
matched onto the \SCET{} at a scale $\mu_h \approxeq m_H$.
Running between $\mu_t$ and $\mu_h$ allows one to sum logarithms of $m_t/m_H$, and finite top-mass 
effects
are included by scaling the result by a correction factor obtained at leading order
(an increase with respect to the {\abbrev EFT} result by a factor of $1.0653$, see 
Eq.~(\ref{massrescaling})).  Terms enhanced by powers of $m_H/m_t$ are thus only included in an 
approximate fashion at \NLO{} and beyond.
The one-step procedure is described in detail in
Appendix~\ref{onestep} and the two-step procedure is described in Appendix~\ref{twostep}.

\begin{figure}
	\centering
\begin{subfigure}[t]{0.48\textwidth}
	\centering
    \includegraphics[width=\textwidth,angle=-90]{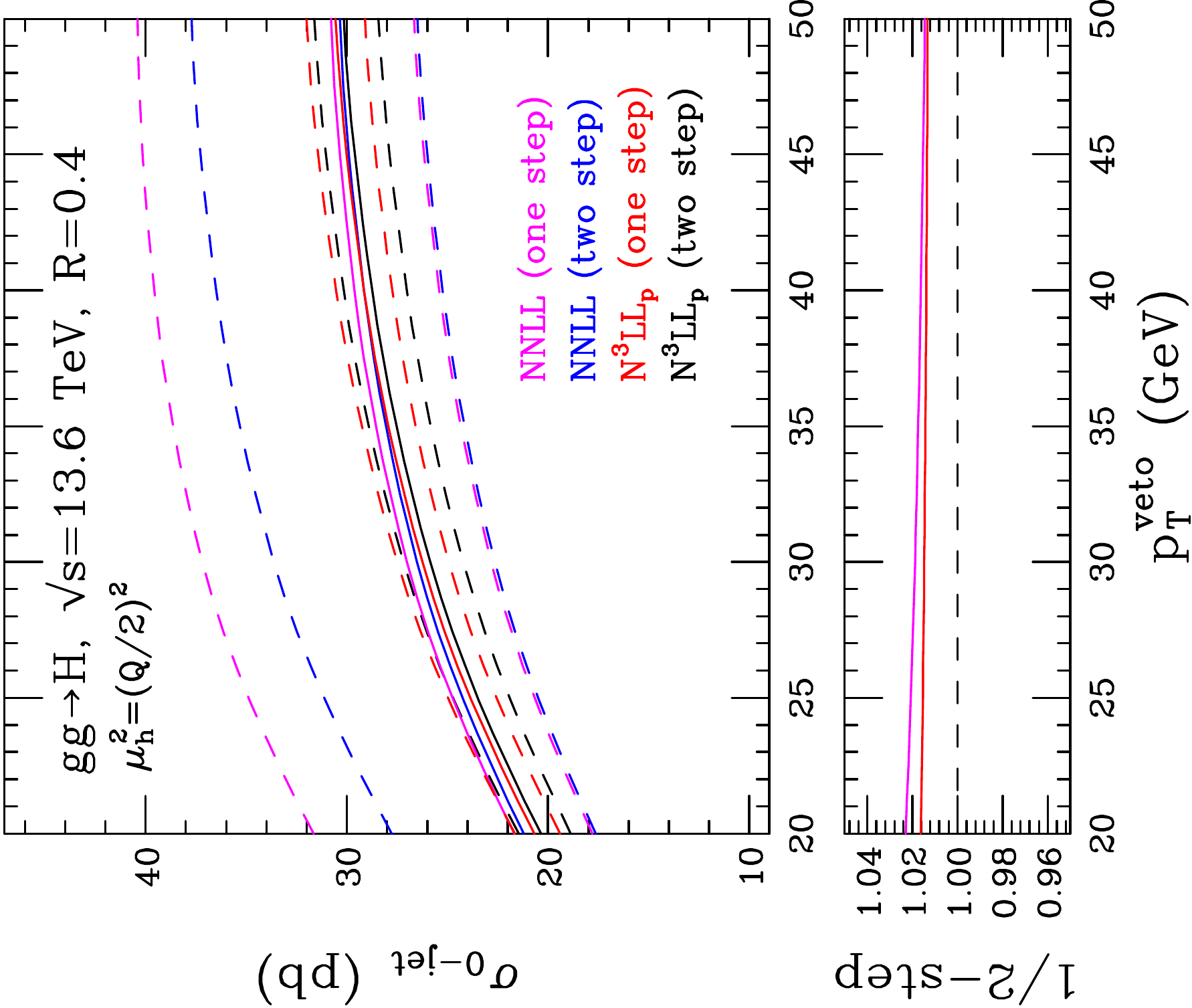}
    \caption{Results in the one- or two-step scheme.
    The lower panel shows the ratio of the
    one-step to the two-step result.
    \label{fig:1511singlestudy}}
\end{subfigure}
\hfill
\begin{subfigure}[t]{0.48\textwidth}
\centering
    \includegraphics[width=\textwidth,angle=-90]{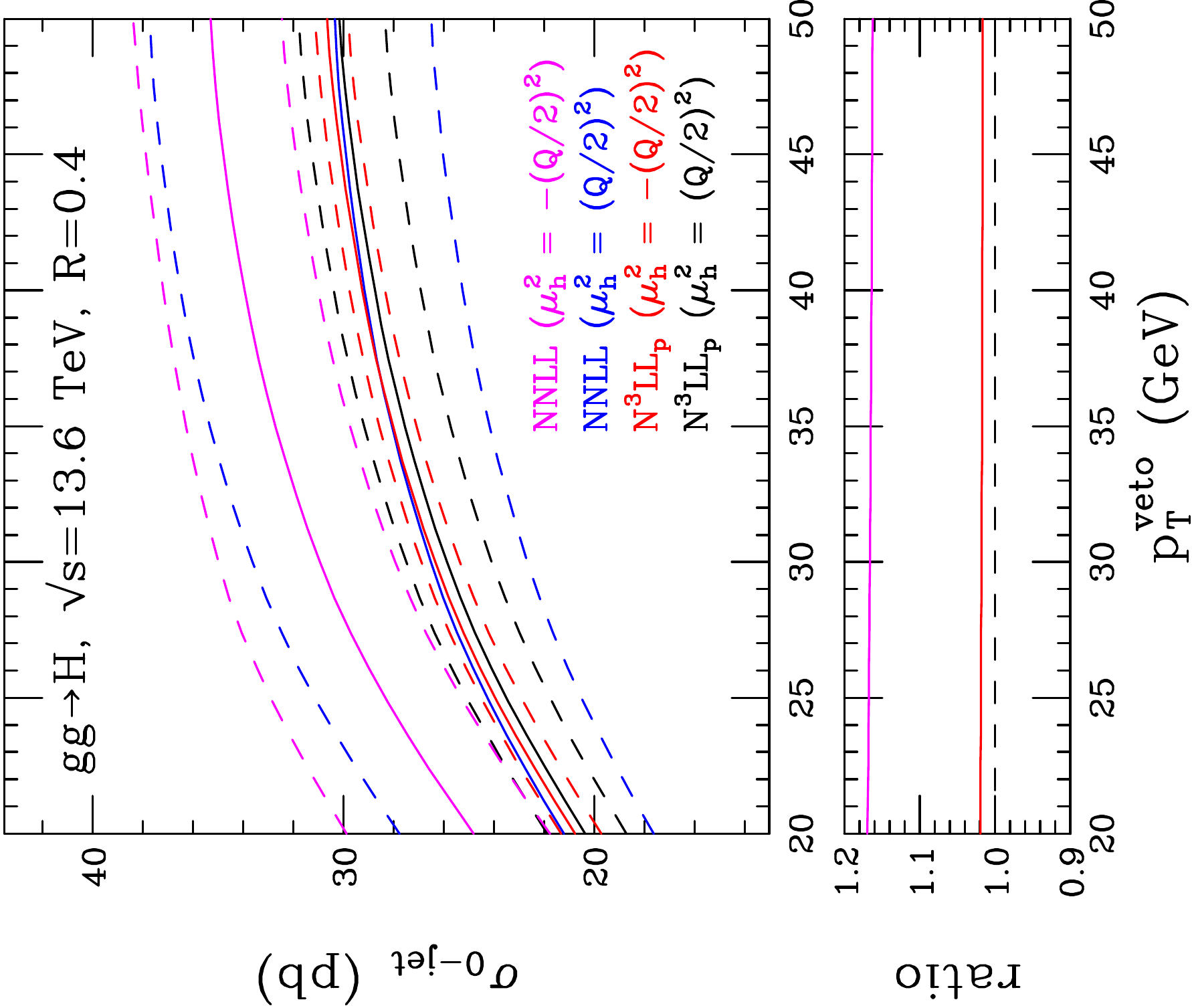}
\caption{Results using a central scale of either $\mu_h^2 = Q^2$
	or $\mu_h^2 = -Q^2$. The lower panel shows the ratio of the result
	for $\mu_h^2 = -Q^2$ to the one for $\mu_h^2 = Q^2$.
	\label{fig:H136timestudy}}
\end{subfigure}
\caption{Comparison of \NNLL{} and \NNNLLpart{} predictions
	for Higgs production at $\sqrt{s}=13.6$~TeV as a function of the jet veto.}
\end{figure}

We compare the numerical difference between the one-
and two-step schemes, computed at $\sqrt s = 13.6$~TeV and
for $R=0.4$ in Fig.~\ref{fig:1511singlestudy}.  Guided 
by fixed-order results, and in accord with previous studies of this process~\cite{Banfi:2015pju},
we set the hard (renormalization) scale using $\mu_h = Q/2$.
We observe that the one-step
scheme results in a cross-section that is about $1.7$--$2.3\%$ larger at \NNLL{} and only
$1.6\%$ larger at \NNNLLpart{}. 
This small difference occurs if one works rigorously at a fixed order of $\alpha_s$.  Working at a fixed order in 
$\alpha_s$ in the component parts of the two-step scheme can lead to larger differences, as described in more 
detail in \cref{sec:Hschemes}.

\subsubsection{Time-like vs. space-like $\mu_h^2$}

We now study the impact of choosing a time-like hard scale for the calculation of the Higgs 
cross-section. To do this, we compare $\mu_h^2 = (Q/2)^2$ (the space-like scale) with $\mu_h^2 = 
-(Q/2)^2$ (the time-like scale). The use of a time-like hard scale allows us to resum certain 
$\pi^2$ terms, by employing a complex strong coupling \cite{Ahrens:2008qu}. For this comparison, we 
consider purely resummed results at \NNLL{} and \NNNLLpart{} accuracy.

Results are shown in Fig.~\ref{fig:H136timestudy}, for the two-step scheme computed at $\sqrt s = 
13.6$~TeV with $R=0.4$. We observe that at \NNLL{}, the resummation of the $\pi^2$ terms 
significantly enhances the cross-section by 17\%. However, at \NNNLLpart{} accuracy, this 
resummation only leads to a small increase of 2\% in the cross-section.

\begin{figure}
	\centering
\begin{subfigure}[t]{0.48\textwidth}
\centering
    \includegraphics[width=\textwidth,angle=-90]{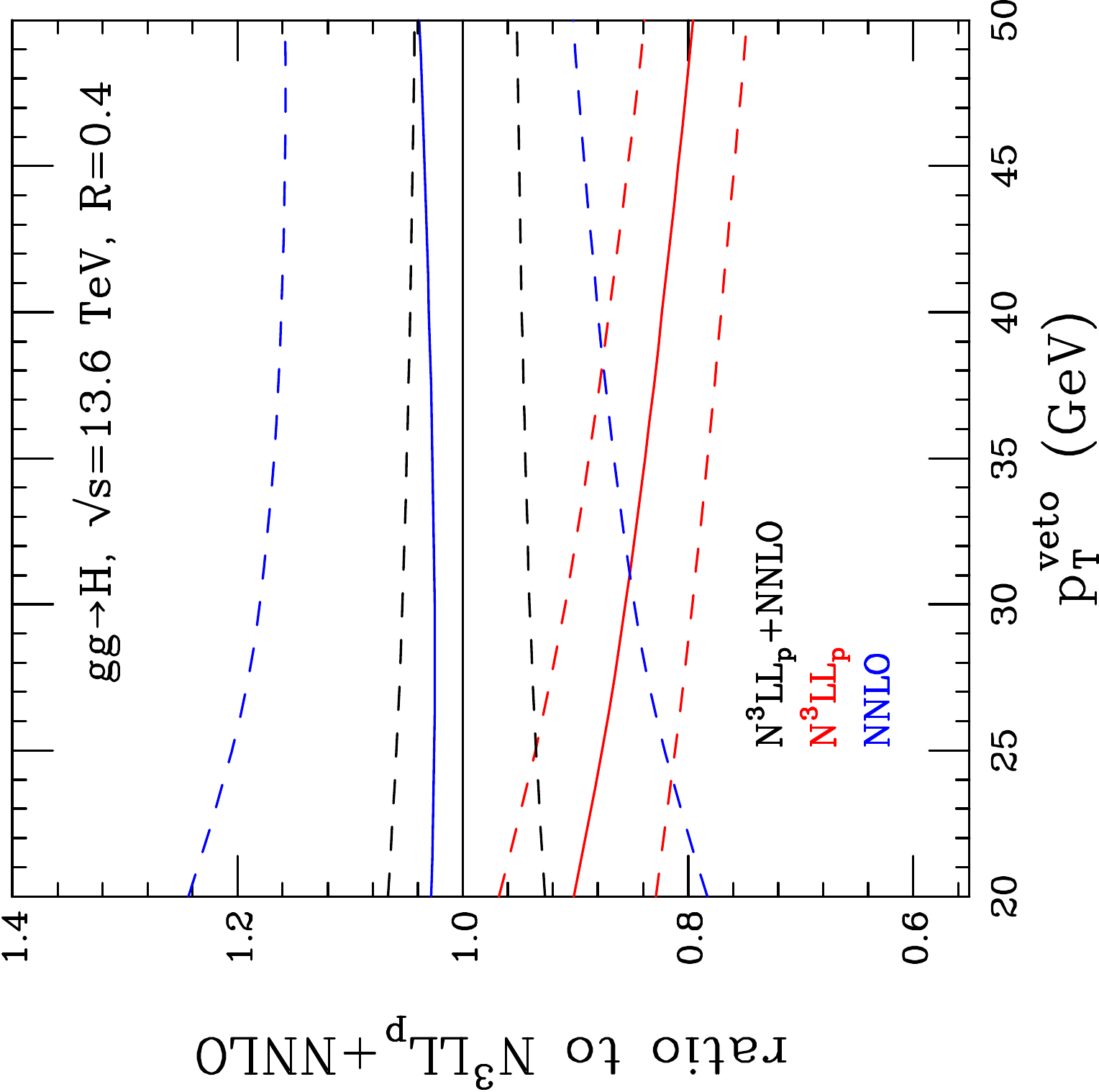}
    \caption{Comparison of \NNLL{}, \NNNLLpart{} and \NNNLLpart{}+\NNLO{} predictions.
    \label{fig:H136vetomatched}}
\end{subfigure}
\hfill
\begin{subfigure}[t]{0.48\textwidth}
	\centering
	\includegraphics[width=\textwidth,angle=-90]{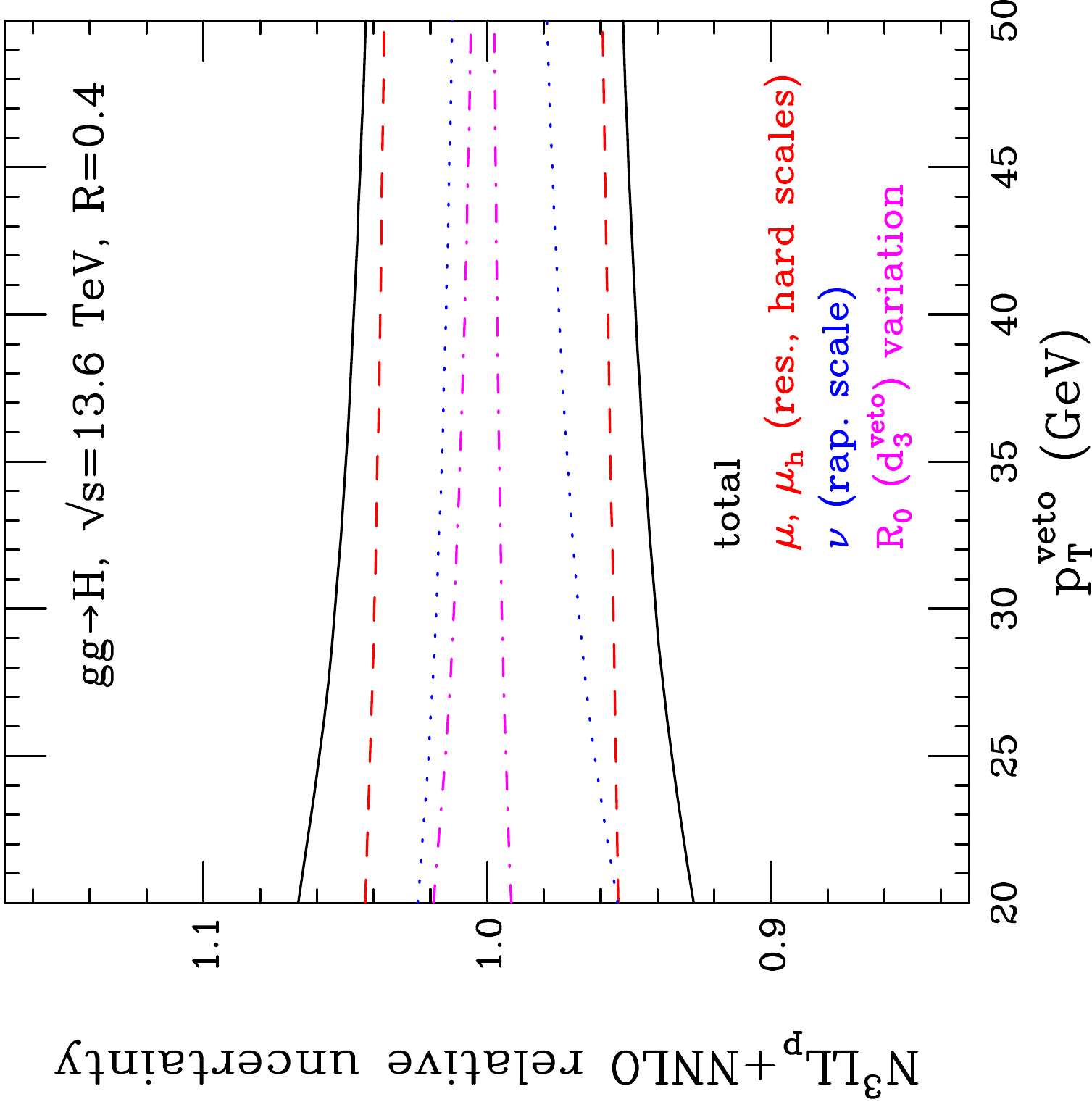}
        \caption{Uncertainty breakdown of the \NNNLLpart{}+\NNLO{}
	prediction.  \label{fig:H136vetomatched_split}}
\end{subfigure}
\caption{Results for Higgs production at $\sqrt{s}=13.6$~TeV as a function of the jet veto.}
\end{figure}

Results for the matched vetoed cross-section are shown in
Fig.~\ref{fig:H136vetomatched}.
After matching, we observe  substantial agreement between the \NNLO{} and \NNNLLpart{}+\NNLO{}
calculations within uncertainties. The central predictions differ
by about 5\% across the range, but the uncertainties are substantially smaller
in the resummed calculation.

The estimated uncertainty on the matched \NNNLLpart{}+\NNLO{} prediction,
broken down into the various sources that we consider, is shown in \cref{fig:H136vetomatched_split}.
Although the uncertainty from the variation of
$\dthreeveto$ reaches 2\% for $\ptveto=20$~GeV, the uncertainty from the variation of the
hard (renormalization) and resummation (factorization) scales dominates across the entire range.

\section{Conclusions}
\label{sec:conclusions}

We have presented a comprehensive study of jet-veto resummation in the production of 
color singlet final states using the most up-to-date theoretical ingredients and achieving 
\NNNLLpart{} accuracy. Our implementation in \MCFM{} improves upon previous public 
\NNLL{} 
calculations by reducing theoretical uncertainties, as demonstrated by comparisons with \ATLAS{} 
and \CMS{} results. Once the one remaining theoretical element, $\dthreeveto$, becomes available, 
it will be simple to upgrade our predictions to full \NNNLL{} accuracy.

The primary motivation for this work comes from the need for reliable and accurate predictions of 
jet-veto cross-sections in processes such as Higgs boson and $W^+W^-$ production, which are 
commonly used to study new physics at the \LHC{}. In these processes, the imposition of a jet veto 
is often necessary to suppress backgrounds and enhance sensitivity to new physics signals.
Experimental results going beyond these two processes 
are much
less frequent. We encourage the experimental collaborations to 
consider measurements of more Standard Model processes with a jet veto, as larger data samples 
become available, to better understand the dependence of these processes on the jet veto 
parameters $\pTveto$ and $R$.

In addition to providing improved predictions for jet-veto cross-sections, our 
work 
also serves as a valuable tool for testing and validation of general purpose shower Monte Carlo 
programs. Our code allows for a detailed investigation of the dependence on the jet parameters 
$\pTveto$ and $R$, providing a benchmark for assessing the logarithmic accuracy and reliability of 
Monte Carlo 
simulations in this important class of processes.

Our analysis shows that at the currently experimentally used values of $\pTveto$ in $W$ and $Z$ 
production, the logarithms are not large enough to justify the use of jet-veto resummation. In 
these cases, fixed-order perturbation theory, which can be used to give the results with a jet veto 
over a limited range of rapidities, is simpler and sufficient. We have also found that attempts to 
resum $\pi^2$ terms using a timelike renormalization point have little numerical importance at 
\NNNLLpart{} if the $\ptveto$ scale is around \SIrange{20}{30}{\GeV}.

The production of a Higgs boson is an exception among single-boson processes. In this case, the 
combination of larger corrections from color factors and slightly larger values of the scale 
($m_H$) appearing in the jet veto logarithms make resummation an important tool for improving the 
accuracy of predictions. In the appendix we have investigated the differences between the one-step 
and two-step procedures for calculating the hard function at the scale of $\pTveto$. 
We find agreement within $2\%$ of these two approaches.

The $W^+W^-$ production process, where the jet veto has experimental importance,
requires both resummation and matching to \NNLO{}. For the $ZZ$ process resummation
is mandatory but the matching to fixed order is less important.  Although this reflects
the expectation that the resummed prediction is more accurate for systems of higher invariant
mass, these findings depend on the exact nature of the cuts for each process.
Our work provides a comprehensive theoretical framework for studying jet vetoes in vector boson 
pair processes, and as data becomes available, a comparative experimental study would be of great 
interest and could help to validate our theoretical predictions.

\section*{Acknowledgments}
RKE would like to thank Simone Alioli, Thomas Becher, Andrew Gilbert, Pier Monni
and Philip Sommer for useful discussions.  In addition, RKE would like to thank TTP
in Karlsruhe for hospitality during the drafting of this paper.
TN would like to thank Robert Szafron for useful discussions.
SS is supported in part by the SERB-MATRICS under Grant No. MTR/2022/000135.
This manuscript has been authored by Fermi Research Alliance, LLC
under Contract No. DE-AC02-07CH11359 with the U.S. Department of
Energy, Office of Science, Office of High Energy Physics.
This research used resources of the Wilson High-Performance Computing Facility at Fermilab.
This research also used resources of the National Energy Research Scientific
Computing Center (NERSC), a U.S. Department of Energy Office of Science
User Facility located at Lawrence Berkeley National Laboratory, operated
under Contract No. DE-AC02-05CH11231 using NERSC award HEP-ERCAP0021890.

\appendix

\section{Reduced beam functions}
\label{sec:reducedbeamfunctions}
We have used the two loop beam function in the presence of a jet veto calculated in 
Ref.~\cite{Abreu:2022zgo}. Their calculation,
together with the corresponding soft function~\cite{Abreu:2022sdc} has been performed in \SCET{}
using the exponential rapidity regulator~\cite{Li:2016axz}. The beam function for quark initiated processes
in the presence of a jet veto has also been presented in Mellin space in Ref.~\cite{Bell:2022nrj}.

The calculation in Ref.~\cite{Abreu:2022zgo} has a perturbative expansion,
\beq
I_{ij}= \sum_{k=0}^{\infty} \Big(\as\Big)^k \, I^{(k)}_{ij} \,.
\eeq
The beam functions with a jet veto are decomposed into a reference observable, the beam function for
the transverse momentum of a color singlet observable and a remainder term accounting for the effects of jet
clustering,
\begin{equation}
\label{eq:twoparts}
I_{ij}(x,Q,\pTveto,R;\mu,\nu) = I^\perp_{ij}(x,Q,\pTveto;\mu,\nu)+\Delta I_{ij}(x,Q,\pTveto,R;\mu,\nu)\; .
\end{equation}
Since the divergence structure of the reference observable is the same as the beam function with a
jet veto, $\Delta I_{ij}$ can be calculated in four dimensions. Results for the reference observable are
available in Refs.~\cite{Luo:2019hmp,Luo:2019bmw}.

The reduced beam function kernels $\Ibar$ as used in our setup
are extracted from the coefficient $I$ as
\beq
\bar{I}_{ij}(z,\pTveto,R,\mu)=e^{-h^A(\ptveto,\mu)} \, I_{ij}(z,\pTveto,R,\mu) \, .
\eeq

They similarly follow a perturbative expansion
\begin{equation}
	\bar{I}_{ik}(z,\pTveto,R,\mu)=\delta_{ik} \, \delta(1-z)
	+ \as \bar{I}^{(1)}_{ik}(z,\pTveto,\mu)
	+ \Big(\as\Big)^2 \bar{I}^{(2)}_{ik}(z,\pTveto,R,\mu) + O(\alpha_s^3)\, .
\end{equation}

\subsection*{Contributions at order $\alpha_s$}
The $\alpha_s$ contributions to $\Ibar$ were first obtained in 
Refs.~\cite{Becher:2011xn,Becher:2012qa} and read,
\begin{equation}\label{barI1exp}
   \Ibar_{ij}(z,\pTveto,R,\mu)
   = \delta(1-z)\,\delta_{ij} +\as
   \left[-2 P_{ij}^{(1)}(z)\,\Lperp + \Rone_{ij}(z) \right]+ {\cal O}(\alpha_s^2) \,,
\end{equation}
where $\Lperp=2\ln(\mu/\pTveto)$.
$R$ is the jet measure used in Eq.~(\ref{jetdef}) and $\Rone (z)$ is a remainder function given below.
At this order there is no dependence on the jet radius, $R$.

Throughout this paper we expand in powers of $\alpha_s/(4\pi)$.
The one exception to this rule are the perturbative {\abbrev DGLAP} splitting functions,
\beq
P_{ij}(z)=\frac{\alpha_s}{2 \pi} \Pone(z) +\Big(\frac{\alpha_s}{2 \pi}\Big)^2 \Ptwo(z) + \ldots
\eeq
Explicit expressions for $\Pone$ and $\Ptwo$ are given in Appendices~\ref{oneloopanomdim}
and \ref{twoloopanomdim}.
The remainder functions at order $\alpha_S$ are~\cite{Becher:2010tm}
\begin{eqnarray}
\label{eq:R1}
   \Rone_{qq}(z)
   &=& C_F \left[ 2(1-z) - \frac{\pi^2}{6}\,\delta(1-z) \right] , \qquad
   \Rone_{qg}(z)
   = 4T_F\,z(1-z)\,,\nn \\
   \Rone_{gg}(z) &=& - C_A\,\frac{\pi^2}{6}\,\delta(1-z) \,, \qquad
   \Rone_{gq}(z) = 2C_F \, z \,.
\end{eqnarray}
where $C_A=3,C_F=\frac{4}{3},T_F=\frac{1}{2}$.

\subsection*{Contributions at order $\alpha_s^2$}
At order $\alpha_s^2$ we have
\begin{align}
  &\bar{I}^{(2)}_{ik}(z,\pTveto,R,\mu)=
    \big[2\*\Pone_{ij}(x) \otimes \Pone_{jk}(y)-\* \beta_0 \* \Pone_{ik}(z)\big] \Lperp^2 \nn \\
    &+\big[-4\*P^{(2)}_{ik}(z)+\beta_0\* \Rone_{ik}(z)
      -2\*\Rone_{ij}(x)\otimes \Pone_{jk}(y)\big] \Lperp + R^{(2)}_{ik}(z,R)  \,.
      \label{I2barform}
\end{align}
In this equation $\otimes$ represents a convolution,
\beq \label{convolution}
f(x) \otimes g(y) = \int_0^1 \,dx\, \int_0^1 \,dy f(x)\, g(y)\, \delta(z-xy)=\int_z^1 \,\frac{dy}{y} f(z/y)\, g(y) \,.
\eeq
Explicit expressions for $\Pone$ and $\Ptwo$ are given in Appendices~\ref{oneloopanomdim} and 
\ref{twoloopanomdim}.
The expressions for $\Pone \otimes \Pone$, $\Rone \otimes \Pone$ are given in appendix \ref{RxP}.

The results from Refs.~\cite{Abreu:2022zgo,Abreu:2022sdc} recast in the language of reduced 
beam functions allow us to extract $R^{(2)}_{ik}(z,R)$.
We have checked that the reduced beam functions have the form predicted 
by \cref{barI1exp,I2barform}.
In addition, we have confirmed the known results for the
$\alpha_s^2$ $R$-dependent contribution to the collinear anomaly exponent. 
The result for the collinear anomaly exponent is given in \cref{sec:collinearanomaly}.

\subsection{Structure of the two-loop reduced beam function}

While a numerical evaluation of the analytical formulas for the reduced beam functions is possible, 
we choose to perform a spline interpolation for improved numerical efficiency.

The reduced beam functions contain distributions of the following structure,
\begin{eqnarray}
\bar{I}^{(2)}_{ij}(z,\pTveto,R,\mu)&=&\bar{I}^{(2)}_{ij,-1}(\pTveto,R,\mu)\, \delta(1-z)
+\bar{I}^{(2)}_{ij,0}(\pTveto,R,\mu) \, D_0(1-z) \nn \\
&+&\bar{I}^{(2)}_{ij,1}(\pTveto,R,\mu) \, D_1(1-z)
+\bar{I}^{(2)}_{ij,2}(z,\pTveto,R,\mu)\, ,
\end{eqnarray}
where,
\beq
D_0(1-z)=\frac{1}{[1-z]_{+}},\;\;\;D_1(1-z)=\bigg[\frac{\ln(1-z)}{(1-z)}\bigg]_{+} \; .
\eeq
$\bar{I}^{(2)}_{ij,2}(z,\pTveto,R,\mu)$ contains terms which are regular at $z=1$.

The analytic results for the beam function of Ref.~\cite{Abreu:2022zgo} are presented as
a power series in $R$ up to powers of $R^8$. The functions themselves contain powers of $1/(1-z)^n$,
in certain cases up to $n=7$ or $8$. However, these singularities at $z=1$ are fictitious as can be 
seen by explicit
expansion. The beam functions require special treatment in this region for numerical stability. 

The dominant region in the convolution of the function $\bar{I}$ with the parton
distributions is precisely the region $z\sim 1$.
If we assume a parton distribution $f(x) \sim 1/x$ we have,
\beq
\bar{I} \otimes f = \int_x^1 \frac{dz}{z} \, \bar{I}(z) \, f(x/z) \sim \frac{1}{x} \int_x^1 \, dz \, \bar{I}(z) \,,
\eeq
showing that all regions of $z$ contribute equally to the integral. However if, as expected, the
parton distribution function falls off more rapidly as $x \to 1$, say $f(x)\sim (1-x)^n/x$,
\beq
\bar{I} \otimes f = \int_x^1 \, \frac{dz}{z} \, \bar{I}(z) \, f(x/z) \sim \frac{1}{x} \, \int_x^1 
\, dz \, \bar{I}(z)\, (1-x/z)^n\,.
\eeq
Thus, it is precisely the large values of $z$ which are crucial for the integral.
In other words, the parton shower process is dominated by cascade from nearby values of $x$. 
Larger cascades from more distant points are suppressed by the fall-off of the parton distributions.
In view of the importance of the region $z=1$, for numerical stability we perform an expansion about $z=1$.

\begin{figure}[h]
	\centering
\begin{subfigure}[t]{0.42\textwidth}
	\centering
    \includegraphics[width=\textwidth,angle=270]{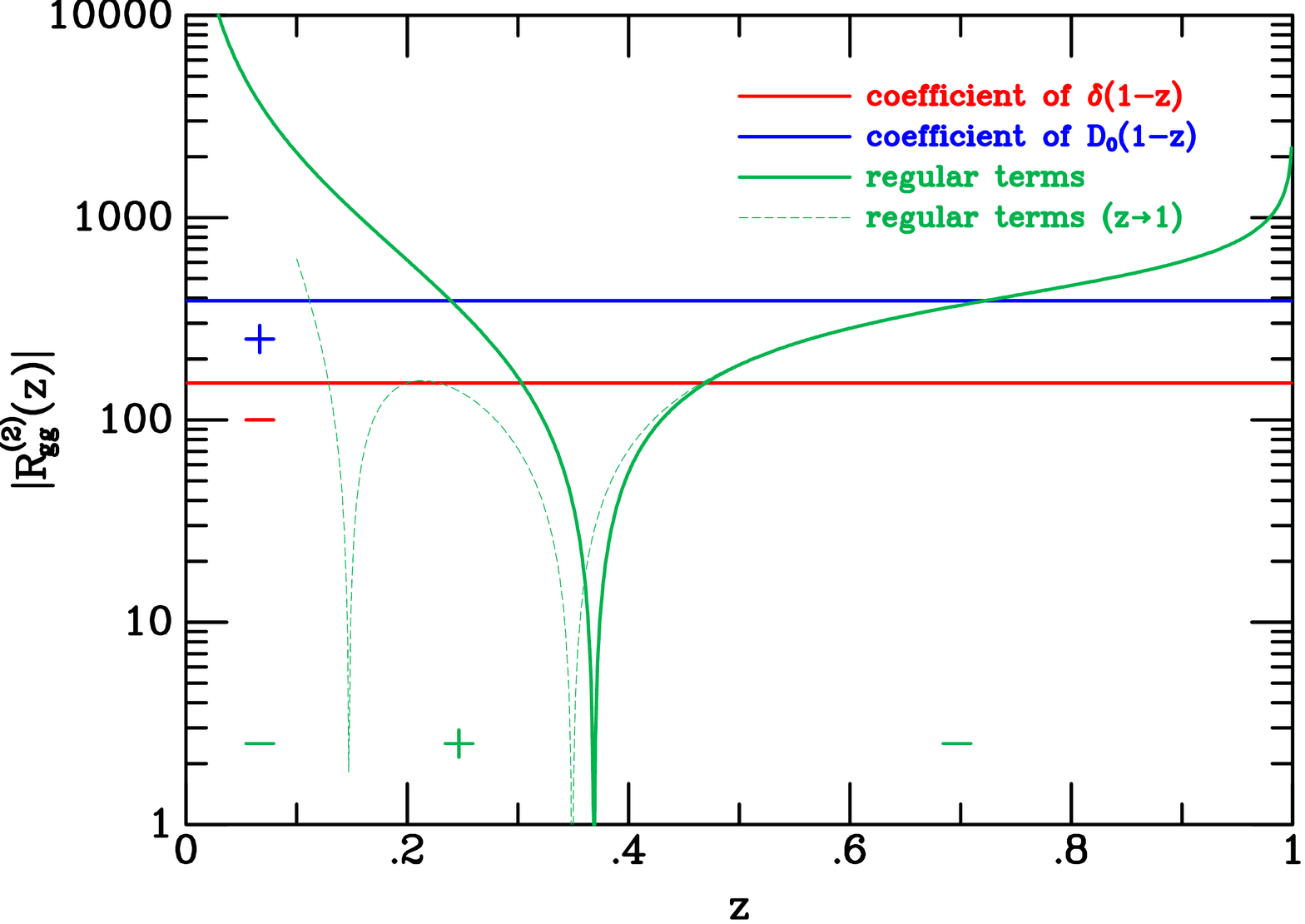}
    \vspace{-3em}
    \caption{$gg$ case.}
\end{subfigure}
	\hfill
\begin{subfigure}[t]{0.42\textwidth}
	\centering
    \includegraphics[width=\textwidth,angle=270]{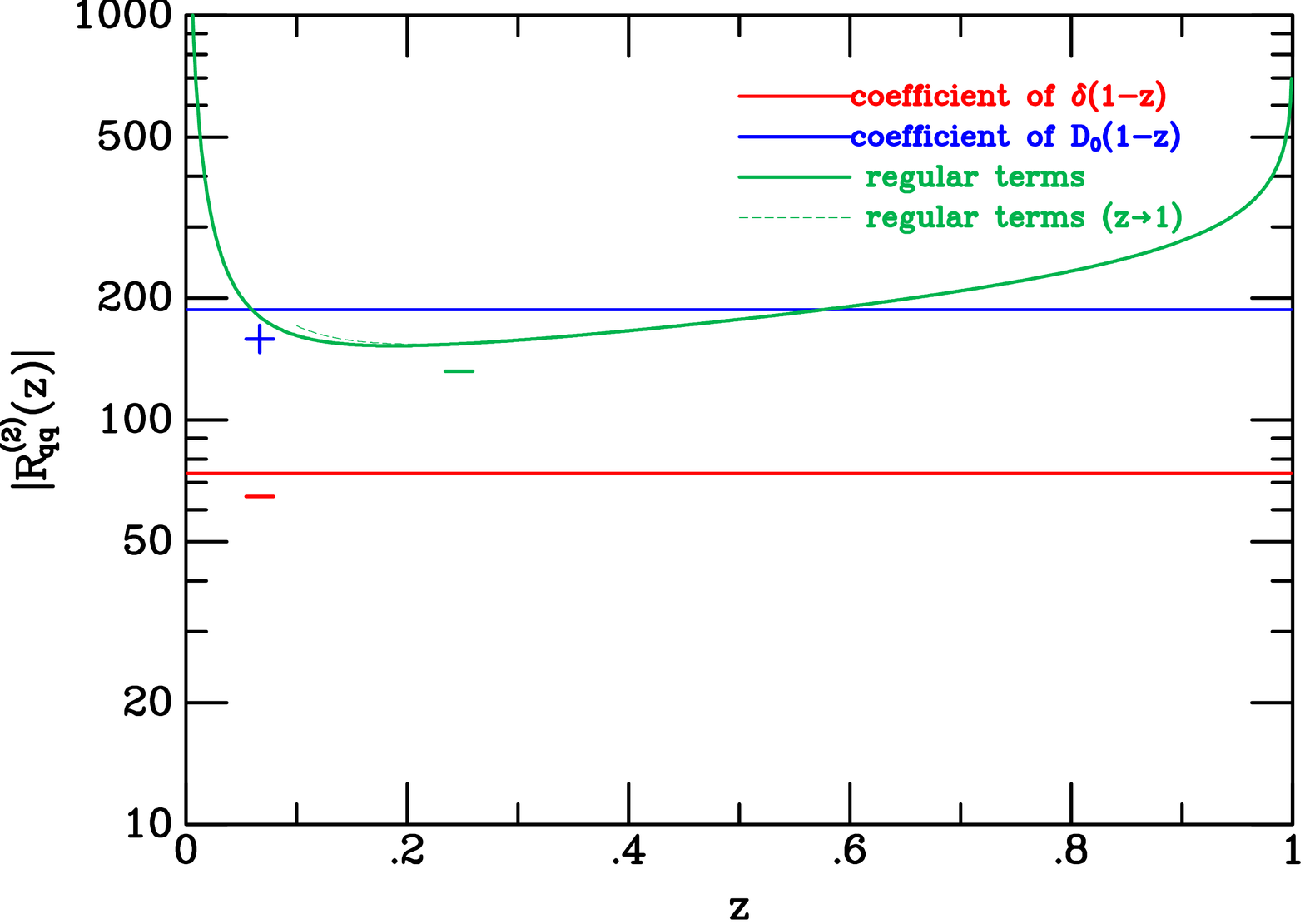}
    \vspace{-3em}
    \caption{$qq$ case}
 \end{subfigure}
\vfill
\begin{subfigure}[t]{0.42\textwidth}
 \centering
 \includegraphics[width=\textwidth,angle=270]{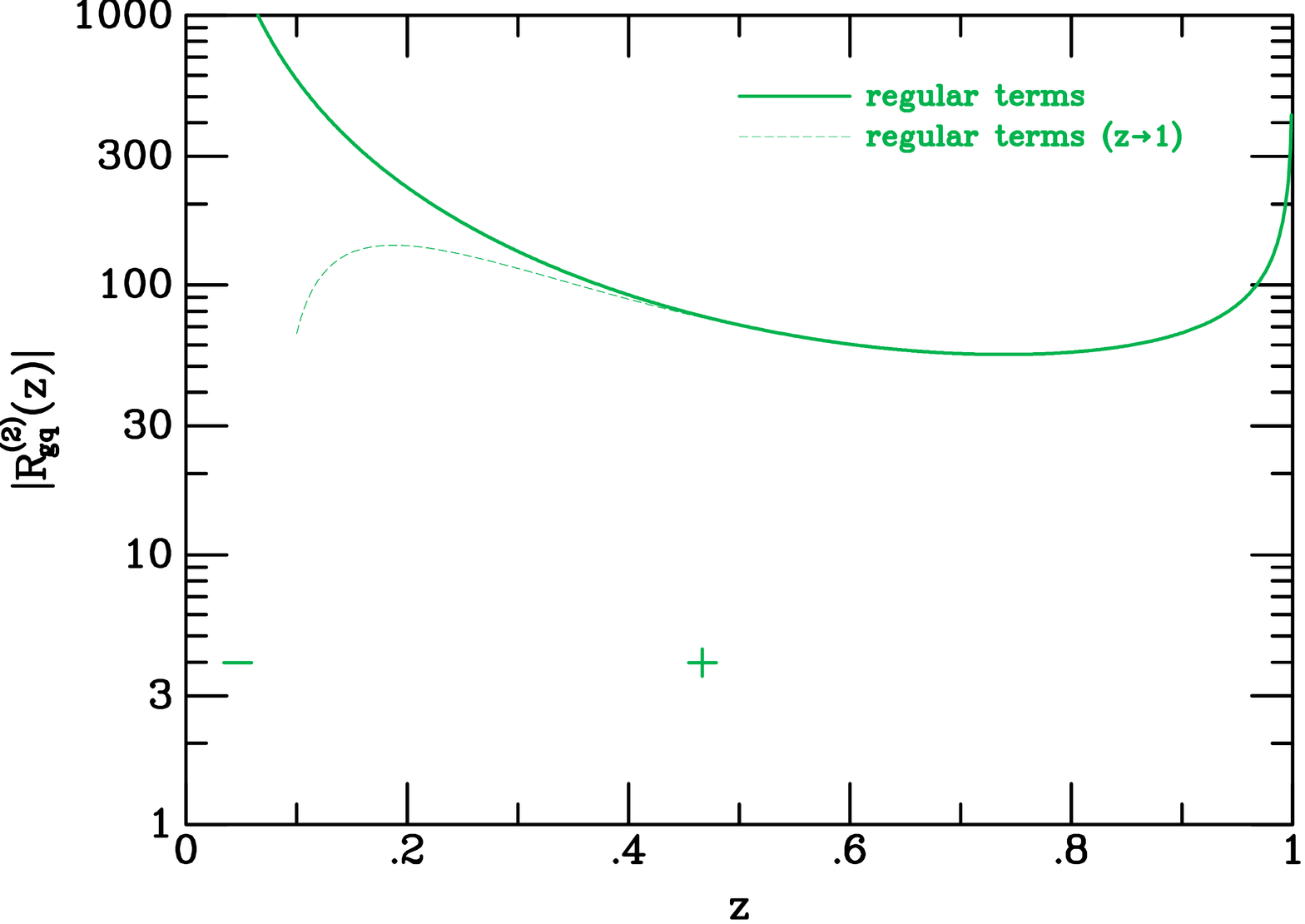}
 \vspace{-3em}
 \caption{$gq$ case.}
\end{subfigure}
\hfill
\begin{subfigure}[t]{0.42\textwidth}
\centering
\includegraphics[width=\textwidth,angle=270]{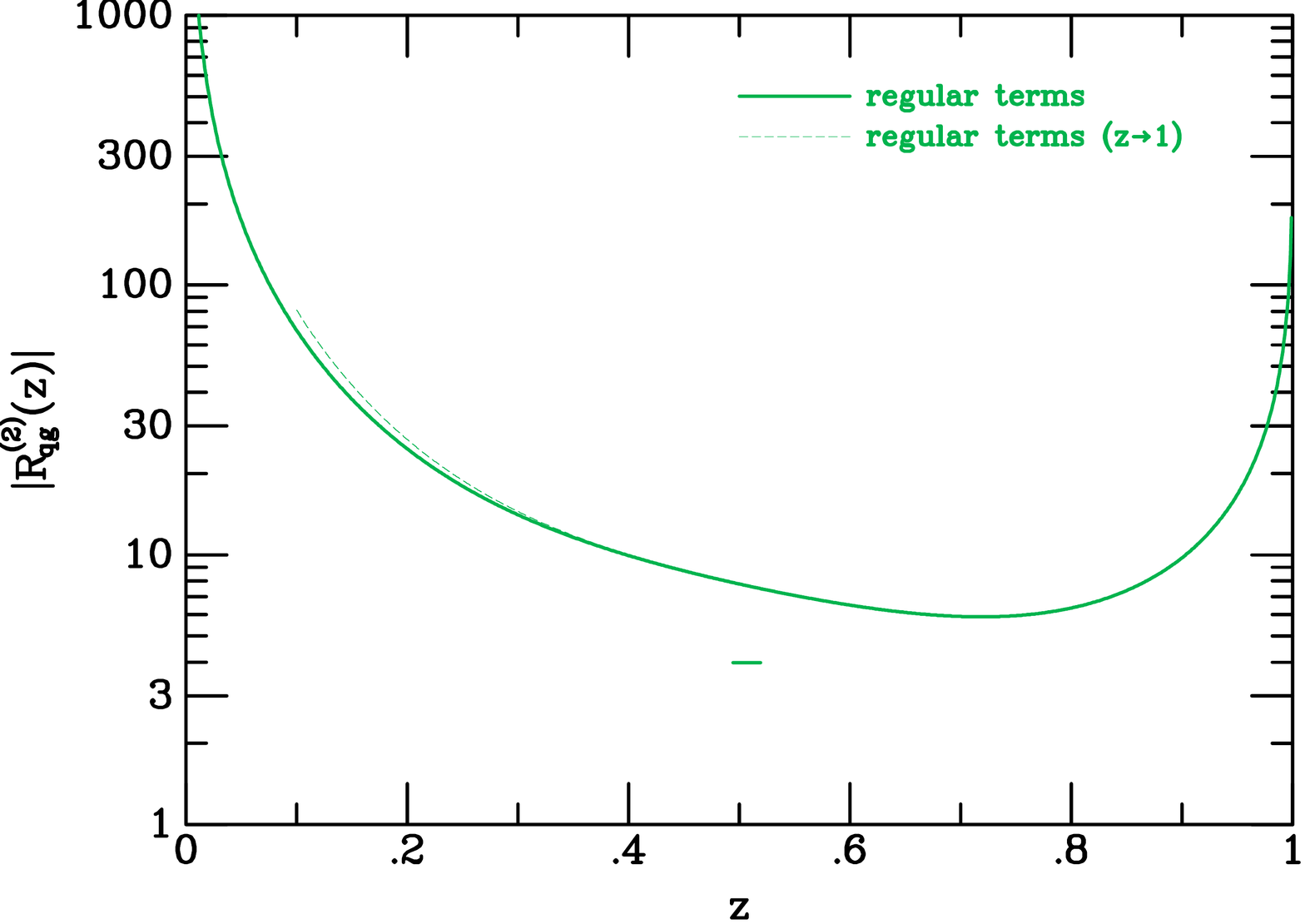}
\vspace{-3em}
\caption{$qg$ case}
\end{subfigure}
\vfill
\begin{subfigure}[t]{0.44\textwidth}
	\centering
    \includegraphics[width=\textwidth,angle=270]{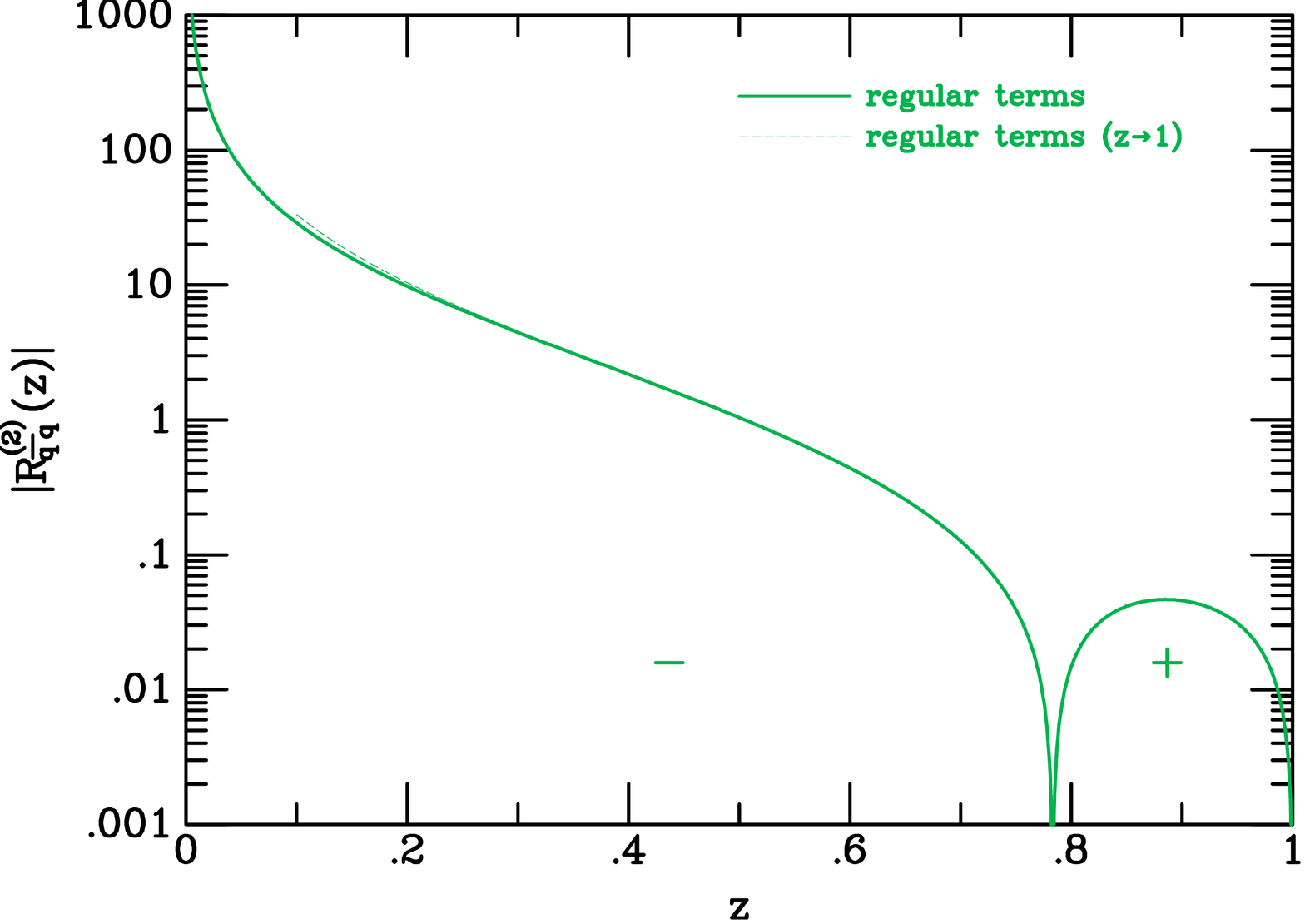}
        \vspace{-3em}
    \caption{$\bar{q}q$ case.}
\end{subfigure}
\hfill
\begin{subfigure}[t]{0.44\textwidth}
	\centering
    \includegraphics[width=\textwidth,angle=270]{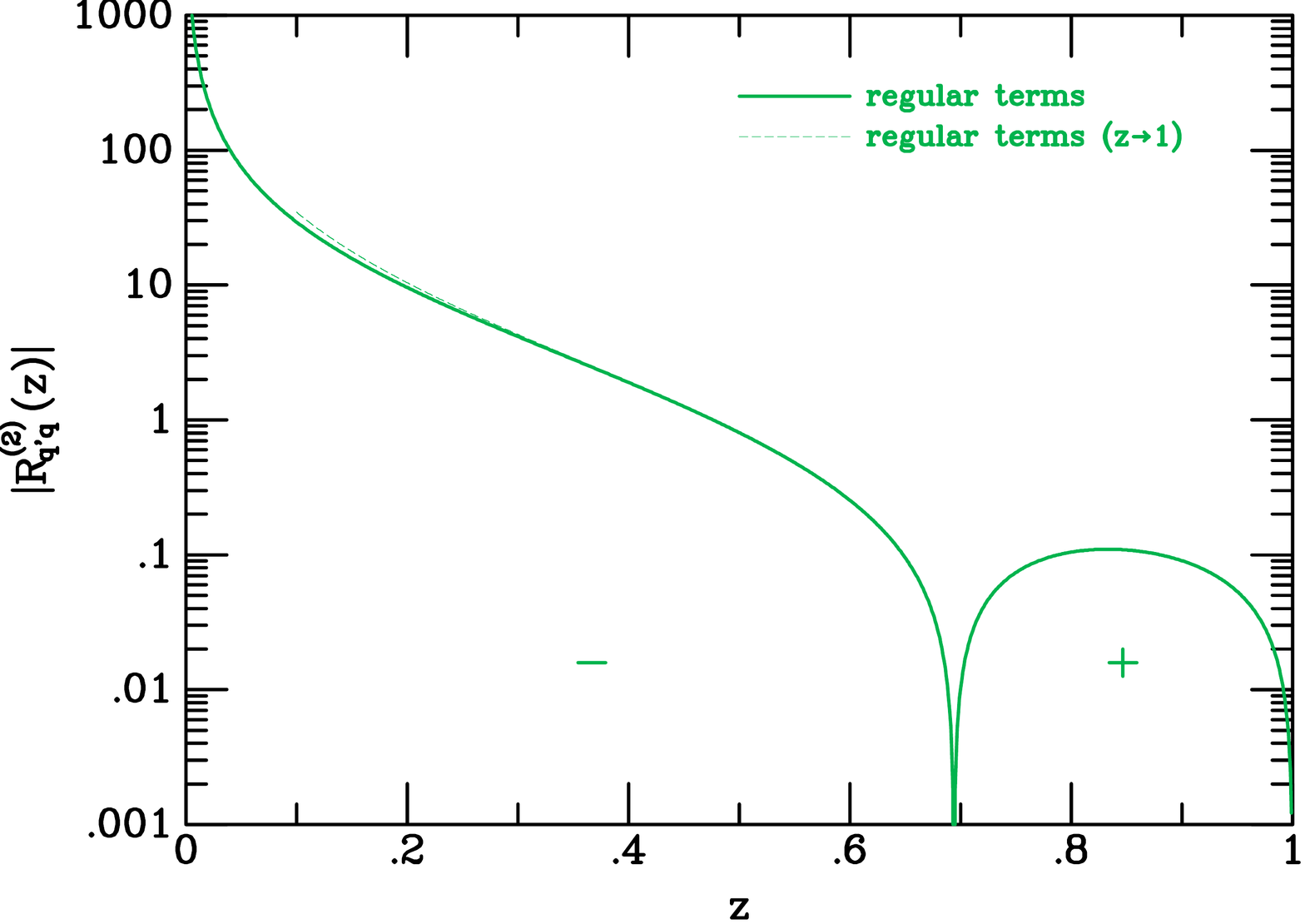}
        \vspace{-3em}
    \caption{$q^\prime q$ case.}
\end{subfigure}
    \caption{Absolute value of $R^{(2)}$ for jet measure $R=0.5$.
      The $\bar{q}^\prime q$ case is the same as the $q^\prime q$ case. The sign of the 
      contribution in the various regions is indicated.}
    \label{fig:allbeams}
\end{figure}

The absolute value of $R^{(2)}$ for the various parton
transitions is shown in \cref{fig:allbeams}. Individual $R$-dependent terms contain expressions of 
the 
form $R^{2n}/(1-z)^k$ where $k$ can be a high power. 
However, the singularity at $z=1$ is only apparent. The resultant limiting forms obtained by series 
expansion about $z=1$
are shown by the dashed lines in the figures. 
In practice, we switch to the expanded form at $z=0.9$, although the figures demonstrate that the 
expanded forms are accurate
down to much smaller values of $z$.

\section{Definition of the beta function and anomalous dimensions}
\label{sec:betaanom}
The coefficients $\beta_n$, $\Gamma_n^A$ and $\gamma_n^g$
have perturbative
expansions in powers of the renormalized coupling. Details are presented below.
\subsection{Expansion of $\beta$-function}
\label{betacoeffs}
The beta function is defined as,
\begin{eqnarray} \label{betafunction}
  \frac{d \alpha_s(\mu)}{d\ln\mu}&=&\beta(\mu) =
  -2\alpha_s(\mu) \sum_{n=0}^\infty\,\beta_n\,\left(\as\right)^{n+1} \nonumber  \\
  &=& -2 \alpha_s(\mu) \, \frac{\alpha_s(\mu)}{4 \pi}
   \Big[ \beta_0
    + \beta_1 \frac{\alpha_s(\mu)}{4 \pi}
    + \beta_2 \Big(\frac{\alpha_s(\mu)}{4 \pi}\Big)^2
    + \beta_3 \Big(\frac{\alpha_s(\mu)}{4 \pi}\Big)^3 +\ldots\Big]\, .
\end{eqnarray}
The coefficients of the \MSbar $\beta$ function to four loops are~\cite{Tarasov:1980au, Larin:1993tp,vanRitbergen:1997va},
{\allowdisplaybreaks
\begin{eqnarray}
\label{eq:betaexp}
\beta_0 &=& \frac{11}{3}\,C_A -\frac{4}{3}\,T_F\,n_f
\,,\nn\\
\beta_1 &=& \frac{34}{3}\,C_A^2  - \Bigl(\frac{20}{3}\,C_A\, + 4 C_F\Bigr)\, T_F\,n_f
\,, \nn\\
\beta_2 &=&
\frac{2857}{54}\,C_A^3 + \Bigl(C_F^2 - \frac{205}{18}\,C_F C_A
 - \frac{1415}{54}\,C_A^2 \Bigr)\, 2T_F\,n_f
 + \Bigl(\frac{11}{9}\, C_F + \frac{79}{54}\, C_A \Bigr)\, 4T_F^2\,n_f^2 \, , \nn \\
\beta_{3} & = &
    C_A^4 \left( \frac{150653}{486} - \frac{44}{9} \zeta_3 \right)
    +  C_A^3 T_F n_f
      \left(  - \frac{39143}{81} + \frac{136}{3} \zeta_3 \right)
\nonumber \\ & &
    + C_A^2 C_F T_F n_f
\left( \frac{7073}{243} - \frac{656}{9} \zeta_3 \right)
   +     C_A C_F^2 T_F n_f
      \left(  - \frac{4204}{27} + \frac{352}{9} \zeta_3 \right)
\nonumber \\ & &
   + 46 C_F^3 T_F n_f
   +  C_A^2 T_F^2 n_f^2
      \left( \frac{7930}{81} + \frac{224}{9} \zeta_3 \right)
    +  C_F^2 T_F^2 n_f^2
      \left( \frac{1352}{27} - \frac{704}{9} \zeta_3 \right)
\nonumber \\ & &
    +  C_A C_F T_F^2 n_f^2
      \left( \frac{17152}{243} + \frac{448}{9} \zeta_3 \right)
  + \frac{424}{243} C_A T_F^3 n_f^3
   + \frac{1232}{243} C_F T_F^3 n_f^3
\nonumber \\ & &
       +  \frac{d_A^{a b c d}d_A^{a b c d}}{N_A }
             \left(  - \frac{80}{9} + \frac{704}{3} \zeta_3 \right)
       + n_f \frac{d_F^{a b c d}d_A^{a b c d}}{N_A }
            \left(   \frac{512}{9} - \frac{1664}{3} \zeta_3 \right)
\nonumber \\ & &
       + n_f^2 \frac{d_F^{a b c d}d_F^{a b c d}}{N_A }
            \left(  - \frac{704}{9} + \frac{512}{3} \zeta_3 \right)\, .
            \label{mainbeta}
\end{eqnarray}
}
For the normalization of the $SU(N)$ generators, the conventions of
Refs.~\cite{vanRitbergen:1997va,vanRitbergen:1998pn}
are,
\begin{eqnarray}
  \label{dsums}
  \frac{d_{A}^{abcd}d_{A}^{abcd}}{N_{A}} &=& \frac{N^2 (N^2+36)}{24} \,,\quad
 \frac{d_{F}^{abcd}d_{A}^{abcd}}{N_{A}} =\frac{N (N^2+6)}{48} \,,\quad    \frac{d_{F}^{abcd}d_{F}^{abcd}}{N_{A}} = \frac{N^4-6 N^2+18}{96 N^2} \,,\nonumber \\
 N_{A} &= &N^2-1 \,,\quad N_{F} = N .
\end{eqnarray}
Numerical values for the $\beta$-function coefficients are,
\begin{eqnarray}
\beta_0 &=&11- \frac{2}{3} \, n_f \, , \nn \\
\beta_1 &=&102- \frac{38}{3} \, n_f \, ,\nn\\
\beta_2 &=&\frac{2857}{2}-\frac{5033}{18}\, n_f+\frac{325}{54}\, n_f^2 \, , \nn \\
\beta_3 &=&\frac{149753}{6}+3564 \zeta_3-\Big(\frac{1078361}{162}+\frac{6508}{27}  \zeta_3\Big) \, n_f
+\Big(\frac{50065}{162}+\frac{6472}{81}  \zeta_3\Big) \, n_f^2\nn\\
&+&\frac{1093}{729} \, n_f^3 \, .
\end{eqnarray}

\subsection{Cusp Anomalous Dimension}
\label{Cusp_Anom_Dim}
The cusp anomalous dimension depends on the label $B$ 
which takes the two values, $B=A,F$ for gluons and quarks, respectively. Its perturbative expansion 
is,
\begin{equation} \label{cusp}
  \Gamma_{\rm cusp}^{B}(\mu) = \sum_{n=0}^\infty\,\Gamma_n^{B}\,\left(\as\right)^{n+1}\,.
\end{equation}
The coefficients up to four loops are \cite{Henn:2019swt,vonManteuffel:2020vjv},
\begin{align}
\label{Cusp}
\Gamma^{B}_{0}&= 4 C_B \, , \\
\Gamma^{B}_{1}&= 16 C_B \bigg\{(C_A   \left(\frac{67}{36}-\frac{\pi ^2}{12}\right)-\frac{5}{9}  n_f 
T_F \bigg\} \, ,\\ 
\Gamma^{B}_{2}&= 64 C_B \bigg\{ C_A^2\left(\frac{11 \zeta_3}{24}+\frac{245}{96}-\frac{67 \pi 
^2}{216}+\frac{11 \pi ^4}{720}\right) 
  + n_f T_F   C_F  \left(\zeta_3-\frac{55}{48}\right)\nn \\
  &+ n_f T_F  C_A \left(-\frac{7 \zeta_3}{6}-\frac{209}{216}+\frac{5 \pi 
  ^2}{54}\right)-\frac{1}{27} (n_f T_F)^2 \bigg\} \, ,\\
\Gamma^{B}_{3}&= 256 C_B \bigg\{
C_A^3 \left( \frac{1309 \zeta_{3}}{432}-\frac{11 \pi ^2 \zeta_{3}}{144}-\frac{\zeta_{3}^2}{16}-\frac{451 \zeta_{5}}{288}+\frac{42139}{10368}-\frac{5525 \pi^2}{7776}
 +\frac{451 \pi ^4}{5760}-\frac{313 \pi ^6}{90720} \right) \nn\\
&+ n_f T_F  C_A^2 \left( -\frac{361 \zeta_{3}}{54}+\frac{7 \pi ^2 \zeta_{3}}{36}+\frac{131 
\zeta_{5}}{72} -\frac{24137}{10368}+\frac{635 \pi ^2}{1944}-\frac{11 \pi ^4}{2160}  \right)  \nn\\
&+  n_f T_F C_F C_A \left(  \frac{29 \zeta_{3}}{9}-\frac{\pi ^2 \zeta_{3}}{6}+\frac{5 
\zeta_{5}}{4}-\frac{17033}{5184}+\frac{55 \pi ^2}{288}-\frac{11 \pi^4}{720} \right)  \nn\\
&+ n_f T_F  C_F^2 \left( \frac{37 \zeta_3}{24}-\frac{5 \zeta_5}{2}+\frac{143}{288} \right) 
+(n_f T_F)^2   C_A \left( \frac{35 \zeta_3}{27}-\frac{7 \pi ^4}{1080}-\frac{19 \pi ^2}{972}+\frac{923}{5184} \right)    \nn\\
&+ (n_f T_F)^2   C_F \left(-\frac{10 \zeta_3}{9}+\frac{\pi ^4}{180}+\frac{299}{648}\right) + (n_f 
T_F)^3  \left(-\frac{1}{81}+\frac{2 \zeta_3}{27}\right)\bigg\}  \nn\\
&+ 256 \frac{d^{abcd}_B d^{abcd}_A}{N_B} \left(  \frac{\zeta_{3}}{6}-\frac{3 
\zeta_{3}^2}{2}+\frac{55 \zeta_{5}}{12}-\frac{\pi^2}{12}-\frac{31 \pi ^6}{7560}   \right) \nn\\
&+ 256 n_f \frac{d^{abcd}_B d^{abcd}_F}{N_B} \left( 
\frac{\pi^2}{6}-\frac{\zeta_3}{3}-\frac{5\zeta_5}{3} \right) 
\,.  
\end{align}

In addition to the relations in Eq.~(\ref{dsums}) we need the related quantities,
\begin{equation}
\frac{d_{F}^{abcd}d_{A}^{abcd}}{N_{F}} =\frac{(N^2-1) (N^2+6)}{48}\, , \quad
\frac{d_{F}^{abcd}d_{F}^{abcd}}{N_{F}} =\frac{(N^2-1) (N^4-6 N^2+18)}{96 N^3}\, .
\end{equation}
\subsection{Non-cusp anomalous dimension}
The non-cusp anomalous dimension has the expansion,
\begin{equation} \label{noncusp}
     \gamma^{q,g}(\mu) = \sum_{n=0}^\infty\,\gamma_n^{q,g}\,\left(\as\right)^{n+1} \,.
\end{equation}
\label{Non_Cusp_Anom_Dim}
We take the coefficients up to three loops from ref.~\cite{Becher:2014oda} Eq.~I.4,
\begin{align}
\gamma_0^q& = -3\*C_F \, ,\\
\gamma_1^q& = 
C_F^2\*\Big(2\*\pi^2-\frac{3}{2}-24\*\zeta_3\Big)+C_F\*C_A\*\Big(26\*\zeta_3-\frac{961}{54}-\frac{11\*\pi^2}{6}\Big)\nonumber
 \\
         &+C_F\*T_F\*n_f\*\Big(\frac{130}{27}+\frac{2\*\pi^2}{3}\Big)\, ,\\
\gamma_2^q&=C_F^3\*\Big( - \frac{29}{2} - 3\*\pi^2 -  \frac{8\*\pi^4}{5} - 68\*\zeta_3 +  
\frac{16\*\pi^2}{3}\*\zeta_3 + 240\*\zeta_5 \Big) \nonumber \\
     & + C_F^2\*C_A\*\Big( -  \frac{151}{4} +  \frac{205\*\pi^2}{9}
      +  \frac{247\*\pi^4}{135} -  \frac{844}{3}\*\zeta_3
      -  \frac{8\*\pi^2}{3}\*\zeta_3 - 120\*\zeta_5 \Big)\nonumber \\
     & + C_F\*C_A^2\*\Big( -  \frac{139345}{2916} -  \frac{7163\*\pi^2}{486}
      -\frac{83\*\pi^4}{90} +  \frac{3526}{9}\*\zeta_3\nonumber 
      -\frac{44\*\pi^2}{9}\*\zeta_3 - 136\*\zeta_5 \Big)\nonumber \\
     & + C_F^2\*T_F\*n_f\*\Big(  \frac{2953}{27} -  \frac{26\*\pi^2}{9} -  \frac{28\*\pi^4}{27} +  
     \frac{512}{9}\*\zeta_3 \Big)\nonumber \\
     & + C_F\*C_A\*T_F\*n_f\*\Big( -  \frac{17318}{729} +  \frac{2594\*\pi^2}{243} +  
     \frac{22\*\pi^4}{45}  -  \frac{1928}{27}\*\zeta_3 \Big)\nonumber \\
      & + C_F\*T_F^2\*n_f^2\*\Big(  \frac{9668}{729} -  \frac{40\*\pi^2}{27} -  
      \frac{32}{27}\*\zeta_3 \Big)\, .
\end{align}

From ref.~\cite{Becher:2009qa}, Eq~A5 we take,
\begin{align}
  \label{gammag}
    \gamma_0^g&=-\beta_0\, ,\\
    \gamma_1^g&=C_A^2\*\Big(\frac{11\*\pi^2}{18}-\frac{692}{27}+2\*\zeta_3\Big)
          +C_A\*T_F\*n_f\*\Big(\frac{256}{27}-\frac{2\*\pi^2}{9}\Big)+4\*C_F\*T_F\*n_f\nonumber\\
              &= ´C_A^2\*\Big(2\*\zeta_3-\frac{59}{9}\Big)+C_A\*\beta_0\*\Big(\frac{\pi^2}{6}-\frac{19}{9}\Big)-\beta_1\, ,\\
      \gamma_2^g&= C_A^3\*\Big( - \frac{97186}{729} +  \frac{6109\*\pi^2}{486} -  
      \frac{319\*\pi^4}{270}
          +  \frac{122}{3}\*\zeta_3 -  \frac{20\*\pi^2}{9}\*\zeta_3 - 16\*\zeta_5 \Big)\nonumber \\
          &+ C_A^2\*T_F\*n_f\*\Big(  \frac{30715}{729} -  \frac{1198\*\pi^2}{243} +  
          \frac{82\*\pi^4}{135} +  \frac{712}{27}\*\zeta_3 \Big)\nonumber \\
          &+ C_A\*C_F\*T_F\*n_f\*\Big(  \frac{2434}{27} -  \frac{2\*\pi^2}{3} -  
          \frac{8\*\pi^4}{45} -  \frac{304}{9}\*\zeta_3 \Big)\nonumber \\
          &- 2\*C_F^2\*T_F\*n_f + C_A\*T_F^2\*n_f^2\*\Big( -  \frac{538}{729} +  
          \frac{40\*\pi^2}{81} -  \frac{224}{27}\*\zeta_3 \Big)
          -\frac{44}{9}\*C_F\*T_F^2\*n_f^2 \, .
\end{align}
Primary references for the calculation of these coefficients can be found in Ref.~\cite{Becher:2014oda}.

We now present results for $\gamma^S$ and $\gamma^t$ which are needed for the implementation of the two-step
calculation of the hard function for Higgs boson production.
Following Ref.~\cite{Ahrens:2009cxz} we have, for
the first three expansion coefficients of the anomalous dimension $\gamma^S$ that enters the evolution
equation of the hard matching coefficient $C^S$ (see also \cite{Idilbi:2005ni,Idilbi:2005er}),
\begin{eqnarray}
  \gamma^S_0 &=& 0 \,, \\
  \gamma^S_1 
  &=& C_A^2 \left( -\frac{160}{27} + \frac{11\pi^2}{9}
   + 4\zeta_3 \right) 
   + C_A T_F n_f \left( -\frac{208}{27} - \frac{4\pi^2}{9} \right) 
   - 8 C_F T_F n_f \,, \\
  \gamma^S_2 
  &=& C_A^3 \left[ \frac{37045}{729} + \frac{6109\pi^2}{243}
   - \frac{319\pi^4}{135} 
   + \left( \frac{244}{3} - \frac{40\pi^2}{9} \right) \zeta_3 
   - 32\zeta_5 \right] \nonumber \\
  &+& C_A^2 T_F n_f \left( -\frac{167800}{729}
   - \frac{2396\pi^2}{243} + \frac{164\pi^4}{135} 
   + \frac{1424}{27}\,\zeta_3 \right) \nonumber \\
  &+& C_A C_F T_F n_f \left( \frac{1178}{27}
   - \frac{4\pi^2}{3} - \frac{16\pi^4}{45} 
   - \frac{608}{9}\,\zeta_3 \right) + 8 C_F^2 T_F n_f \nonumber \\  
  &+& C_A T_F^2 n_f^2 \left( \frac{24520}{729}
   + \frac{80\pi^2}{81} - \frac{448}{27}\,\zeta_3 \right) 
   + \frac{176}{9} C_F T_F^2 n_f^2 \,.
   \label{eq:gammaS}
\end{eqnarray}

The function $\gamma^t$ is given by,
\begin{equation} \label{eq:gammatdef}
\gamma^t(\alpha_s) = \alpha_s^2 \, \frac{d}{d\alpha_s}
 \Bigg(\frac{\beta(\alpha_s)}{\alpha_s^2}\Bigg)
  = -2\beta_1 \left(\frac{\alpha_s}{4\pi}\right)^2
    -4\beta_2 \left(\frac{\alpha_s}{4\pi}\right)^3
    -6\beta_3 \left(\frac{\alpha_s}{4\pi}\right)^4 + {\cal O}(\alpha_s^5)\,.
\end{equation}
As shown in Eq.~(\ref{mu_independence_for_Hbar}) $\mu$ independence provides the constraint, 
\beq
2\gamma^g(\alpha_s)=\gamma^t(\alpha_s)+\gamma^S(\alpha_s)+\beta(\alpha_s)/\alpha_s\, ,
\eeq
leading to the simple relationship between the coefficients in $\gamma^g$ and $\gamma^S$,
\begin{equation}
\gamma_0^S = 2\gamma_0^g + 2\beta_0\,, 
\gamma_1^S = 2\gamma_1^g + 4\beta_1\,,
\gamma_2^S = 2\gamma_2^g + 6\beta_2,\, 
\gamma_3^S = 2\gamma_3^g + 8\beta_3\, . 
\end{equation}

\section{Definitions for beam function ingredients}
\label{beam_function_ingredients}
\subsection{Exponent $h$}
\label{exponenth}
We define the auxiliary functions $h^B$ for $B=F,A$ which, when combined with the hard function and
the collinear anomaly factor, will yield
a renormalization group invariant hard function. $h^{F/A}$ is defined to satisfy the \RGE{} 
equation,
\begin{equation}\label{eq:hevol}
\frac{d}{d\ln\mu}\,h^{F/A}(\pTveto,\mu)
   = 2\,\Gamma_{\rm cusp}^{F/A}(\mu)\,\ln\frac{\mu}{\pTveto} - 2\,\gamma^{q/g}(\mu) \,, \\
\end{equation}
The factor $h$ removes logarithms from the beam function and has a perturbative expansion in terms of the renormalized coupling,
\begin{equation}
  h^B(\pTveto,\mu)=\as h^B_{0}
  +\left(\as\right)^2  h^B_{1}
  +\left(\as\right)^3  h^B_{2}
  +\left(\as\right)^4  h^B_{3} + \ldots \, .
\end{equation}
Thus for the particular case $B=F$ we have that,
\begin{eqnarray}\label{hexpand}
h^F_{0}(\pTveto,\mu)&=&\frac{1}{4}\*\Gamma^{F}_{0}\*\Lperp^2-\gamma_0^{q}\*\Lperp\,, \nonumber \\
h^F_{1}(\pTveto,\mu)&=&\frac{1}{12}\*\Gamma^{F}_{0}\*\beta_0\*\Lperp^3+\frac{1}{4}\*(\Gamma^{F}_{1}-2\*\gamma_0^{q}\*\beta_0)\*\Lperp^2-\gamma_1^{q}\*\Lperp\,, \nonumber \\
h^F_{2}(\pTveto,\mu)&=&\frac{1}{24}\Gamma^{F}_{0}\*\beta_0^2\*\Lperp^4
 +(\frac{1}{12}\*\Gamma^{F}_{0}\*\beta_1+\frac{1}{6}\*\Gamma^{F}_{1}\*\beta_0-\frac{1}{3}\*\gamma_0^{q}\*\beta_0^2)\*\Lperp^3 \nonumber \\
 &+&(\frac{1}{4}\*\Gamma^{F}_{2}-\frac{1}{2}\*\gamma_0^{q}\*\beta_1-\gamma_1^{q}\*\beta_0)\*\Lperp^2-\gamma_2^{q}\*\Lperp\,, \nonumber \\
h^F_{3}(\pTveto,\mu)&=&
 +\frac{1}{40} \*\Gamma^{F}_{0}\*\beta_0^3\*\Lperp^5
 +(\frac{5}{48}\*\Gamma^{F}_{0}\*\beta_0\*\beta_1+\frac{1}{8}\*\Gamma^{F}_{1}\*\beta_0^2-\frac{1}{4}\*\gamma_0^{q}\*\beta_0^3)\*\Lperp^4 \nonumber  \\
 &+&(\frac{1}{12}\*\Gamma^{F}_{0}\*\beta_2+\frac{1}{6}\*\Gamma^{F}_{1}\*\beta_1+\frac{1}{4}\*\Gamma^{F}_{2}\*\beta_0-\frac{5}{6}\*\gamma_0^{q}\*\beta_0\*\beta_1
          -\gamma_1^{q}\*\beta_0^2)\*\Lperp^3\nonumber  \\
&+&(\frac{1}{4}\*\Gamma^{F}_{3}-\frac{1}{2}\*\gamma_0^{q}\*\beta_2-\gamma_1^{q}\*\beta_1-\frac{3}{2}\*\gamma_2^{q}\*\beta_0)\*\Lperp^2 -\gamma_{3}^{q}\*\Lperp \,,
\end{eqnarray}
where $\Lperp=2 \ln (\mu/\pTveto)$.
The corresponding result for $B=A,q=g$, (i.e.~for incoming gluons) is given by a similar expression {\it mutatis mutandis}.
The expansion coefficients of the $\beta$-function, $\Gamma_{\rm cusp}^{F/A}$ and $\gamma^{q/g}$, used in Eq.~(\ref{hexpand}),
are as given in Appendices~\ref{betacoeffs},\ref{Cusp_Anom_Dim} and \ref{Non_Cusp_Anom_Dim}.

\subsection{One loop splitting functions}
\label{oneloopanomdim}
The one-loop {\abbrev DGLAP} splitting functions as defined in \cite{Altarelli:1977zs} are
\begin{eqnarray}\label{APkernels}
   P_{qq}^{(1)}(z)
   &=& C_F \left( \frac{1+z^2}{1-z} \right)_+ ,\\
   P_{qg}^{(1)}(z)
   &=& T_F \left[ z^2 + (1-z)^2 \right]\,,\\
   P_{gg}^{(1)}(z)
   &=& 2 C_A \left[ \frac{z}{\left(1-z\right)_+} + \frac{1-z}{z} + z(1-z) \right]
    + \frac{\beta_0}{2}\,\delta(1-z) \,, \\
   P_{gq}^{(1)}(z)
   &=& C_F\,\frac{1+(1-z)^2}{z}\,,
\end{eqnarray}
\subsection{Two loop splitting functions}
\label{twoloopanomdim}
Now we turn to the two-loop anomalous dimensions that contribute 
at sub-leading log level to the transitions between parton types.
In the quark sector there are four independent transitions that we must produce values for
(viz. $q^\prime \leftarrow q$,$\bar{q}^\prime \leftarrow q$,$q \leftarrow q$ and $\bar{q} \leftarrow q$).
They are expressed in terms of four functions,
\beq
P^{(2)}_{q^\prime q}=P^{S(2)}_{qq}\, , \;
P^{(2)}_{\bar{q}^\prime q}=P^{S(2)}_{\bar{q}q}\, , \;
P^{(2)}_{qq}=P^{V(2)}_{qq}+P^{S(2)}_{qq}\, , \;
P^{(2)}_{\bar{q}q}=P^{V(2)}_{\bar{q}q}+P^{S(2)}_{\bar{q}q}\, .
\eeq
At next-to-leading order, the functions $P^S_{qq}$ and $P^S_{\bar{q}q}$ 
are non-zero, but we have the additional relation, $P^S_{qq}=P^S_{\bar{q}q}$.
To facilitate the presentation we define the auxiliary functions,
\beqn
\label{pqq}
p_{qq}(z)&=&\frac{2}{1-z}-1-z\, ,\;\; p_{qq}^{(r)}(z)=-1-z\,, \\
\label{pqg}
p_{qg}(z)&=&z^2+(1-z)^2\, ,  \\
\label{pgq}
p_{gq}(z)&=&\frac{1+(1-z)^2}{z}\, ,  \\
\label{pgg}
p_{gg}(z)&=& \frac{1}{1-z}+\frac{1}{z} -2 +z(1-z), \;\;
p_{gg}^{(r)}(z)=\frac{1}{z} -2 +z(1-z) \, .
\eeqn
The two valence functions needed for the quark sector are,
~\cite{Curci:1980uw,Furmanski:1980cm,Ellis:1996mzs},
{\allowdisplaybreaks
\beqn \label{pqq2}
&&P_{qq}^{V(2)}(z)=
  C_F^2 \Bigg\{ -\Bigg[2 \ln z \ln(1-z)+\frac{3}{2} \ln z  \Bigg] p_{qq}(z) 
 \nonumber \\ && 
 -\Bigg(\frac{3}{2}+\frac{7}{2} z\Bigg)\ln z
      -\frac{1}{2} (1+z) \ln^2 z -5 (1-z)\Bigg\}
\nonumber \\ &&
+C_F C_A \Bigg\{ (1+z) \ln z+\frac{20}{3} (1-z)
 +\Bigg[\frac{1}{2} \ln^2 z+\frac{11}{6} \ln z \Bigg] p_{qq}(z)
\nonumber \\ &&
        +\Bigg[\frac{67}{18}-\frac{\pi^2}{6} \Bigg]  \Big(\frac{1}{(1-z)_{+}} +p_{qq}^{(r)}(z)\Big) \Bigg\}
\nonumber \\ &&
      -C_F T_F n_f \Bigg\{\frac{4}{3}  (1-z) + \frac{2}{3} p_{qq}(z) \ln z 
      +\frac{10}{9} \Big(\frac{1}{(1-z)_{+}}  +p_{qq}^{(r)}(z)\Big)\Bigg\} 
\nonumber \\
      +&&\Bigg\{
   C_F^2 \Bigg[\frac{3}{8}-\frac{\pi^2}{2}+6 \zeta_3 \Bigg]
   +C_F C_A \Bigg[\frac{17}{24}+\frac{11\pi^2}{18}-3 \zeta_3\Bigg]
\nonumber \\ && 
 -C_F T_F n_f \Bigg[\frac{1}{6}+\frac{2 \pi^2}{9}\Bigg]\Bigg\}\delta(1-z) \; ,\\
&&P_{\bar{q}q}^{V(2)}(z) = C_F \Bigg(C_F-\frac{C_A}{2}\Bigg) 
\Big\{2 p_{qq}(-z) S_2(z)  +2 (1+z) \ln z+4 (1-z) \Big\} \; ,
\eeqn
}
and for the singlet function we have,
\beqn 
&&P_{qq}^{S(2)}=C_F T_F \Bigg\{\frac{20}{9\*z}-2+6\*z-\frac{56}{9}\*z^2+(1+5\*z+\frac{8}{3}\*z^2)\ln z
 -(1+z)\ln^2z \Bigg\}\, .
 \label{pqg2}
\eeqn

The other three transitions are simply given by,
{\allowdisplaybreaks
\beqn 
&&P_{qg}^{(2)}=C_F T_F
       \Bigg\{2-\frac{9}{2} z-(\frac{1}{2}-2 z) \ln z-(\frac{1}{2}- z) \ln^2z +2 \ln(1-z)        
\nonumber \\ &&
      +\Bigg[ \ln^2\Bigg(\frac{1-z}{z}\Bigg)-2 \ln\Bigg(\frac{1-z}{z}\Bigg)
      -\frac{\pi^2}{3}  +5 \Bigg] p_{qg}(z)\Bigg\} \nonumber \\ &&      
      +C_A T_F
       \Bigg\{\frac{91}{9}+\frac{7}{9} z+\frac{20}{9 z}
  +\Bigg(\frac{68}{3} z-\frac{19}{3}\Bigg) \ln z
\nonumber \\ &&
 -2 \ln(1-z)       -(1+4 z) \ln^2z   + p_{qg}(-z) S_2(z) 
\nonumber \\ && 
       +\Bigg[-\frac{1}{2}\ln^2z+\frac{22}{3} \ln z- \ln^2(1-z)+2 \ln(1-z)+\frac{\pi^2}{6}-\frac{109}{9}\Bigg] p_{qg}(z)
\Bigg\}\, ,
\eeqn
}
{\allowdisplaybreaks
\beqn \label{pgq2}
&&P_{gq}^{(2)}(z)=C_F^2
\Bigg\{-\frac{5}{2}-\frac{7 z}{2}+\Bigg(2+\frac{7}{2} z\Bigg) \ln z-\Bigg(1-\frac{1}{2} z\Bigg)\ln^2z  \nonumber \\
 &-&2 z \ln(1-z)-\Bigg[ 3 \ln(1-z)+\ln^2(1-z)\Bigg] p_{gq}(z)\Bigg\}
 \nonumber \\ &&
      +C_F C_A \Bigg\{
  \frac{28}{9}+\frac{65}{18} z+\frac{44}{9} z^2
 -\Bigg(12+5 z+\frac{8}{3} z^2 \Bigg) \ln z 
   \nonumber \\ &&
      +(4+z) \ln^2z+2 z \ln(1-z) +S_2(z) p_{gq}(-z)  \nonumber \\ &&
      +\Bigg[\frac{1}{2}-2 \ln z \ln(1-z)+\frac{1}{2} \ln^2z
 +\frac{11}{3} \ln(1-z)+\ln^2(1-z)-\frac{\pi^2}{6}\Bigg] p_{gq}(z)
\Bigg\} \nonumber \\ &&
      +C_F T_F n_f \Bigg\{
    -\frac{4}{3} z-\Bigg[\frac{20}{9}+\frac{4}{3} \ln(1-z) \Bigg] p_{gq}(z)
    \Bigg\} \, , \\
\label{pgg2}
&&P_{gg}^{(2)}(z)=C_F T_F n_f
 \Bigg\{-16+8 z+\frac{20}{3}z^2 +\frac{4}{3 z}-(6+10 z) \ln z-(2+2 z) \ln^2 z \Bigg\}
  \nonumber \\ &&
      +C_A T_F n_f
       \Bigg\{ 2-2 z+\frac{26}{9} \Bigg(z^2-\frac{1}{z}\Bigg)
       -\frac{4}{3} (1+z) \ln z\nonumber \\
       &&-\frac{20}{9} \Big(\frac{1}{(1-z)_{+}} + p^{(r)}_{gg}(z) \Big) \Bigg\}\nonumber \\ && 
      +C_A^2 \Bigg\{ \frac{27}{2} (1-z)+\frac{67}{9} \Bigg(z^2-\frac{1}{z}
      \Bigg) 
      -\Bigg(\frac{25}{3}-\frac{11}{3} z+\frac{44}{3} z^2\Bigg) \ln z \nonumber \\ && 
       +4 (1+z) \ln^2 z +2 p_{gg}(-z) S_2(z) 
\nonumber \\ && 
+ \Bigg[ \ln^2 z-4 \ln z \ln(1-z)\Bigg] p_{gg}(z)
 + \Bigg[ \frac{67}{9}-\frac{\pi^2}{3}\Bigg] \Big(\frac{1}{(1-z)_{+}} + p^{(r)}_{gg}(z) \Big) \Bigg\} \nonumber \\
&+&\Bigg\{C_A^2 \Big[\frac{8}{3}+3 \zeta_3\Big]-C_F T_F n_f -\frac{4}{3} C_A T_F n_f\Bigg\}\delta(1-z) \; .
\eeqn
}
The function $S_2(z)$ is defined by
\beq \label{chdis_s2}
S_2(z)= \int_{\frac{z}{1+z}}^{\frac{1}{1+z}} \frac{dy}{y} 
\ln \left(\frac{1-y}{y}\right) \; .
\eeq
In terms of the dilogarithm function 
\beq
\mbox{Li}_2(z) = - \int_0^{z}\; \frac{dy}{y} \ln (1-y)\, ,
\eeq
we have
\beq\label{S2Li2}
S_2(z) = -2\,\mbox{Li}_2(-z)+\frac{1}{2}\ln^2 z -2\ln z\ln(1+z)
-\frac{\pi^2}{6}\;.
\eeq
\subsection{$\Pone \otimes \Pone$ and $\Rone \otimes \Pone$}
\label{RxP}
We give here expressions for the convolutions of functions appearing in the beam functions.
The convolutions are defined as in Eq.~(\ref{convolution}).
Similar expressions have been given in \cite{Berger:2010xi,Becher:2013xia}
The convolutions of the one-loop {\abbrev DGLAP} kernels from Eqs.~(\ref{APkernels}) are,
{\allowdisplaybreaks
\begin{eqnarray}
  \Pone_{qq}&\otimes & \Pone_{qg}  =   C_F\* T_F\* \Big(2\*z-\frac{1}{2}+(2\* z-4\* z^2-1)\* \ln z
 +(2-4\* z\* (1-z)) \* \ln(1-z) \Big)\, ,\\
\Pone_{qg}&\otimes & \Pone_{gg}  =   C_A\* T_F\*
\Big(2\* (1+4\* z)\* \ln z +\frac{4}{ 3\*z}+1+8\* z-\frac{31}{3}\* z^2\Big)\nonumber \\
&+&\big(2\*C_A \ln(1-z) + \frac{\beta_0}{2} \big)\Pone_{qg}(z)\, ,\\
\Pone_{gq}&\otimes & \Pone_{qq}  =   C_F^2\* \Big(2-\frac{1}{2}\* z+(2-z)\* \ln z\Big)
+2\* C_F \Pone_{gq}(z)\*\ln(1-z)\Big)\, ,\\
\Pone_{gg}&\otimes & \Pone_{gq}  =  C_A\* C_F\* \Big (8+z +\frac{(4\*z^3-31)}{3 \*z}-\frac{4\* (1+z+z^2)}{z} \* \ln z \Big)\nonumber \\
&+&\big(2\* C_A \ln(1-z)+\frac{\beta_0}{2} \big)\Pone_{gq}(z)\, , \\
\Pone_{qg}&\otimes & \Pone_{gq}  =   C_F\* T_F\* \Big( 2\* (1+z)\* \ln z+1-z +\frac{4}{3} \* \frac{(1-z^3)}{z}\Big)\, ,\\
\Pone_{qq}&\otimes & \Pone_{qq}  =  C_F^2 \Big( 8\* \Big[\frac{\ln(1-z)}{(1-z)}\Big]_{+}- 4 \* (1+z)\* \ln(1-z)-2\*(1-z) \nonumber \\
&+&\big(3 +3\*z-\frac{4}{(1-z)}\big)\ln z \Big)
+3\* C_F\* \Pone_{qq}(z)-C_F^2\* (\frac{9}{4}+4\*\zeta_2) \delta(1-z) \, ,\\
\Pone_{gg}&\otimes & \Pone_{gg}  = 4 C_A^2\* \Big(
2\* \Big[\frac{\ln(1-z)}{(1-z)}\Big]_{+} + 2\* (\frac{(1-z)}{z}+z\* (1-z)-1)\ln(1-z)
+3\* (1-z) \nonumber \\
&-&(\frac{1}{1-z}+\frac{1}{z}-z^2+3\*z)\ln z 
-\frac{11\*(1-z^3)}{3\*z}\Big)\nonumber \\
&+&\beta_0 \Pone_{gg}(z)-(\frac{\beta_0^2}{4}+4\*C_A^2 \zeta_2) \delta(1-z)\, .
\end{eqnarray}}
The convolutions of lowest order {\abbrev DGLAP} kernels, Eq.~(\ref{APkernels}) with the one-loop 
finite terms in the beam functions, Eq.~(\ref{eq:R1}) are,
{\allowdisplaybreaks
\begin{eqnarray}
\Rone_{gg} \otimes \Pone_{gg}   &=& -C_A \* \zeta_2 \* \Pone_{gg}(z)\, , \\
\Rone_{gq} \otimes \Pone_{qg}   &=& 2\* C_F\* T_F\* \Big((1-z)\* (1+2\* z)+2\* z\* \ln z \Big) \, , \\
\Rone_{qq} \otimes \Pone_{qq}   &=& C_F \* \Big(C_F\* (1-z) (4 \ln(1-z)-2\* \ln z-1)-\zeta_2 \* \Pone_{qq}(z)\Big)\, ,\\
\Rone_{qg} \otimes \Pone_{gq}   &=& -4\* C_F\* T_F\* \Big(1+z\* \ln z-\frac{(1+2\*z^3)}{3 \*z}\Big) \, , \\
\Rone_{qg} \otimes \Pone_{gg} &=& -C_A\* T_F\* (16\* z\* \ln z-\frac{68}{3}\* z^2+20\* z+4-\frac{4}{3 z})\nonumber \\
&+&(2\* C_A\* \ln(1-z) +\frac{\beta_0}{2})\* \Rone_{qg}(z) \, ,\\
\Rone_{qq} \otimes \Pone_{qg} &=&  C_F\* T_F\* (2\* z^2+2\* z-4-(2+4\* z)\* \ln z)-C_F\*\zeta_2 \Pone_{qg}(z)\, , \\
\Rone_{gq} \otimes \Pone_{qq} &=& -C_F^2\*(2\*z\*\ln z-4\*z\*\ln(1-z)-z-2)\, , \\
\Rone_{gg} \otimes \Pone_{gq} &=& -C_A\*\zeta_2 \* \Pone_{gq}(z) \, .
\end{eqnarray}}

\section{Rapidity anomalous dimension}
\label{app:rapidityanomdim}
Solving the collinear anomaly \RG{} equation (\cref{eq:Fggevol}) as an expansion in $\alpha_s$ 
(\cref{eq:Fevolexp}) we have that,
\begin{eqnarray}
	\label{Fggexpansion}
	F_{gg}^{(0)}(\pTveto,\mu_h)&=& \Gamma_0^A \Lperp+d_1^\veto(R,A) \, , \nn \\
	F_{gg}^{(1)}(\pTveto,\mu_h)&=& \frac{1}{2} 
	\Gamma_0^{A}\*\beta_0\*\Lperp^2+\Gamma_1^{A}\*\Lperp+\dtwoveto(R,A) \, , \nn \\
	F_{gg}^{(2)}(\pTveto,\mu_h)&=& \frac{1}{3}\Gamma_0^{A}\*\beta_0^2\*\Lperp^3+\frac{1}{2} 
	(\Gamma_0^{A}\*\beta_1+2\*\Gamma_1^{A}\*\beta_0)\*\Lperp^2 \nn\\
	&+&(\Gamma_2^{A}+2\*\beta_0\*\dtwoveto(R,A))\*\Lperp+\dthreeveto(R,A) \nn \, ,\\
	F_{gg}^{(3)}(\pTveto,\mu_h)&=&
	\frac{1}{4}\*\beta_0^3\*\Gamma_0^{A}\*\Lperp^4+(\Gamma_1^{A}\*\beta_0^2+\frac{5}{6}\*\Gamma_0^{A}\*\beta_0\*\beta_1)\*\Lperp^3\nn
	 \\
	&+&(\frac{1}{2}\*\Gamma_0^{A}\*\beta_2+\Gamma_1^{A}\*\beta_1+\frac{3}{2}\*\Gamma_2^{A}\*\beta_0+3\*\dtwoveto(R,A)\*\beta_0^2)\*\Lperp^2
	 \nn \\
	&+&(\Gamma_{3}^{A}+3\*\dthreeveto(R,A)\*\beta_0+2\*\dtwoveto(R,A)\*\beta_1)\Lperp+\dfourveto(R,A)
	 \, . 
\end{eqnarray}
where $\Lperp=2 \ln (\mu_h/\pTveto)$.
The corresponding result for $F_{qq}$ is given in Eq.~(\ref{Fqqexpansion}).
Because $F_{gg}$ appears in the exponent, we see that $\doneveto$ contributes in \NLL{},
$\dtwoveto$ in \NNLL{}, and $\dthreeveto$ in \NNNLL{}.

\subsection{$\dtwoveto$ expansion}

The expansion coefficients for $\dtwoveto$, which is defined in Eq.~(\ref{fRBexpansion}),
are given by~\cite{Banfi:2012yh,Banfi:2012jm,Becher:2013xia},
\begin{eqnarray}
   c_L^A &=& \frac{131}{72} - \frac{\pi^2}{6} - \frac{11}{6} \ln 2 = - 1.096259 \,, \nn \\
   c_0^A &=& -\frac{805}{216} + \frac{11 \pi^2}{72} + \frac{35 }{18}\ln 2 + \frac{11}{6} \ln^2 2+\frac{\zeta_3}{2} = 0.6106495\,,\nn \\
   c_2^A &=& \frac{1429}{172800} + \frac{\pi^2}{48} + \frac{13}{180} \ln 2 = 0.263947 \,, \nn \\
   c_4^A &=& - \frac{9383279}{406425600} - \frac{\pi^2}{3456} + \frac{587}{120960} \ln 2 = - 0.0225794 \,, \nn \\
   c_6^A &=& \frac{74801417}{97542144000} - \frac{23}{67200} \ln 2 = 5.29625\cdot 10^{-4} \,, \nn \\
   c_8^A &=& - \frac{50937246539}{2266099089408000} - \frac{\pi^2}{24883200}+ \frac{28529}{1916006400} \ln 2= - 1.25537\cdot 10^{-5} \,, \nn \\
   c_{10}^A &=& \frac{348989849431}{243708656615424000} - \frac{3509}{3962649600} \ln 2 = 8.18201\cdot 10^{-7} \,.
\end{eqnarray}
and
\begin{eqnarray}
   c_L^f &=& - \frac{23}{36} + \frac{2}{3} \ln 2 = - 0.1767908 \,, \nn \\
   c_0^f &=& \frac{157}{108} - \frac{\pi^2}{18} -\frac{8}{9}\ln 2 - \frac{2}{3} \ln^22 = - 0.03104049 \,, \nn \\
   c_2^f &=& \frac{3071}{86400} - \frac{7}{360} \ln 2 = 0.0220661 \,, \nn \\
   c_4^f &=& - \frac{168401}{101606400} + \frac{53}{30240} \ln 2 = -4.42544\cdot 10^{-4} \,, \nn \\
   c_6^f &=& \frac{7001023}{48771072000} - \frac{11}{100800} \ln 2 = 6.79076\cdot 10^{-5} \,, \nn \\
   c_8^f &=& - \frac{5664846191}{566524772352000} + \frac{4001}{479001600} \ln 2 = - 4.20958\cdot 10^{-6} \,, \nn \\
   c_{10}^f &=& \frac{68089272001}{83774850711552000} - \frac{13817}{21794572800} \ln 2 = 3.73334\cdot 10^{-7} \,,
\end{eqnarray}
We see that for values of the jet radius  $R<1$ the terms $c_6,c_8$ and $c_{10}$ can be dropped.

For the gluon case the expansion of the function in numerical form is,
\begin{align}
	f(R,A) &= - \left( 1.0963\,C_A + 0.1768\,T_F n_f \right) \ln R
	+ \left(  0.6106\,C_A -  0.0310\,T_F n_f \right) \nonumber \\
	&+ \left( -0.5585\,C_A + 0.0221\,T_F n_f \right) R^2 \nonumber \\
	&+ \left( 0.0399\,C_A - 0.0004\,T_F n_f \right) R^4 + \dots \; ,
	\label{fRgluon}
\end{align}
whereas for the quark case we have
\begin{align}
	f(R,F) &= - \left( 1.0963\,C_A + 0.1768\,T_F n_f \right) \ln R
	+ \left(  0.6106\,C_A -  0.0310\,T_F n_f \right) \nonumber \\
	&+ \left( - 0.8225\,C_F +0.2639\,C_A + 0.0221\,T_F n_f \right) R^2 \nonumber \\
	&+ \left(0.0625\,C_F -0.02258 \,C_A - 0.0004\,T_F n_f \right) R^4 + \dots \;.
	\label{fRquark}
\end{align}

\section{Renormalization Group Evolution}
\label{sec:rengroup}
The evolution equation matching for a generic hard matching coefficient $C$ has the form,
\begin{equation}\label{eq:evolCV}
   \frac{d}{d\ln\mu}\,\ln C(Q^2,\mu)
   = \left[ \Gamma_{\rm cusp}(\alpha_s(\mu))\,\ln\frac{Q^2}{\mu^2}
   + \gamma(\alpha_s(\mu)) \right] \,.
\end{equation}
Following ref.~\cite{Becher:2006mr} the solution to the evolution equation Eq.~(\ref{eq:evolCV}) is,
\begin{eqnarray}\label{CVsol1}
   C(Q^2,\mu) &=& \exp\left[ 2S(\mu_h,\mu) - a^{\gamma}(\mu_h,\mu) \right]
   \left( \frac{Q^2}{\mu_h^2} \right)^{-a^\Gamma(\mu_h,\mu)}\, C(Q^2,\mu_h) \, , \\
   \label{CVsol2}
   \ln C(Q^2,\mu) &=&
    2S(\mu_h,\mu)-a^{\gamma}(\mu_h,\mu) -a^{\Gamma}(\mu_h,\mu) \ln \left( \frac{Q^2}{\mu_h^2} \right)+\ln C(Q^2,\mu_h)\, ,
\end{eqnarray}
where $\mu_h\sim Q$ is a hard matching scale at which the Wilson coefficient 
$C$ is calculated using fixed-order perturbation theory.
The Sudakov exponent $S$ and the exponents $a^\gamma,a^\Gamma$
are the solutions to the auxiliary differential equations,
\begin{eqnarray}\label{dgl}
   \frac{d}{d\ln\mu}\,S(\nu,\mu)
   &=& - \Gamma_{\rm cusp}\big(\alpha_s(\mu)\big)\,\ln\frac{\mu}{\nu} \, ,\\
   \frac{d}{d\ln\mu}\,a^\Gamma(\nu,\mu)
   &=& - \Gamma_{\rm cusp}\big(\alpha_s(\mu)\big)  \, ,\\
   \frac{d}{d\ln\mu}\,a^{\gamma}(\nu,\mu)
   &=& - \gamma\big(\alpha_s(\mu)\big) \, . 
\end{eqnarray}
with the boundary conditions $S(\nu,\nu)=a^\Gamma(\nu,\nu)=a^{\gamma}(\nu,\nu)=0$ at $\mu=\nu$.
Differentiating Eq.~(\ref{CVsol2}) we recover Eq.~(\ref{eq:evolCV}).

The solutions to the evolution equation are conveniently expressed in terms of the running coupling,
\begin{eqnarray}\label{RGEsolgamma}
   a^\Gamma(\nu,\mu) &=& - \int\limits_{\alpha_s(\nu)}^{\alpha_s(\mu)}\!
    d\alpha\,\frac{\Gamma_{\rm cusp}(\alpha)}{\beta(\alpha)} \,, \\
    \label{RGEsolGamma}
    S(\nu,\mu) &=& - \int\limits_{\alpha_s(\nu)}^{\alpha_s(\mu)}\!
    d\alpha\,\frac{\Gamma_{\rm cusp}(\alpha)}{\beta(\alpha)}
    \int\limits_{\alpha_s(\nu)}^\alpha
    \frac{d\alpha'}{\beta(\alpha')} \, .
    \label{RGEsolS}
\end{eqnarray}
Substituting the values for the beta function coefficients in the \MSbar scheme
given in Appendix~\ref{betacoeffs}
and the values for cusp anomalous dimension given in Appendix~\ref{Cusp_Anom_Dim} into Eq.~(\ref{RGEsolgamma}) we obtain,
\begin{equation}
a^\Gamma(\mu_h,\mu) = a^\Gamma_0 + a^\Gamma_1 + a^\Gamma_2 + a^\Gamma_3 \, ,
\end{equation}
where the coefficients in the expansion are,
\begin{eqnarray}
a^\Gamma_0 &=& \frac{\Gamma_0 \ln (r)}{2 \beta_0}\,, \quad\quad r=\alpha_s(\mu)/\alpha_s(\mu_h) \, ,
\\
a^\Gamma_1 &=&  \frac{\alpha_s(\mu_h) (r-1) (\beta_0 \Gamma_1-\beta_1
    \Gamma_0)}{8 \pi  \beta_0^2}\,,
\\
a^\Gamma_2 &=& \frac{\alpha_s^2(\mu_h) (r^2-1)\left(-\beta_0 \beta_1
    \Gamma_1+\beta_0 (\beta_0 \Gamma_2-\beta_2
    \Gamma_0)+\beta_1^2 \Gamma_0\right)}{64 \pi ^2
    \beta_0^3}\,,
\\
a^\Gamma_3 &=& -\alpha^3_s(\mu_h) \left(r^3-1\right) \nonumber \\
  &\times& \frac{\left(\beta_0^2
    (-\beta_0 \Gamma_3+\beta_2 \Gamma_1+\beta_3
    \Gamma_0)-\beta_0 \beta_1^2 \Gamma_1+\beta_0
    \beta_1 (\beta_0 \Gamma_2-2 \beta_2
    \Gamma_0)+\beta_1^3 \Gamma_0\right)}
     {384 \pi ^3 \beta_0^4}\, .
\end{eqnarray}
The solution for $a^{\gamma}$ follows from the one for $a^\Gamma$ by making
the replacement $\Gamma_k \to \gamma_k$. The non-cusp anomalous dimensions $\gamma$
are given in Appendix~\ref{Non_Cusp_Anom_Dim}.

Evaluating Eq.~(\ref{RGEsolGamma}) to obtain the evolution for $S$ we get,
\begin{equation}
  S(\mu_h,\mu) = S^0 + S^1 + S^2
\, .
\end{equation}
with,
\begin{eqnarray}
\label{S0}
S^0&=&\frac{1}{8\beta_0^3}\Biggl(
\frac{8\pi\beta_0\Gamma_0(r+r(-\ln
(r))-1)}{\alpha_s(\mu_h)r}+2(r-1)(\beta_1
\Gamma_0-\beta_0\Gamma_1)\nonumber\\&&
+\ln(r)(2\beta_0
\Gamma_1+\beta_1\Gamma_0\ln(r)-2\beta_1
\Gamma_0)
\Biggr)\, ,
\\
S^1&=&-\frac{\alpha_s(\mu_h)}{32\pi\beta_0^4}\Biggl(
2\ln(r)\left(-\beta_0\beta_1
\Gamma_1r+\beta_0\beta_2\Gamma_0+\beta_1^2
\Gamma_0(r-1)\right)\nonumber\\
&+&(r-1)\left(-\beta_0\beta_1
\Gamma_1(r-3)+\beta_0(\beta_0(r-1)
\Gamma_2-\beta_2\Gamma_0(r+1))+\beta_1^2
\Gamma_0(r-1)\right)
\Biggr)\, ,
\\
S^2&=&\frac{\alpha_s^2(\mu_h)}{256\pi^2\beta_0^5}\Biggl(
2\ln(r)\Big(\beta_1r^2
\left(-\beta_0\beta_1\Gamma_1+\beta_0(\beta_0
\Gamma_2-\beta_2\Gamma_0)+\beta_1^2
\Gamma_0\right)\nonumber\\
&-&\Gamma_0\left(\beta_0^2\beta_3-2
\beta_0\beta_1\beta_2+\beta_1^3\right)\Big)\nonumber\\
&+&(r-1)
\Bigl(\beta_0^2(2(\beta_0(r+1)\Gamma_3-2\beta_2
\Gamma_1)-\beta_3\Gamma_0(r+1))+\beta_0
\beta_1^2\Gamma_1(r+5)\nonumber\\
&+&\beta_0\beta_1(\beta_2\Gamma_0(r+5)-3\beta_0(r+1)\Gamma_2)-4\beta_1^3\Gamma_0\Bigr)
\Biggr) \, .
\end{eqnarray}

\subsection{Recovery of the double log formula}
As we have seen $S$ satisfies a \RGE{} given by Eq.~(\ref{dgl}) with a solution given by 
Eq.~(\ref{RGEsolS}).
The leading term in $S_0$, Eq.~(\ref{S0}) is
\beq
S_0 \approx \frac{\pi \Gamma_0}{\beta_0^2 {\alpha_s(\mu_h)}}\Big( 1+\ln\big(\frac{1}{r}\big)-\frac{1}{r}\Big) \, ,
\eeq
where $r=\alpha_s(\mu)/\alpha_s(\mu_h)$. In this form the presence of a double log is obscured. We can easily recover 
the double log by retaining only the leading terms.
The leading expression for $r$ is given
by solving the equation for the beta function,
\beq
\frac{1}{r}=1-\frac{\alpha_s(\mu_h)}{2\pi}\beta_0\ln\big(\frac{\mu_h}{\mu}\big)\, ,
\eeq
\beq
S_0 \approx \frac{\pi \Gamma_0}{\beta_0^2 {\alpha_s(\mu_h)}}
\Big[\frac{\alpha_s(\mu_h)}{2\pi}\beta_0 \ln\big(\frac{\mu_h}{\mu}\big)+\ln\big(1-\frac{\alpha_s(\mu_h)}{2\pi}\beta_0\ln\big(\frac{\mu_h}{\mu}\big)\big)\Big]\, .
\eeq
Expanding for small  $\alpha_s(\mu_h) \ln (\mu_h/\mu)$ we get,
\beq
S(\mu_h,\mu) \approx -\frac{\Gamma_0}{2} \frac{\alpha_S(\mu_h)}{ 4 \pi} \ln^2\big(\frac{\mu_h}{\mu}\big) \, .
\eeq
This gives the expected log squared with a negative sign.

\section{The hard function for the Drell-Yan process}
\label{hardDY}
The form factors of the vector current have been presented several places in the
literature~\cite{Kramer:1986sg,Matsuura:1987wt,Matsuura:1988sm,Moch:2005tm,Moch:2005id,Gehrmann:2010ue}.
The bare form factor is given as,
\begin{equation}
  {F}^{q,{\rm bare}}(q^2,\mu^2) = 1 + \left(\frac{\alpha_s^{\rm bare}}{4\pi}\right)\,(\Delta)^\ep
  {\cal F}^q_1 + \left(\frac{\alpha_s^{\rm bare}}{4\pi}\right)^2\,(\Delta)^{2\ep} {\cal F}^q_2
+ {\cal O} (\alpha_s^3)\, ,
\end{equation}
where,
\beq
\Delta=4 \pi e^{-\gamma_E} \Bigg(\frac{\mu^2}{-q^2-\img  0}\Bigg)\, .
\eeq
In the following we will drop $4 \pi e^{-\gamma_E}$, so that all poles should be understood in the 
$\overline{\text{MS}}$ sense.
The values found for the bare coefficients are,
\begin{eqnarray}
{\cal F}^q_1 &=& C_F\Biggl[
-\frac{2}{\ep^2}-\frac{3}{\ep}+\zeta_2-8
+\ep\left(\frac{3\zeta_2}{2}+\frac{14\zeta_3}{3}-16\right)\nonumber \\
&+&\ep^2\left(\frac{47\zeta_2^2}{20}+4\zeta_2+7\zeta_3-32\right)\Biggr] +O(\ep^3)\,, \\
{\cal F}^q_2 &=& C_F^2 \Biggl[
\frac{2}{\ep^4}+\frac{6}{\ep^3}-\frac{1}{\ep^2}\left(2\zeta_2-\frac{41}{2}\right)-\frac{1}{\ep}\left(\frac{64\zeta_3}{3}-\frac{221}{4}\right)\nonumber \\
&-&\left(13\zeta_2^2-\frac{17\zeta_2}{2}+58\zeta_3-\frac{1151}{8}\right)\Biggr]\nonumber \\
&+&C_FC_A\Biggl[
-\frac{11}{6\ep^3}
+\frac{1}{\ep^2}\left(\zeta_2-\frac{83}{9}\right)
-\frac{1}{\ep}\left(\frac{11\zeta_2}{6}-13\zeta_3+\frac{4129}{108}\right)\nonumber \\
&+&\left(\frac{44\zeta_2^2}{5}-\frac{119\zeta_2}{9}+\frac{467\zeta_3}{9}-\frac{89173}{648}\right)\Biggr]\nonumber \\
&+&C_F n_f\Biggl[
\frac{1}{3\ep^3}
+\frac{14}{9\ep^2}
+\frac{1}{\ep}\left(\frac{\zeta_2}{3}+\frac{353}{54}\right)
+\left(\frac{14\zeta_2}{9}-\frac{26\zeta_3}{9}+\frac{7541}{324}\right)\Biggr]+O(\ep)\,.
\end{eqnarray}
The renormalized form factor can then be written as,
\begin{equation} \label{RenormFormFactor}
  {F}^{q}(\mu^2,q^2,\ep) = 1 + \left(\frac{\alpha_s(\mu)}{4\pi}\right)  F^{q}_{1}(\mu^2,q^2,\ep)
  + \left(\frac{\alpha_s(\mu)}{4\pi}\right)^2 F^{q}_{2}(\mu^2,q^2,\ep) 
+ {\cal O} (\alpha_s^3)\;.
\end{equation}
where,
\beqn
F^q_1(\mu^2,q^2,\ep)&=&\Delta^\ep {\cal F}^q_1 \, , \nonumber \\
F^q_2(\mu^2,q^2,\ep)&=&\Delta^{2\ep} {\cal F}^q_2 -\frac{\beta_0}{\ep} \Delta^\ep  {\cal F}^q_1 \, .
\eeqn
In the full theory the matrix element between on-shell
massless quark and gluon states, after charge renormalization is given by $F^q(\mu^2,q^2,\ep)$.
Charge renormalization has removed the {\abbrev UV} poles, but the renormalized form factor still 
contains {\abbrev IR} poles.

The matrix element in the effective theory involves only scaleless, dimensionally regulated integrals
and hence is equal to zero. This vanishing can be interpreted as a cancellation between ultra-violet and
infrared poles:
\beq
\frac{1}{\epsilon_{\text{IR}}}-\frac{1}{\epsilon_{\text{UV}}}\, .
\eeq
After matching, the {\abbrev IR} poles in the on-shell matrix element are effectively transformed 
into {\abbrev UV} poles and need to
be renormalized as follows,
\begin{eqnarray}
  C^V(\alpha_s(\mu^2),\mu^2,q^2) &=& \lim_{\epsilon \to 0} \big(Z^V(\epsilon,\mu^2 q^2)\big)^{-1} \, F^q(\mu^2,q^2,\epsilon)\, ,\nonumber \\
\ln \big[C^V(\alpha_s(\mu^2),\mu^2,q^2)\big] &=& \ln \big[F_q(\mu^2,q^2,\epsilon)\big]-\ln \big[Z^V(\epsilon,\mu^2,q^2)\big] \, .
\label{CVexpression}
\end{eqnarray}
The renormalization constant, $Z^V$ contains only pure pole terms,
\begin{eqnarray}
  &&\ln Z^V(\epsilon,\mu^2,q^2)=\big(\as\big) \Bigg[-\frac{\Gamma_0^F}{2 \ep^2}
    + \frac{1}{2 \ep} \Big(\Gamma^F_0 L+2 \gamma^q_0\Big)\Bigg]\nonumber \\
  &+&\big(\as\big)^2  \Bigg[ \frac{3 \Gamma^F_0 \beta_0}{8 \ep^3} 
    -\frac{1}{\ep^2} \Big[ \frac{\Gamma^F_0 \beta_0}{4} L -C_F\big(C_A (\frac{16}{9}+\zeta_2\big)+\frac{4}{9} n_f)\Big]
    + \frac{1}{4 \ep} \Big(\Gamma^F_1 L+2 \gamma^q_1\Big)\Bigg],
\end{eqnarray}
where $L=\ln((-q^2-\img 0)/\mu^2)$.
       
The matching coefficients have a perturbative expansion in terms of the renormalized coupling,
\begin{eqnarray}
{C}^V(\alpha_s(\mu^2), \mu^2,q^2) &=& 1 + \sum_{n=1}^{\infty} \left( \frac{\alpha_s (\mu^2)}{4\pi}\right)^n  \, {C}^V_{n} (\mu^2,q^2).
\end{eqnarray}
The matching coefficients, which are known to two loop order~\cite{Idilbi:2005ky,Idilbi:2006dg} (and beyond~\cite{Gehrmann:2010ue})
for Drell-Yan production,  can be obtained from Eq.~(\ref{CVexpression}):
{\allowdisplaybreaks
\begin{eqnarray}
C^V_1 &=& C_F \bigg(- L^2 + 3 L - 8 + \zeta_2  \bigg) , \\
C^V_2 &=&
C_F^2 \bigg(
\frac{1}{2}L^4-3L^3+\bigg(\frac{25}{2}-\zeta_2\bigg)L^2+\bigg(-\frac{45}{2}+24\zeta_3-9\zeta_2\bigg)L\nonumber \\
&+&\frac{255}{8}-30\zeta_3+21\zeta_2-\frac{83}{10}\zeta_2^2 \bigg)\nonumber \\ &&
+C_FC_A \bigg(
\frac{11}{9}L^3+\bigg(-\frac{233}{18}+2\zeta_2\bigg)L^2+\bigg(\frac{2545}{54}-26\zeta_3+\frac{22}{3}\zeta_2\bigg)L\nonumber \\
&-&\frac{51157}{648}+\frac{313}{9}\zeta_3-\frac{337}{18}\zeta_2+\frac{44}{5}\zeta_2^2 \bigg)\nonumber \\ 
&+&C_F n_f \bigg(
-\frac{2}{9}L^3+\frac{19}{9}L^2+\bigg(-\frac{209}{27}-\frac{4}{3}\zeta_2\bigg)L+\frac{4085}{324}+\frac{2}{9}\zeta_3+\frac{23}{9}\zeta_2 \bigg) \, ,
\end{eqnarray}
where $L=\ln((-q^2-\img 0)/\mu^2)$. $C^V$ satisfies the renormalization group equation,
\begin{equation}
\frac{d}{d \ln \mu}
  \ln[{C}^V(\alpha_s(\mu^2), \mu^2,q^2)] = \Gamma^F_{\cusp}(\mu)\ln\Big(\frac{-q^2-\img 0}{\mu^2}\Big)+2\gamma^q(\mu) \, ,
\end{equation}
with the anomalous dimensions as given in Appendix~\ref{Cusp_Anom_Dim} and Appendix~\ref{Non_Cusp_Anom_Dim}.

The derivation of the hard function for boson pair processes has been described in Ref.~\cite{Campbell:2022gdq}.
\section{The hard function for Higgs production}
\subsection{Implementation of one-step procedure}
\label{onestep}
The one-step procedure \cite{Berger:2010xi,Stewart:2013faa}
is based on the observation that the ratio $m_t/m_H$ is not large.
For an on-shell Higgs boson the parameter,
$m_H^2/m_t^2\approx \frac{1}{2}$ whereas $\alpha_s \ln(m_t^2/m_H^2) \approx 0.65 \alpha_s$,
indicating that power corrections should be more important than resumming logarithms. 
The matching is performed at a scale $\mu_h$ by integrating out the top quark and all gluons
and light quarks with off-shellness above $\mu_h$.

The hard Wilson coefficient so defined satisfies the \RGE{},
\begin{equation} \label{eq:C_RGE}
  \mu \frac{d}{d\mu} \ln C^{H}(m_t^2, q^2, \mu^2) = \Gamma^A_\cusp(\alpha_s(\mu))\,
  \ln\frac{- q^2-\img 0}{\mu^2} + 2 \gamma^g[\alpha_s(\mu)]\, ,
\end{equation}
where $\Gamma_{\cusp}$ and $\gamma^g$ are given in Eqs.~(\ref{cusp}) and (\ref{noncusp}).
As a consequence of Eq.~(\ref{eq:C_RGE}) the Wilson coefficient has the following structure,
\begin{align} \label{eq:CH}
C^H(m_t^2, q^2, \mu_h^2)
&= \alpha_s(\mu_h) F_0^H\Bigl(\frac{q^2}{4m_t^2}\Bigr)
\biggl\{1 + \frac{\alpha_s(\mu_h)}{4\pi}
\biggl[ C^H_1\Bigl(\frac{-q^2 -\img 0}{\mu_h^2} \Bigr)
+ F^H_1\Bigl(\frac{q^2}{4m_t^2}\Bigr) \biggr]
\nn \\ & \quad
+ \Bigg(\frac{\alpha_s(\mu_h)}{(4\pi)}\Bigg)^2
\biggl[ C^H_2\Bigl(\frac{-q^2 -\img 0}{\mu_h^2}, \frac{q^2}{4m_t^2} \Bigr)
+ F^H_2\Bigl(\frac{q^2}{4m_t^2}\Bigr) \biggr]
\biggr\}
\,,\end{align}
The finite terms can be derived from Ref.~\cite{Davies:2019wmk},
\begin{align} \label{eq:Ftop0}
F^H_0(z) &= \frac{3}{2z} - \frac{3}{2z}\Bigl\lvert 1 - \frac{1}{z}\Bigr\rvert
\begin{cases}
\arcsin^2(\sqrt{z})\, , & 0 < z \leq 1 \,,\\
\ln^2[-\img(\sqrt{z} + \sqrt{z-1})] \,, \quad & z > 1
\,,\end{cases}\\
\label{eq:Ftop0expand}
& \approx 1+\frac{7\*z}{30}+\frac{2\*z^2}{21}+\frac{26\*z^3}{525}+\frac{512\*z^4}{17325}+O(z^5), \quad z<1
\,.\end{align}
For the values of $m_t$ and $m_H$ in Table~\ref{inputparameters},
\beq \label{massrescaling}
|F^H_0(z_0)|^2=1.0653\, ,\;\;\; z_0=\frac{m_H^2}{4 m_t^2}\,.
\eeq
The coefficients $C_1^H$ and $C_2^H$ are fixed by the Eq.~(\ref{eq:C_RGE}).
\begin{eqnarray}
C^H_1(L) &=& C_A \Bigl(-L^2 + \frac{\pi^2}{6} \Bigr)\, , \\
C^H_2(L, z)&=& \frac{1}{2} C_A^2 L^4 + \frac{1}{3} C_A \beta_0 L^3 +
  C_A\Bigl[\Bigl(-\frac{4}{3} + \frac{\pi^2}{6} \Bigr) C_A  - \frac{5}{3}\beta_0 - F_1(z) \Bigr] L^2 \nn \\
  &+& \Bigl[\Bigl(\frac{59}{9} - 2\zeta_3 \Bigr) C_A^2  + \Bigl(\frac{19}{9}-\frac{\pi^2}{3}\Bigr) C_A \beta_0 - F_1(z) \beta_0 \Bigr] L\,.
\end{eqnarray}
where $z=q^2/4/m_t^2$ and $L=\ln[(-q^2-\img 0)/\mu_h^2]$.

The full analytic $m_t$ dependence of the virtual two-loop corrections to $gg\to H$ in terms of harmonic polylogarithms were obtained in
Refs.~\cite{Spira:1995rr,Harlander:2005rq, Anastasiou:2006hc}.
For our purposes the results expanded in $m_H^2/m_t^2$ from Refs.~\cite{Harlander:2009bw, Pak:2009bx,Davies:2019wmk}
will be sufficient.
The functions $F^H_1(z),F^H_2(z)$ which, together with $F^H_0(z)$ in Eq.~(\ref{eq:Ftop0expand})
encode the $m_t$ dependence of the hard Wilson coefficient in Eq.~(\ref{eq:CH}). Following the procedure described in
Appendix~\ref{hardDY} they are easily extracted from Ref.~\cite{Davies:2019wmk},
\begin{eqnarray} \label{eq:Ftop1}
 &&F^H_1(z) =
\Bigl(5-\frac{38}{45}\,z-\frac{1289}{4725}\,z^2-\frac{155}{1134}\,z^3-\frac{5385047}{65488500}\,z^4\Bigr)C_A\nn\\
 &+&\Bigl(-3+\frac{307}{90}\,z+\frac{25813}{18900}\,z^2+\frac{3055907}{3969000}\,z^3+\frac{659504801}{1309770000}\,z^4\Bigr)C_F+\ord{z^5}\\
\label{eq:Ftop2}
 &&F^H_2(z) =\bigl(7 C_A^2 + 11 C_A C_F - 6 C_F \beta_0 \bigr) \lnz+ \Bigl(-\frac{419}{27} + \frac{7\pi^2}{6} + \frac{\pi^4}{72} - 44 \zeta_3 \Bigr) C_A^2\nn \\
 &+& \Bigl(-\frac{217}{2} - \frac{\pi^2}{2} + 44\zeta_3 \Bigr) C_A C_F+\Bigl(\frac{2255}{108} + \frac{5\pi^2}{12} + \frac{23\zeta_3}{3} \Bigr) C_A \beta_0- \frac{5}{6} C_A T_F
\nn \\
 &+& \frac{27}{2} C_F^2+ \Bigl(\frac{41}{2} - 12 \zeta_3\Bigr) C_F \beta_0-\frac{4}{3} C_F T_F \nn \\
 &+&z \Big[C_A^2 \Big(\frac{11723}{384} \zeta_3-\frac{404063}{14400}-\frac{223}{108} \lnz-\frac{19}{135} \pi^2\Big) \nn\\
 &+&C_F C_A \Big(\frac{2297}{16} \zeta_3-\frac{1099453}{8100}-\frac{242}{135} \lnz-\frac{953}{540} \pi^2+\frac{28}{15} \pi^2 \ln2\Big) \nn\\
 &+&C_F^2 \Big(\frac{13321}{96} \zeta_3-\frac{36803}{240}+\frac{7}{3} \pi^2-\frac{56}{15} \pi^2 \ln2\Big) \nn\\
 &+&C_F \Big(\frac{77}{12} \zeta_3-\frac{4393}{405}-\frac{7337}{2700} \beta_0+\frac{39}{10} \lnz \beta_0+\frac{28}{45} \pi^2+\frac{7}{15} \pi^2 \beta_0\Big) \nn \\
 &+&C_A \Big(\frac{77}{384} \zeta_3-\frac{64097}{129600}-\frac{269}{75} \beta_0+\frac{2}{15} \lnz-\frac{31}{180} \lnz \beta_0\Big)\Big] \nn \\
 &+&z^2 \Big[C_A^2 \Big(\frac{110251}{9216} \zeta_3-\frac{3084463261}{254016000}-\frac{2869}{4536} \lnz-\frac{1289}{28350} \pi^2\Big)\nn \\
 &+&C_F C_A \Big(\frac{2997917}{23040} \zeta_3-\frac{55535378557}{381024000}-\frac{18337}{28350} \lnz-\frac{128447}{113400} \pi^2+\frac{1714}{1575} \pi^2 \ln2\Big)\nn\\
 &+&C_F^2 \Big(\frac{36173}{192} \zeta_3-\frac{95081911}{453600}+\frac{857}{630} \pi^2-\frac{3428}{1575} \pi^2 \ln2\Big) \nn\\
 &+&C_A \Big(\frac{265053121}{1524096000}-\frac{16177}{92160} \zeta_3-\frac{45617}{47250} \beta_0+\frac{16}{315} \lnz-\frac{623}{5400} \lnz \beta_0\Big) \nn\\
 &+&C_F \Big(\frac{21973}{7680} \zeta_3-\frac{8108339}{1555200}-\frac{509813}{3969000} \beta_0-\frac{8}{15} \lnz+\frac{29147}{18900} \lnz \beta_0 \nn\\
 &+&\frac{1714}{4725} \pi^2+\frac{857}{3150} \pi^2 \beta_0\Big)\Big] + \ord{z^3}\,.
\end{eqnarray}
We can assess the quality of the expansion in $z$ by numerical evaluation,
\beqn \label{eq:CHnmer}
&&C^H(m_t^2, q^2, q^2)=\alpha_s(q) F_0(z)\Big[1+15.9348\*\as\*(1+0.0158 (8\*z)+.00098312 (8\*z)^2)\nn \\
 &+&97.0371\*\Big(\as\Big)^2\*(1+0.1883 (8\*z)+0.0120 (8\*z)^2) \nn \\
  &+&143.466\*\Big(\as\Big)^2\*\frac{\ln(-8z-\img 0)}{\pi}\*(1+0.0288 (8\*z)+0.001462 (8\*z)^2) \Big]\, .
 \eeqn
In the vicinity of the Higgs boson pole ($8 z \approx 1$) subsequent terms in the $z$ expansion are expected to contribute below the percent level.
\subsection{Implementation of the two-step procedure}
\label{twostep}
In the two-step procedure of Refs.~\cite{Idilbi:2005er,Idilbi:2005ni,Ahrens:2009cxz,Mantry:2009qz} 
one first integrates out the top quark
at a scale $\mu_t\approxeq m_t$ and subsequently matches from the \QCD{} effective Lagrangian onto 
\SCET{} at $\mu_h \approxeq m_H$.
Running between $\mu_h$ and $\mu_t$ allows one to sum logarithms of $m_t/m_H$, but one neglects power of $m_H/m_t$. 
\subsubsection{$C^t(m_t^2,\mu_t^2)$}
For a heavy top quark the effective Lagrangian for the production of a top quark is
given by,
\begin{equation}
  \label{eq:Leff}
        {\cal L}_{\rm eff}=C^t(m_t^2,\mu_t^2)\, \frac{H}{v} \, \frac{\alpha_s(\mu_t^2)}{12\pi}\, G_{\mu\nu\,a}G^{\mu\nu}_a  \, ,
\end{equation}
where $v\approx 246$~GeV is the Higgs boson vacuum expectation value. 
The hard matching scale $\mu_t$ at which the Wilson coefficient can be computed perturbatively is of order $m_t$.
The short distance coefficient $C^t(m_t^2,\mu^2)$ obeys the RGE,
\beq \label{eq:Ctevol}
\frac{d}{d\ln \mu} C^t(m_t^2,\mu^2)=\gamma^t(\alpha_s)  \, C^t(m_t^2,\mu^2),\;\;\;
\gamma^t(\alpha_s)=\alpha_s^2 \, \frac{d }{d \alpha_s} \Big(\frac{\beta(\alpha_s)}{\alpha_s^2}\Big)\, .
\eeq
The expressions for the short-distance coefficient $C^t(m_t^2,\mu_t^2)$ at \NNLO{} is,
\begin{equation}
C^t(m_t^2,\mu_t^2) = 1 + \frac{\alpha_s(\mu_t)}{4 \pi} \, C^t_1
 + \left(\frac{\alpha_s(\mu_t)}{4 \pi}\right)^2 \, C^t_2(m_t^2,\mu_t^2) + \ldots \, ,
\end{equation}
where (c.f. Eq. (12) of Ref.~\cite{Ahrens:2009cxz}),
\begin{eqnarray}
C^t_1 &=& 5 C_A - 3C_F \nn \\
C^t_2(m_t^2,\mu_t^2) &=& \frac{27}{2}C_F^2 + \left( 11\ln\frac{m_t^2}{\mu_t^2}
 -\frac{100}{3} \right) C_F C_A - \left(7 \ln \frac{m_t^2}{\mu_t^2}
 -\frac{1063}{36} \right) C_A^2 \nn \\
 && \quad -\frac{4}{3} C_F T_F - \frac{5}{6} C_A T_F
  -\left(8\ln\frac{m_t^2}{\mu_t^2} + 5 \right) C_F T_F n_f
  -\frac{47}{9}C_A T_F n_f\, .
\end{eqnarray}
The evolution of these coefficients to the resummation scale $\mu$ is described in Appendix A of Ref.~\cite{Becher:2012qa}.
The solution to the evolution equation Eq.~(\ref{eq:Ctevol}) for $C^t$ at scale $\mu$ is,
\begin{equation}
C^t(m_t^2,\mu^2) = \frac{\beta(\alpha_s(\mu))}{\alpha_s^2(\mu)} \, \frac{\alpha_s^2(\mu_t)}{\beta(\alpha_s(\mu_t))} \,
C^t(m_t^2,\mu_t^2)\, .
\end{equation}
The result at NNLO for the square of the coefficient function is,
\begin{eqnarray}
\label{eq:Ctexp}
  \big[C^t(m_t^2,\mu^2)\big]^2 &=& 1+\Big(\as\Big) \Big[2 C^t_1+2 (r_t-1) \frac{\beta_1}{\beta_0}\Big]\nonumber \\
  &+&\Big(\as\Big)^2 \Big[(C^t_1)^2+2 C^t_2(m_t^2,\mu_t^2) 
+\frac{(2 \beta_2 \beta_0+\beta_1^2)}{\beta_0^2} (r_t-1)^2 \nonumber \\
&+&2 \frac{(2 \beta_2 \beta_0+2 \beta_1 \beta_0 C^t_1-\beta_1^2)}{\beta_0^2} (r_t-1)\Big] \, ,
\end{eqnarray}
where $r_t=\alpha_s(\mu)/\alpha_s(\mu_t)$.
This extends the NLO result in Eq.~(2) of Ref.~\cite{Becher:2012qa}.

\subsubsection{$C^S(-q^2,\mu_h)$}
$C^S$ is the Wilson coefficient matching the two gluon operator in Eq.~(\ref{eq:Leff}) to an 
operator in \SCET{}
in which all the hard modes have been integrated out. 
The result for the matching 
coefficient $C^S$ from Eqs.(16) and (17) of Ref.~\cite{Ahrens:2009cxz}.
It is given by,
\begin{equation}
  \label{eq:CSdef}
  C^S(-q^2,\mu_h^2) = 1 + \sum_{n=1}^\infty\,C^S_{n}(L)
   \left( \frac{\alpha_s(\mu_h^2)}{4\pi} \right)^n \,.
\end{equation}
The coefficient $C^S$ obeys the renormalization equation,
\begin{equation}\label{eq:CSevol}
   \frac{d}{d\ln\mu}\,C^S(-q^2-i\epsilon,\mu^2)
   = \left[ \Gamma_{\rm cusp}^A(\alpha_s)\,
   \ln\frac{-q^2-i\epsilon}{\mu^2} + \gamma^S(\alpha_s) \right] 
   C^S(-q^2-i\epsilon,\mu^2) \, ,
\end{equation}
with $L=\ln(-q^2-\img 0)/\mu_h^2$ and $\gamma^S$ is given in Eq~(\ref{eq:gammaS}).

The logarithmic terms are determined by Eq.~(\ref{eq:CSevol}).
The full results for the one- and two-loop coefficients are,
\begin{eqnarray}
   C^S_1 &=& C_A \Big( -L^2 + \frac{\pi^2}{6} \Big) , \\
   C^S_2 &=& C_A^2 \Big[ \frac{L^4}{2} + \frac{11}{9}\,L^3
    + \Big( -\frac{67}{9} + \frac{\pi^2}{6} \Big) L^2
    + \Big( \frac{80}{27} - \frac{11\pi^2}{9} - 2\zeta_3 \Big) L \nn\\
   &+&  \frac{5105}{162} + \frac{67\pi^2}{36}
    + \frac{\pi^4}{72} - \frac{143}{9}\,\zeta_3 \Big]
    + C_F T_F n_f \Big( 4L - \frac{67}{3} + 16\zeta_3 \Big) \nn \\
  &+& C_A T_F n_f \Big[ -\frac{4}{9}\,L^3
   + \frac{20}{9}\,L^2
   + \Big( \frac{104}{27} + \frac{4\pi^2}{9} \Big) L
   - \frac{1832}{81} - \frac{5\pi^2}{9} - \frac{92}{9}\,\zeta_3
   \Big] \, .
\end{eqnarray}
The full result for the renormalization group invariant hard function in the two-step scheme is,
\begin{eqnarray}\label{eq:Hbardef}
   \bar H(m_t,m_H,\pTveto)
   &=& \left( \frac{\alpha_s(\mu)}{\alpha_s(\pTveto)} \right)^2 (C^t(m_t^2,\mu))^2
   \left| C^S(-m_H^2,\mu) \right|^2 \nn \\
   &\times & \left( \frac{m_H}{\pTveto} \right)^{-2F_{gg}(\pTveto,\mu)}\,
    e^{2 h^A(\pTveto,\mu)} \, .
\end{eqnarray}
The $\mu$-independence of this hard function can be used to constrain $\gamma^S$,
\beq \label{mu_independence_for_Hbar}
\frac{d}{d \ln \mu} \bar{H}(m_t,m_H,\pTveto)=0\, .
\eeq
Using Eqs.~(\ref{betafunction},\ref{eq:Ctevol},\ref{eq:CSevol},\ref{eq:Fggevol},\ref{eq:hevol})
we can derive the relation between the collinear anomalous dimensions,
\beq 
2\gamma^g(\alpha_s)=\gamma^t(\alpha_s)+\gamma^S(\alpha_s)+\beta(\alpha_s)/\alpha_s\, .
\eeq

This relation could be cast in a more transparent form by noting that the quantity
$(\alpha_s C^S)$ obeys a similar evolution equation to Eq.~(\ref{eq:CSevol}),
\begin{eqnarray}
   && \frac{d}{d\ln\mu}\,\left[ \alpha_s(\mu) C^S(-m_H^2-i\epsilon,\mu^2) \right]
   = \nn \\ && \alpha_s(\mu) \left[ \Gamma_{\rm cusp}^A(\alpha_s)\,
   \ln\frac{-m_H^2-i\epsilon}{\mu^2} + \gamma^S(\alpha_s) \right]
   C^S(-m_H^2-i\epsilon,\mu^2)
   + \beta(\alpha_s) C^S(-m_H^2-i\epsilon,\mu^2) \nn \\
   &=& \left[ \Gamma_{\rm cusp}^A(\alpha_s)\,
   \ln\frac{-m_H^2-i\epsilon}{\mu^2} + \gamma^{S^\prime}(\alpha_s) \right]
   \left[ \alpha_s(\mu) C^S(-m_H^2-i\epsilon,\mu^2) \right]\, ,
\end{eqnarray}
but with anomalous dimension
$\gamma^{S^\prime}(\alpha_s) = \gamma^S(\alpha_s) + \beta(\alpha_s)/\alpha_s$.
We then have the relation $2\gamma^g(\alpha_s)=\gamma^t(\alpha_s)+\gamma^{S^\prime}(\alpha_s)$.
This indicates that after the second matching, the evolution down to a lower scale satisfies
the same renormalization equation in both the one-step and the two-step schemes.

\subsection{Assessment of the two schemes for the Higgs hard function}
\label{sec:Hschemes}

The two schemes for the calculation of the hard function have application in jet veto
resummation but also in the resummation of the Higgs boson transverse momentum. A complete
discussion of the error budget for Higgs boson production including scale dependence, parton
distribution dependence, the influence of loops of $b$-quarks and electroweak corrections is
beyond the scope of this paper. Here we shall simply compare and contrast the one-step and
the two-step scheme, in the Higgs on shell region where $m_H^2\approx m_t^2/2$.

It is easy to check the internal consistency of the two schemes in the limit where we drop terms of order $q^2/(4 m_t^2)$.
Setting $z=0$ in Eq.~(\ref{eq:CH}) and evaluating
all coefficient functions at a common scale $\mu$, we have that,
\begin{equation}
\alpha_s(\mu) \, C^t(m_t^2,\mu^2) \, C^S(-q^2,\mu^2)=C^{H}(m_t^2,q^2,\mu^2)_{z=0}+O(\alpha_s^4)\, .
\end{equation}
We can test this equivalence numerically.  We start by fixing $\mu^2=q^2$ and
consider the quantities that enter the calculation of the cross-section, i.e.
the square of the absolute values.  In the two-step scheme we have,
\begin{eqnarray}
|C^t(m_t^2,q^2)|^2 &=& 1 + 0.1957 + 0.0204 \, , \nn \\
|C^s(-q^2,q^2)|^2 &=& 1 + 0.6146 + 0.2155 \, ,
\end{eqnarray}
where the second and third terms represent the ${\mathcal O}(\alpha_s)$
and  ${\mathcal O}(\alpha_s^2)$ terms respectively, evaluated using
$\alpha_s(q^2) = 0.1118$.
In the one-step case we get,
\begin{eqnarray}
|C^H_{z=0}(m_t^2,q^2,q^2)/\alpha_s(q)|^2 = 1 + 0.8104 + 0.3563\, .
\end{eqnarray}
Performing a strict fixed-order truncation of the product of the two-step
result we  have,
\begin{equation}
\left[ |C^t(m_t^2,q^2)|^2 |C^s(-q^2,q^2)|^2 \right]_\text{expanded} = 1 + 0.8104 + 0.3563\, ,
\end{equation}
which is in perfect agreement with the one-step case.  This indicates that
the numerical implementation of the two procedures is correct.  If we instead evaluate the
product after the individual expansions have been performed, a choice
of equal formal accuracy, we have,
\begin{equation}
|C^t(m_t^2,q^2)|^2_\text{expanded} \, |C^s(-q^2,q^2)|^2_\text{expanded}
 = 1 + 0.9306 + 0.2953\, .
\end{equation}
This results in a significant difference.  We therefore work with with the strict fixed-order
truncation throughout this paper.

We now restore the $z$-dependence
in $F_1^H$ and $F_2^H$ in Eq.~(\ref{eq:CH}), but still keep $z=0$ in the overall factor
$F_0^H(z)$. We then find that the ratio of the one-step to the two-step becomes
$1.0028$ at \NLO{} and $1.0053$ at \NNLO{}, i.e. these corrections are very small.
Now we allow the matching scale for the top quark, $\mu_t$ to take its natural value,
$\mu_t = m_t$ and find one/two-step ratios of $1.0054$ at \NLO{} and $1.0073$ at \NNLO{},
again a small effect.
Finally, we reinstate the hard evolution down to the resummation scale and find that
the ratio of the one-step to the two-step (at $p_T^\text{veto}=25$~GeV) is
$1.0177$ at \NLO{} and $1.0125$ at \NNLO{}.  The cumulative effect at this point is noticeable
but still small.  However, we note that we have so far kept $z=0$ in the overall factor
$F_0^H(z)$.  The one-step procedure is recovered by re-instating $F_0^H(z)$.
This implies that, in order to obtain the level of agreement quoted above between the
two schemes, the overall factor of $F_0^H(z)$ must also be applied to give a modified version of the two-step scheme.
Neglecting this step would result in a significant difference, since $|F_0^H(z)|^2 = 1.0653$ see Eq.(\ref{massrescaling}). 

Our overall conclusion on the two schemes is in line with the known result that Higgs boson
production has substantial corrections. Accounting for the most important mass effects by rescaling
the two-step result by the exact result at leading order, the one-step procedure gives a larger result than
the two-step procedure for $\pTveto=25$~GeV at the level of $1.3\%$.
Any substantial difference between the two methods beyond this level is most likely due to uncontrolled
higher order effects.

\bibliography{unified}
\bibliographystyle{JHEP}

\end{document}